\documentclass[
11pt, 
oneside, 
english, 
singlespacing, 
headsepline, 
]{MastersDoctoralThesis} 

\usepackage[utf8]{inputenc} 
\usepackage[T1]{fontenc} 

\usepackage{mathpazo} 
\usepackage{amsmath}
\usepackage{amssymb}
\usepackage[backend=bibtex,style=numeric,natbib=true, sorting=none]{biblatex} 
\usepackage{array}
\usepackage{siunitx}
\DeclareSIUnit{\rad}{rad}
\usepackage[autostyle=true]{csquotes} 
\usepackage{subcaption}
\usepackage{textgreek} 
\usepackage{multicol} 
\usepackage{fixltx2e} 
\usepackage{listings} 
\usepackage{chemformula} 
\usepackage{placeins}
\usepackage{upgreek}
\usepackage{floatrow} 
\newfloatcommand{capbtabbox}{table}[][\FBwidth]
\usepackage[bottom]{footmisc}
\usepackage{float}
\definecolor{light-gray}{gray}{0.96} 
\lstdefinestyle{customc}{
	belowcaptionskip=1\baselineskip,
	breaklines=true,
	frame=lines,
	language=Python,
	basicstyle=\footnotesize,
	keywordstyle=\bfseries\color{green!40!black},
	commentstyle=\itshape\color{purple!40!black},
	backgroundcolor=\color{light-gray},
	identifierstyle=\color{black},
	numbers=left,
	stringstyle=\color{orange},
		escapechar=\&
}
\usepackage{xcolor}
\usepackage{colortbl}
\usepackage{multirow }
\usepackage{rotating }
\usepackage{graphicx }
\makeatletter
\renewcommand\float@endH{\@endfloatbox\vskip\intextsep
	\if@flstyle\setbox\@currbox\float@makebox\columnwidth\fi
	\box\@currbox\vskip\intextsep\relax\@doendpe}
\makeatother

\thesistitle{Deployment of the readout electronics for the BESIII Cylindrical GEM Inner Tracker} 
\supervisor{Prof. Michela  \textsc{Greco}} 
\degree{PhD thesis} 
\author{Alberto \textsc{Bortone}} 

\subject{Physics} 
\university{\href{https://www.unito.it/}{Università degli studi di Torino}} 
\department{\href{https://fisica.campusnet.unito.it/do/home.pl}{Dipartimento di Fisica}} 
\group{\href{http://researchgroup.university.com}{INFN Torino}} 
\faculty{\href{http://faculty.university.com}{Faculty Name}} 

\AtBeginDocument{
\hypersetup{pdftitle=\ttitle} 
\hypersetup{pdfauthor=\authorname} 
\hypersetup{pdfkeywords=\keywordnames} 
\hypersetup{colorlinks=false}
}
\graphicspath{{./img/}}
\begin{document}

\frontmatter 

\pagestyle{plain} 

\thispagestyle{empty}
\begin{center}
\begin{large}
Universit\`a degli Studi di Torino \\

{\bf Dottorato di Ricerca in Fisica} \\
\end{large}
\end{center}
\hrulefill

\vspace{2cm}
\begin{figure}[h!]
    \centering
    \includegraphics[width=0.28\linewidth]{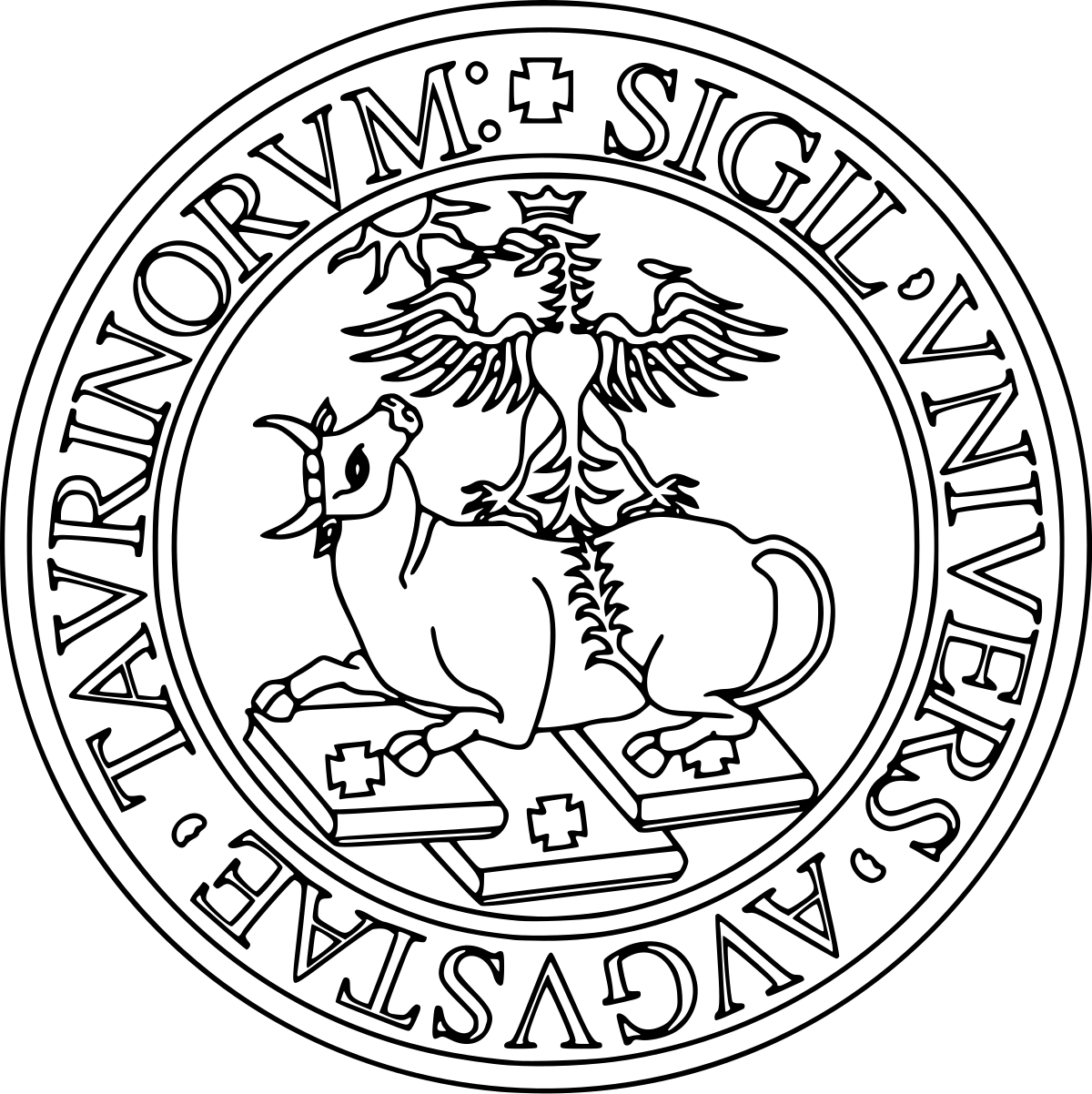}
\end{figure}
\vspace{3cm}
\begin{center}
\huge{\bf\ttitle}

\end{center}

\vspace{2cm}

\large{\bf \authorname}
\newpage
\begin{center}
\begin{large}
Universit\`a degli Studi di Torino \\
{\bf Dipartimento di Fisica}

\end{large}
\end{center}
\hrulefill
\begin{center}
\begin{large}
{\bf Dottorato di Ricerca in Fisica}\\
{\bf Ciclo XXXIV}
\end{large}
\end{center}

\vspace{2cm}
\begin{center}
    
\huge{\bf \ttitle}
\end{center}

\vspace{5cm}
\large{\bf\authorname}
\vspace{5cm} \newline
\large{\bf Tutor:  Prof. Michela Greco}
\vspace{0.5cm}\newline
\large{\bf PhD coordinator: Prof. Paolo Gambino}
\vspace{0.5cm}\newline
\large{\bf Academic years: 2018-2021}
\vspace{0.5cm}\newline
\large{\bf SSD: FIS/01}

\begin{abstract}
\addchaptertocentry{\abstractname} 
The Beijing Spectrometer (BESIII) is an experimental setup located at the Beijing Electron Positron Collider (BEPCII).
The recently approved ten-year extension of data acquisition for BESIII prompted an upgrade program for both the BEPCII collider that houses the experiment and some of the sub-detectors that make up the spectrometer. The current inner drift chamber is suffering from aging and it has been proposed to replace it with a detector based on the Cylindrical Gas Electron Multiplier (CGEM) technology.

The CGEM Inner Tracker (CGEM-IT) consists of three coaxial layers of triple GEM. The lightweight tracker is designed to recover efficiency and improve z-determination and secondary vertex position reconstruction, resulting in a better performance and longer lifetime.\\ 
A novel system has been deeloped  for the readout of the CGEM detector, including a new ASIC, called TIGER (Torino Integrated GEM Electronics for Readout), designed to amplify and digitize the CGEM signals. The data output from TIGER are collected and processed by a first FPGA-based module, GEM Read Out Card, which is responsible for configuring and controlling the front-end ASICs. A second FPGA-based module, GEM Data Concentrator (GEM DC), builds the event packets and transmits them to the BESIII data acquisition system.\\

In this thesis I will describe the entire custom electronics chain designed for the CGEM-IT to achieve the desired performance. The work I did as part of my PhD involved all steps of the acquisition chain: characterization and testing of the on-detector electronics, debugging and optimization of the off-detector electronics, and development of the acquisition and analysis software to test the detectors and the electronics in different configurations.\\
The structure of my work is presented along with the results of  the measurements performed on the detector and the analysis of the data obtained with the detection of cosmic rays and under test beam.
\end{abstract}

\tableofcontents 

%
%
%
%
%




\mainmatter 

\pagestyle{thesis} 

\chapter{Introduction}
A new GEM cylindrical inner tracker (CGEM-IT) was developed as part of the BESIII experiment.\\ Such a detector requires a custom readout chain to meet the requirements and constraints of the experiment. The work presented here addresses the development of this readout chain and of the tools used to study and characterize the electronics and the detector.\\ In this chapter, the BESIII experiment is described, as well as the operating principles of the GEM technology and the CGEM-IT design are presented. A brief introduction to the physical processes underlying the GEM detectors can be found  in appendix \ref{working_principles}.
\section{BESIII}
The "Beijing Electron Positron Collider" (BEPCII \cite{bepcII2009}) is a double ring e$^+$ e$^-$ collider as well as a synchrotron radiation source of the Institute of High Energy Physics (IHEP) in the People's Republic of China. BEPCII is designed as a tau-charm factory, with an energy in the center of mass between 2 and \SI{4.95}{\giga \electronvolt} and a luminosity up to \SI{e33}{\cm^{-2}  \second ^{-1}}. BEPCII was recently upgraded by introducing a top-up beam injection mode, that allows to inject new particles in the storage rings at a higher circulating current, effectively increasing the integrated luminosity by \SI{30}{\percent}. This upgrade required consistent improvement of the feedback systems, accelerator steering and BESIII trigger system. \\
A \SI{202}{\meter} long LINAC injects particles with an energy of  1-\SI{2.47}{\giga \electronvolt} in two \SI{237.5}{\meter} long storage rings, where the beams can be stored for ~\SI{2.7}{\hour}. The particle bunches collide at the interaction point with a horizontal crossing angle of \SI{11}{\milli \radian} and a bunch spacing of \SI{8}{\nano \second}. Each ring holds 93 particle bunches with a beam current of \SI{910}{\milli \ampere}. \\
At the BEPCII interaction point is the "Beijing Spectrometer III" (BESIII \cite{Ablikim2010}). BESIII is a high energy physics experiment that has been running successfully since 2008 and provides an extensive physics program for study. In particular, the high J/$\Psi$ production rate makes it suitable for the study of exotic hadrons consisting of light quarks and gluons, which are essential for understanding the nature of the strong interaction. The experiment is expected to continue taking data until at least 2030, taking advantage of current and future upgrades to the accelerator and detectors.\\
\subsection{BESIII detector}
\label{sec:BESIII_detector}
\begin{figure}
    \centering
    \includegraphics[width=0.9\textwidth]{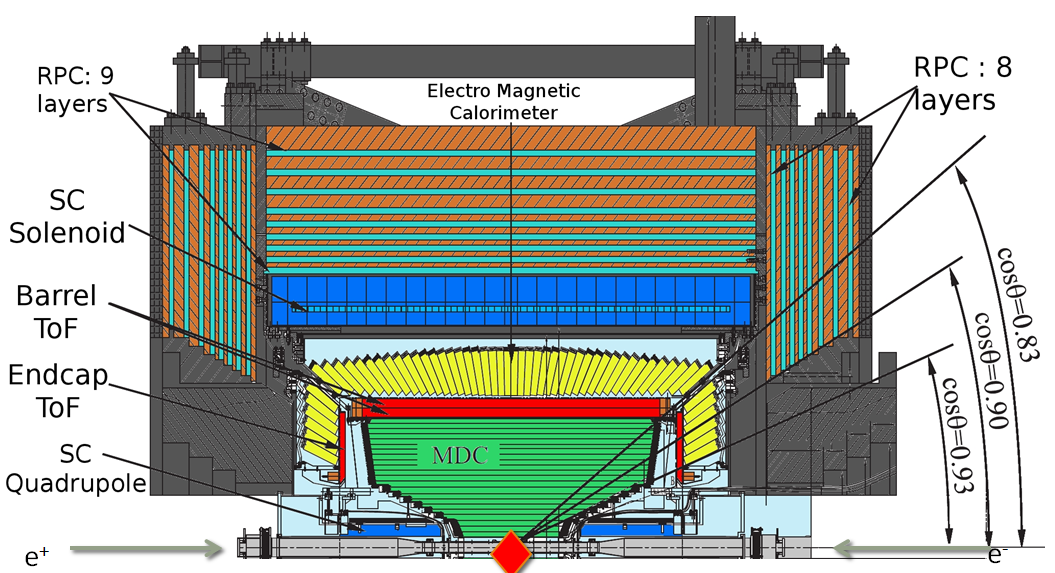}
    \caption{Cutaway drawing of the BESIII detector. The drawing shows the top half of the detector and the various sub-detectors described in section \ref{sec:BESIII_detector}. The red square indicates the interaction point and the arrows show the angular coverage in different directions.}
    \label{fig:BESIII_det}
\end{figure}

BESIII measures the trajectories and momenta of charged and neutral particles using a series of sub-detectors and a solenoid magnet that generates a \SI{1}{\tesla} magnetic field.  The detector is located at the interaction point of the two accumulation rings and has a geometrical acceptance of \SI{93}{\percent}.\\
BESIII can be divided into four sub-detectors. The structure of the detector is depicted in figure \ref{fig:BESIII_det}:
    \subsubsection*{Multi-layer Drift Chamber (MDC)} The MDC is the main BESIII tracking detector and provides the accurate measurement of the position and  momentum of the charged particles produced in the collisions. The MDC is a cylindrical chamber with 43 sense wire layers, whose first eight compose the inner MDC. It operates at \SI{2200}{\volt} with a \ch{He} / \ch{C3H8} 60:40 gas mixture. The average gas gain is \num{3e4}.  The detector provides a spatial resolution of $\upsigma_{r\phi} \approx $ \SI{130}{\micro \meter} and $\upsigma_{z} \approx $ \SI{2}{\milli \meter}, a momentum resolution of $\upsigma_{p}/p \approx $ \SI{0.5}{\percent} at \SI{1}{\giga \electronvolt / c}.
    \subsubsection*{Time Of Flight (TOF)} 
     The TOF detector consists of two different detector architectures. In the barrel region, it consists of two layers of 88 plastic scintillation bars (BC404), read by fine-mesh PMTs, while 36 overlapping trapezoidal-shaped Multi-gap Resistive Plate Chambers are mounted in each end-cap. The TOF detector extracts the timing information of the crossing particle with a resolution of  \SI{70}{\pico \second} in the barrel region and \SI{60}{\pico \second} in the endcap. Its particle identification capability is determined by the time-of-flight difference for particles of different species and the time resolution of the detector.
    \subsubsection*{Electromagnetic Calorimeter (EMC)}
    The calorimeter consists of one barrel and two end-cap sections covering \SI{93}{\percent} of 4π. 
    It consists of 44 rings of 120 crystals in the barrel and six layers of crystals in the end-caps. The entire calorimeter consists of 6272 \ch{CsI(Tl)}  crystals with a total weight of \SI{24}{\tonne}. It provides the energy information with a resolution of \SI{2.5}{\percent} in the barrel and \SI{5}{\percent} in the end-caps.
    \subsubsection*{Muon Chamber}
    The BESIII muon chamber is a gaseous detector based on Resistive Plate Chambers (RPCs). It consists of nine layers in the barrels and eight layers in the end-caps. It allows to identify muons and separate them from charged pions, other hadrons and background radiation. The chosen gas mixture is Ar (\SI{50}{\percent}) - F134a (\SI{42}{\percent}) - HC(CH$_3$)$_3$ (\SI{8}{\percent}).\\
    
\subsection{BESIII data acquisition and trigger system}
\begin{figure}
\centering
\includegraphics[width=0.9\textwidth]{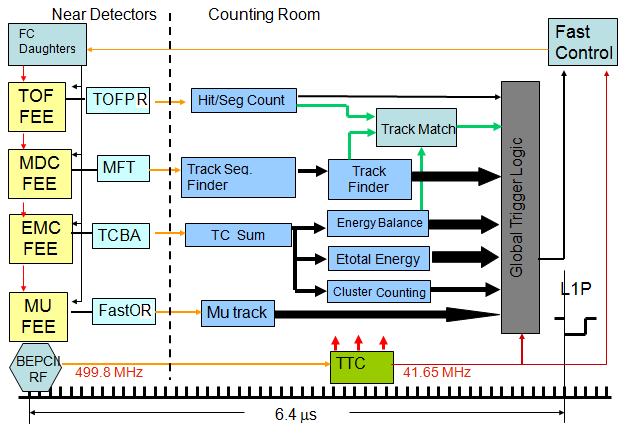}
\caption{Block diagram of L1 trigger system. The data acquired by the on-detector electronics are first processed at the sub-detector level and then combined by the global trigger logic to generate the trigger signal to be distributed. The trigger generation takes \SI{6.4}{\micro \second}, while the latency is stretched to \SI{8.6}{\micro \second}.}
\label{fig:besIII_trigger}
\end{figure}

The purpose of the trigger system is to select physics events of interest from the enormous background and suppress the background to a level that the Data Acquisition (DAQ) system can sustain. The BESIII DAQ system is safely designed for a maximum throughput of \SI{4}{\kilo \hertz} of events with an average event size of 12 kbytes.\\
BEPCII collides electron and positron bunches at a frequency of \SI{125}{\mega \hertz}. The main background events in BESIII are caused by lost beam particles and their interaction with the detector. The rate of background events is estimated to be about \SI{13}{\mega \hertz} \cite{Ablikim2010}. In comparison, the signal rate at the $J/\uppsi$ resonance is about \SI{2}{\kilo \hertz}. Thus, the task of the trigger system is to suppress the background by more than three orders of magnitude while maintaining high efficiency for signal events. Careful monitoring of trigger efficiency is important to avoid losing events due to inefficient triggers \cite{Trigger_eff}.\\
In the design of the trigger system, one of the main goals is to keep the total dead time as low as possible, which corresponds to  an acceptable loss of luminosity. To achieve this, the system has been developed on two levels: an hardware level (L1, synchronized with the \SI{41.65}{\mega \hertz} system clock), that reduces much of the background without causing any dead time, and a software level (L2), that filters out even more of the background using analysis algorithms. Due to the small time between the interactions (\SI{8}{\nano \second}) with respect to the signal processing time ($\sim$ \SI{1.5}{\micro \second} for the calorimeter), a pipelined approach must be used for the L1 trigger  \cite{trigger}.\\
To generate the L1 trigger, the signals from sub-detectors are first processed by four trigger sub-systems and then transmitted via optical links to the Global Trigger Logic crates, where they are analyzed to determine hit patterns, track segmentation, clusters, and total energy (figure \ref{fig:besIII_trigger}). The L1 trigger has a latency of \SI{8.6}{\micro \second}, an average frequency of \SI{4}{\kilo \hertz},  and a dead time of \SI{3}{\micro \second}.\\
The L1 trigger signals are transmitted along with the system clock and other commands to the Fast Control System and then distributed to the readout electronics of each sub-detector via optical links. A more detailed explanation of the Fast Control System can be found in \ref{backend}.\\
The DAQ system is based on VME crates and a commercial server farm. It is designed to manage large amounts of data coming in from the front-end electronics system and to record valid data in the storage. It uses  multi-level  buffering,  parallel  processing,  high-speed  optical data transfer and high-speed Ethernet technologies. When an L1 trigger is received, the event data stored in the buffers of the  VME crates are transmitted to the online computer farm, where they are combined into complete events and then filtered (L3 trigger).\\
\section{CGEM-IT}
\label{section:CGEM_it}
\begin{figure}
\centering
 \includegraphics[width=0.6\linewidth]{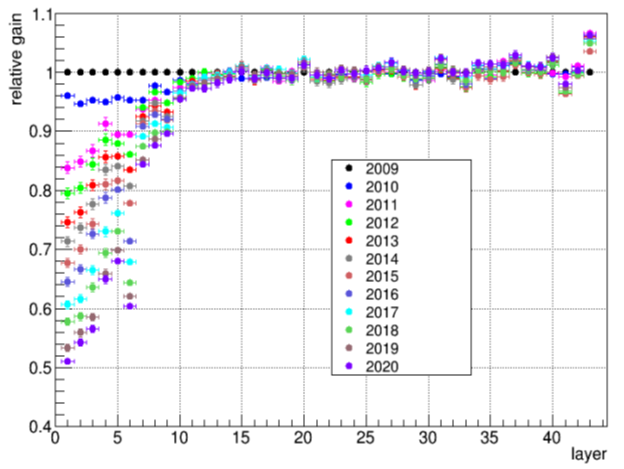}
  \caption{Gain loss of the MDC compared to 2009. The inner MDC consists of the eight innermost layers. Their gain loss per year is about \SI{4}{\percent}.}.
\label{fig:gain_loss}
\end{figure}

BESIII started to collect physics data in 2009 and will run at least until 2030.  Due to the high luminosity\footnote{From 2009 to 2019 the integrated charge of the innermost layers of the MDC is about \SI{170}{\milli \coulomb / \centi \meter.}}, the current MDC tracker is degrading with a gain loss of about \SI{4}{\percent} per year for the innermost layers, as shown in figure \ref{fig:gain_loss}. To partially compensate for the gain loss, the voltage was increased and about 0.2\%  water  vapor was added  to the MDC  gas  mixture  to  solve  the  cathode  aging problem.\\ 
Despite these precautions, the innermost layers are operated at a lower relative gain, resulting in a degradation of the detector performance. The gain compared to the status in 2009 has been reduced by up to \SI{60}{\percent}, which  affects the event reconstruction and degrades the detection efficiency in the innermost layers by up to \SI{50}{\percent} \cite{Dong_2016}. \\
An innovative solution using a lightweight tracker based on Cylindrical
Gas Electron Multiplier (CGEM) technology was proposed by the Italian collaboration at BESIII and boosted by an INFN (Italian National Institute for Nuclear Physics) - IHEP
(Chinese Institute of High Energy Physics) network\footnote{The project was funded by the European Commission RISE Project 645664-BESIIICGEM, RISE-MSCA-H2020-2014 and in the RISE Project  872901-FEST, H2020-MSCA-RISE-2019 and included INFN and IHEP as well as institutes from Mainz and Uppsala Universities.}.\\
\subsection{Gas Electron Multiplier (GEM)}
\begin{figure}[h!]
    \centering
    \includegraphics[width=0.6\textwidth]{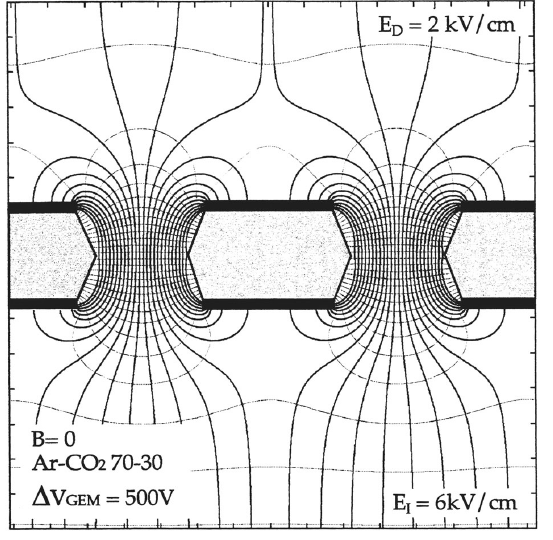}
    \caption{GEM electric field near the holes under typical operating conditions \cite{Bachmann:1999xc}.}
    \label{fig:gem_field}
    \end{figure}

Since the 1990s, the techniques of photolithography, selective etching, and laser processing, originally introduced in the semiconductor industry, have been applied to the development of new gas-filled devices collectively known as micropattern gas detectors. This technology allows the implementation of fine-scale details in the structure of the multiplication and readout stages. Over the years, various structures and designs emerged that exhibit improvements over conventional gas detectors in terms of spatial resolution, time resolution and tolerated rate per unit area.\\    
\begin{figure}[h!]
    \centering
    \includegraphics[width=0.6\textwidth]{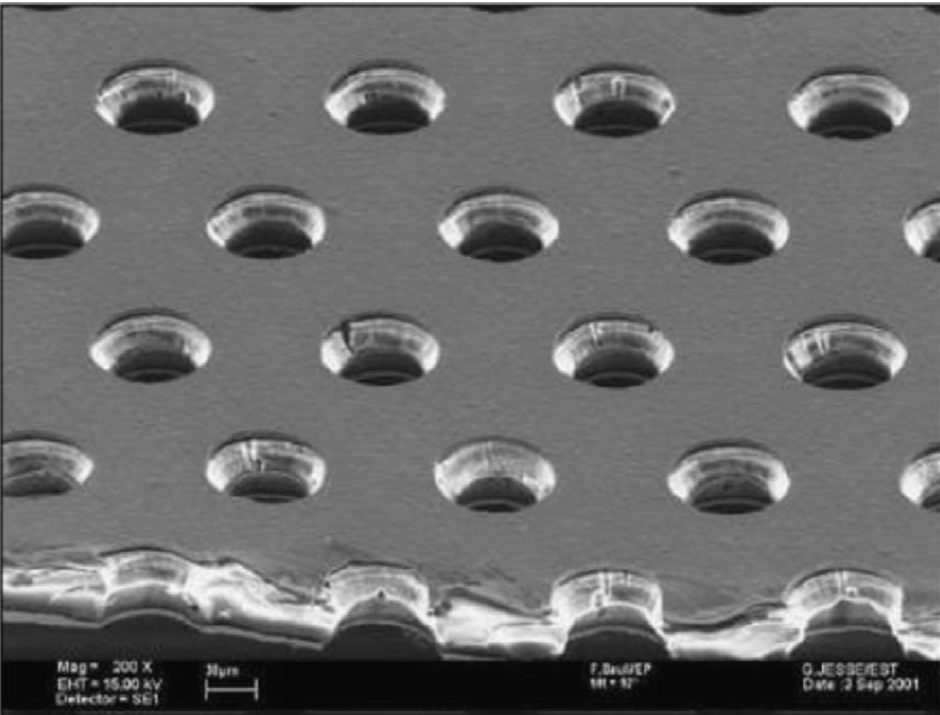}
    \caption{SEM image of a part of a GEM foil. The hole pitch is \SI{140}{\micro \meter} and the diameter is \SI{70}{\micro \meter} \cite{Altunbas:2002ds}.}
    \label{fig:GEM_foil}
    \end{figure}
Gas Electron Multipliers (GEM) \cite{SAULI2002, Sauli2016} were introduced in 1997 as part of the development frame of micropattern gas detectors. They overcome some limitations of earlier technologies and excel in the field of high rate radiation detectors for high energy physics.\\
Figure \ref{fig:GEM_foil} shows a  microscope image of a GEM electrode. A GEM foil consists of a thin ($\sim$ \SI{50}{\micro \meter}) polyimide sheet coated with metal on both sides and pierced with a high density of holes (\SIrange[]{50}{100}{\per \milli \meter \squared}). By applying a few hundred  volts (\SIrange[]{200}{300}{\volt}) between the copper foils, an electric field of \SIrange[]{50}{100}{\kilo \volt \per \centi \meter} is generated (figure \ref{fig:gem_field}) and the electrons accelerated into the holes are thereby multiplied. Most of the electrons generated during the multiplication transfer into the next region. The GEM foil  then acts as a charge preamplifier and almost completely preserves  the original ionization pattern. The high rate capability results from the high density of holes: each hole acts as an independent proportional counter, shielded from the neighbours, and the gain is not affected by the space charge up to very high radiation fluxes.\\
Since the multiplication takes place almost entirely in the high dipole field within the holes, the gain is only slightly affected by external fields and is therefore insensitive to the shape of the foil, allowing the foil to be bent in various shapes. The charge collection and readout plane, which is separate from the multiplying electrode, can be patterned as desired with strips, pads, or a combination of both.\\
The GEM manufacturing method is a refinement of double-sided printed circuit technology\footnote{Developed by Rui de Olivera at CERN \cite{Pinto_2009}.}. The metal-clad polymer is engraved on both sides with the desired hole pattern using photolithography and a controlled immersion in a polymer-specific solvent opens the
holes in the insulator. \\
\begin{figure}[h!]
\centering
\includegraphics[width=0.9\textwidth]{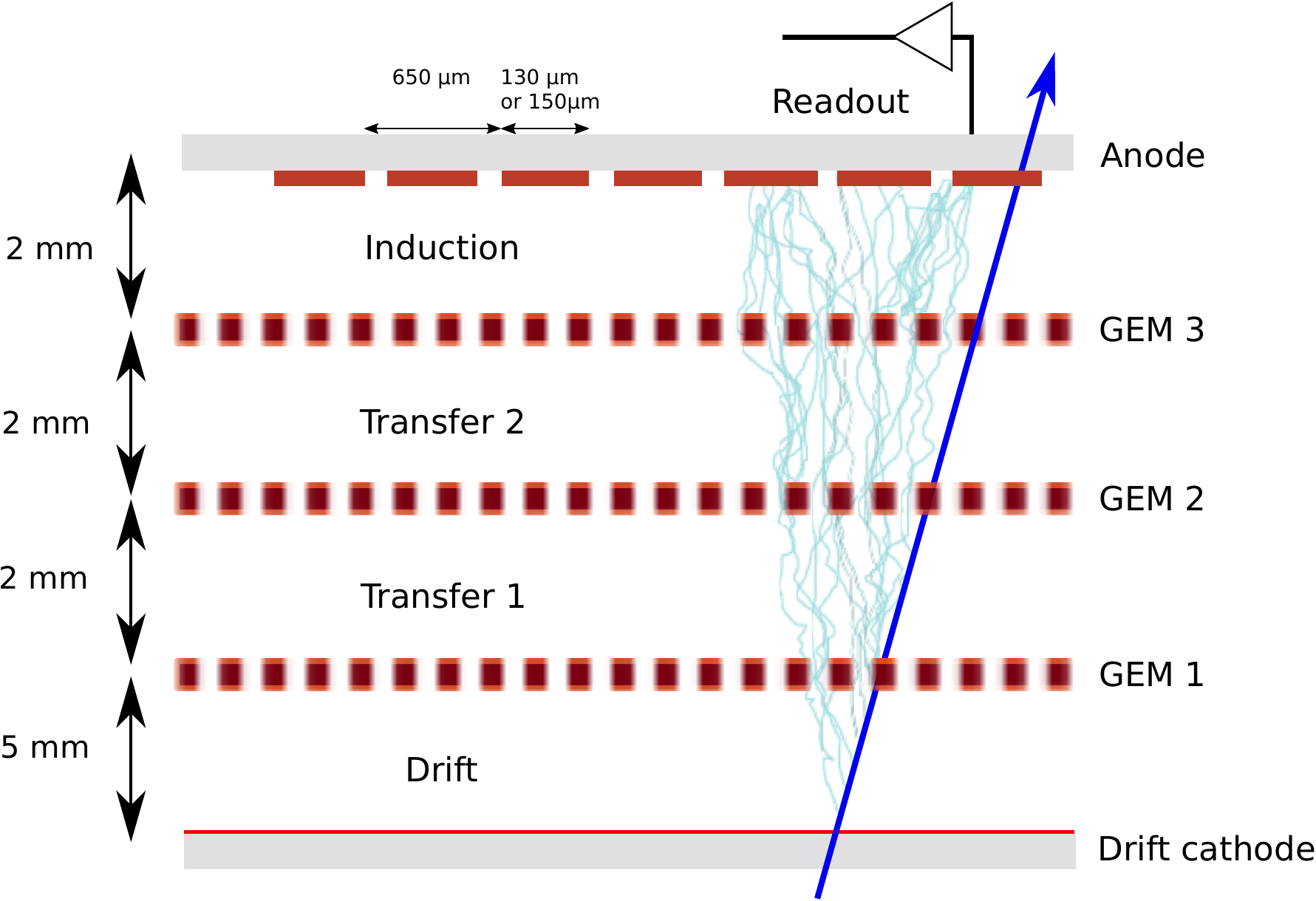}
\caption{Triple-GEM structure. The blue arrow represents a charged particle passing through the detector. The dimensions shown in the figure are those of the CGEM-IT (section \ref{section:CGEM_it}).}
\label{fig:triple_gem}
\end{figure}
\begin{figure}[h!]
\centering
\includegraphics[width=0.9\textwidth]{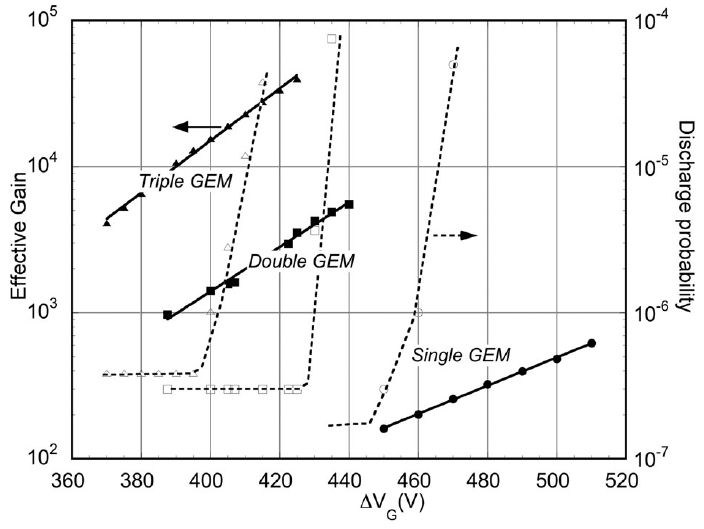}
\caption{Gain (solid lines) and discharge probability (dashed lines) of single, double, and triple-GEM detectors as a function of each GEM voltage \cite{Bachmann:2002}.}
\label{fig:gain_disc}
\end{figure}
Multiple GEM foils can be stacked in a cascade \cite{Buttner:1997qh, Bouclier:1996im}, to achieve more gain while maintaining the discharge risk low (figure \ref{fig:gain_disc}). For a triple-GEM, the discharge probability increases at a gain of $\sim 3 \times 10^4$, which is higher than the gain required for fast particle detection. Figure \ref{fig:triple_gem} shows an example of a triple-GEM structure. The electrons generated by ionization in the Drift region are multiplied by the three GEM layers and induce a signal at the readout plane. The secondary ionization in the other volumes of the gas is not multiplied by the three GEM layers and therefore provides a negligible contribution to the signal. \\
Depending on the requirements, such detectors can be operated with a wide range of gas fillings. The cross section of the gas for the ionization process directly affects the detector efficiency and the collected charge. In addition, the drift and diffusion velocities of the electrons are determined by the gas mixture, which greatly affects the space and time resolution. The gain also depends on the gas mixture: those with a low electron attachment coefficient and a high cross section for electron secondary ionization must be selected. Noble gasses (usually argon) are functional for these purposes. The addition of poly-atomic gasses, such as isobutane, limits the size of the avalanche and constrain the gain in the proportional mode, because they can absorb the UV photons emitted by the excited noble gas atoms. This is a property of most organic compounds from the family of hydrocarbons and alcohols, as well as some compounds such as Freon, \ch{CO2}, \ch{CF4}.\\
Thanks to the absence of thin anodes or wires, the GEM technology is robust to damage from polymerization to very high integrated fluxes.

\FloatBarrier
\subsection{CGEM-IT design}
\label{design}
\begin{figure}
\centering
    \begin{subfigure}{.6\textwidth}
  \centering
    \includegraphics[width=0.9\linewidth]{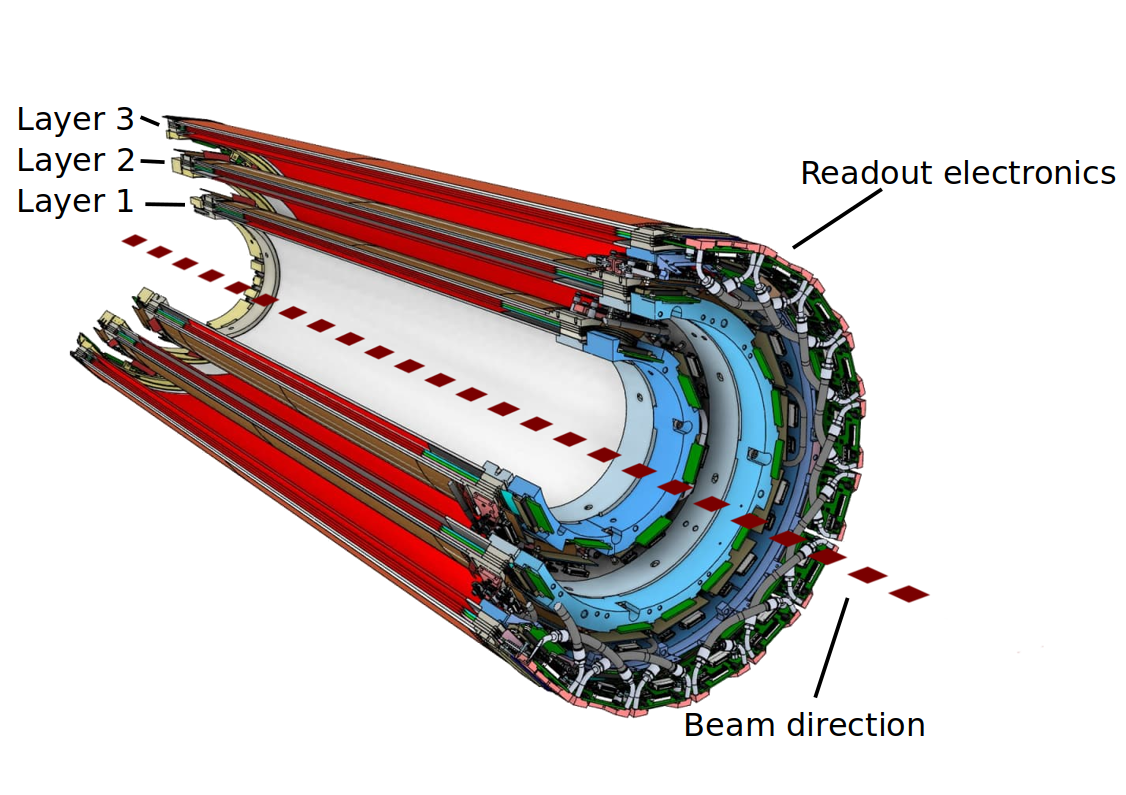}
        \end{subfigure}%
 \begin{subfigure}{.4\textwidth}
  \centering
      \includegraphics[width=0.8\linewidth]{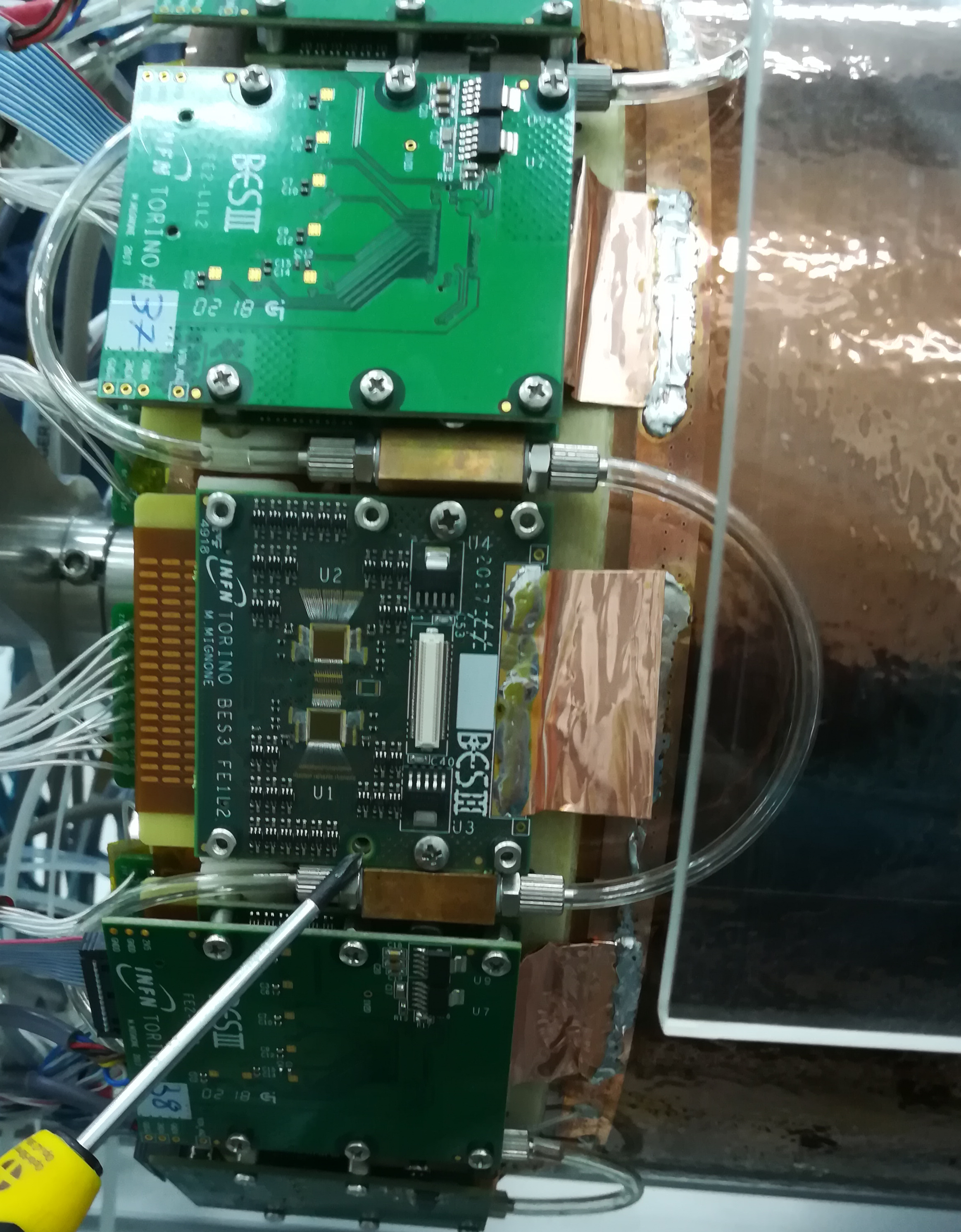}
   \end{subfigure}%
      \caption{Left: a cut-off of the technical sketch of the CGEM detector. The readout electronics is located on both sides of the detector. Right: front-end boards mounted on the detector. One board is removed to show the TIGER ASICs. A copper foil connects the ground plane of the front-end boards to the ground plane of the detector.}
         \label{fig:CGEM_cad}
\end{figure}
\begin{figure}
\begin{floatrow}
\ffigbox{%
  \includegraphics[width=1\linewidth]{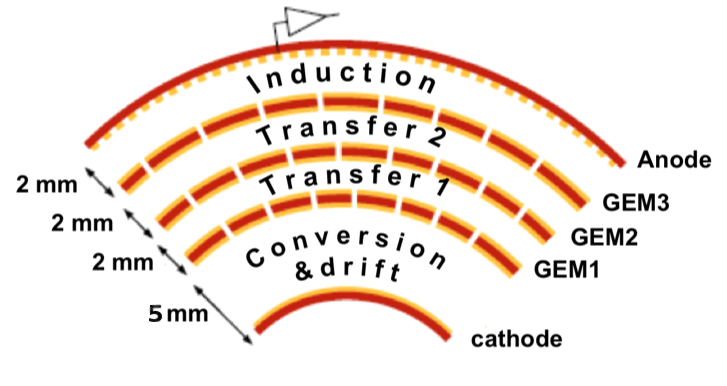}
  \caption{Triple-GEM layout of a single layer. Each detector layer consists of a cathode, an anode and three GEM multiplication foils.}
 \label{fig:CGEM_spac}%
}

\capbtabbox{%
  \begin{tabular}{cc|cc}
\multicolumn{2}{c|}{GEM voltages} & \multicolumn{2}{c}{Fields} \\ \hline               &                & Induction           & \SI{5}{\kilo \volt \per \centi \meter}       \\ \hline
G3             & \SI{275}{\volt}           &                 &          \\ \hline
               &                & Transfer 2      & \SI{3}{\kilo \volt \per \centi \meter}       \\ \hline
G2             & \SI{280}{\volt}              &                 &          \\ \hline
               &                & Transfer 1      & \SI{3}{\kilo \volt \per \centi \meter}        \\ \hline
G1             & \SI{280}{\volt}              &                 &          \\ \hline
               &                & Drift       & \SI{1.5}{\kilo \volt \per \centi \meter}     
\end{tabular}
}{%
\caption{ Voltage across the GEM foils and transfer fields. The total GEM voltage of \SI{835}{\volt} corresponds to a gain of about 13500.}
    \label{table:cgem_field}
    }
\end{floatrow}
\end{figure}
\begin{figure}
\centering
\begin{subfigure}{.4\textwidth}
  \centering
    \includegraphics[width=0.9\linewidth]{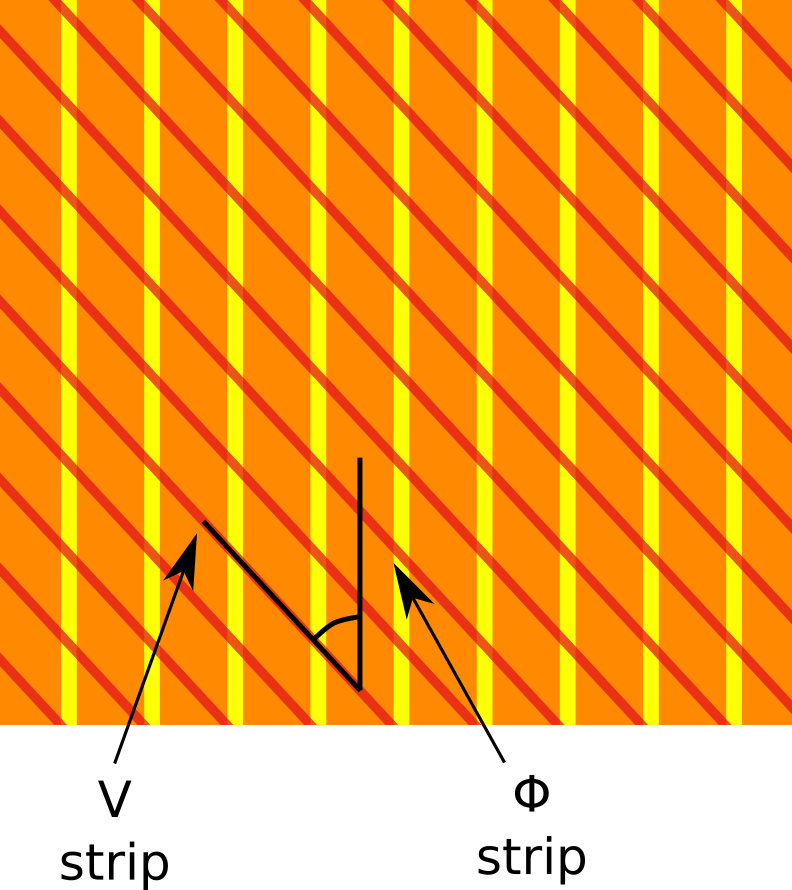}
   \caption{}
   \label{fig:strips_square}
   \end{subfigure}%
    \begin{subfigure}{.6\textwidth}
  \centering
    \includegraphics[width=0.9\linewidth]{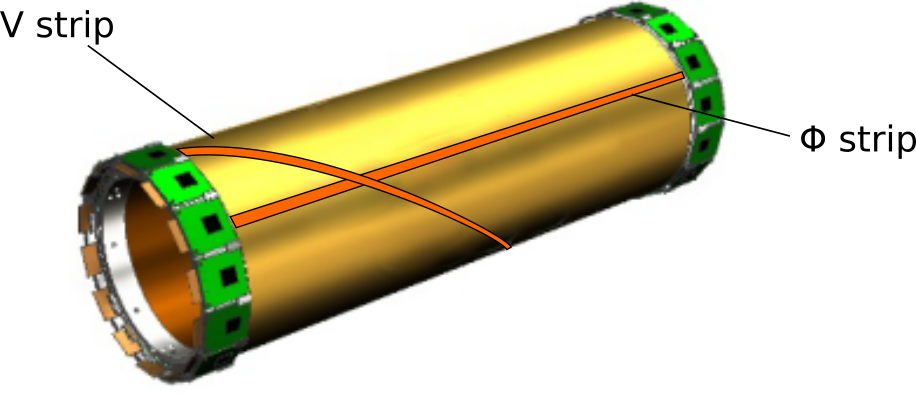}
      \caption{}
       \label{fig:strips_GEM}
        \end{subfigure}%
  \caption{Strip layout. (a): the anode structure for layer 1, with the $\Upphi$ strips in orange, the V strips in red and the \SI{46.7}{\degree} angle between them in black; (b): a sketch of the orientation of the anode strips on the cylindrical structure.}
 \label{fig:strips}
\end{figure}

\begin{table}
\begin{tabular}{cccccc}
Layer & Inner diameter (mm) & Outer diameter (mm) & Active area length (mm)\\ \hline
1     & 153.8               & 188.4               & 532 \\
2     & 242.8               & 243.4               & 690  \\
3     & 323.8               & 358.5               & 847  
\end{tabular}
\caption{Dimensions of CGEM-IT layers. The length of the active area is equivalent to the length of the $\Upphi$ strips.}
    \label{table:cgem_sizes}
\end{table}
The CGEM-IT consists of three coaxial layers of triple-GEM (figure \ref{fig:CGEM_cad} and table \ref{table:cgem_sizes}). Each cylindrical detector layer is independently assembled, it has an autonomous gas enclosure and can be operated stand-alone. The front-end electronics, electrode connections, and gas inlet and outlet are located on both sides of the cylinders.\\
A section through the triple-GEM cylindrical structure of the CGEM-IT is shown in figure \ref{fig:CGEM_spac}.\\
The CGEM-IT uses a mixture of argon and isobutane (90\%-10\%). The GEM voltages and field settings are listed in table \ref{table:cgem_field}. To set the electric fields in the gaps, seven different electrodes must be biased: the two copper planes of each GEM and the cathode (section \ref{HV}).
The readout circuit consists of a \SI{3}{\micro \meter} copper cladding over \SI{50}{\micro \meter} polyimide. To obtain a 2D readout, two arrays of strips are used: \SI{570}{\micro \meter} wide $\Upphi$ strips, that are parallel to the detector axis, and \SI{130}{\micro \meter} wide V strips, that have a stereo angle with respect to the $\Upphi$ strips (see figure \ref{fig:strips}). The stereo angle is maximized according to the active area of each layer: \SI{46.7}{\degree} for the innermost layer (layer 1), \SI{-31.0}{\degree} for the central layer (layer 2), and \SI{32.9}{\degree} for the outer layer (layer 3). The pitch, \SI{650}{\micro \meter}, is the same for both arrays and all layers.
\begin{table} 
\centering
    \begin{tabular}{|c|c|}\hline
    $r\phi$ resolution             &  \SI{130}{\micro \meter} \\ 
    z resolution             &  \SI{<1}{\milli \meter} \\ 
    $\sigma_{p_t}/p_t$         &  \SI{0.5}{\%} at 1 GeV/c\\ 
    Material budget        &  \SI{1.5}{\%} X$_0$ \\ 
    Maximum rate      &  \SI{e7}{\hertz \per \centi \meter^2} \\ \hline
    \end{tabular}
    \caption{Design goals of the CGEM-IT detector.}
    \label{table:cgem_perf}
\end{table}
The whole system consists of about 10,000 electronics channels.
Since it is a tracker, the first goal of the CGEM-IT is a precise position determination, with an expected resolution of \SI{130}{\micro \meter} in the \textit{$r\phi$} plane and \SI{300}{\micro \meter} on the \textit{z} coordinate along the beam direction \cite{marcello2018}. The main design goals are listed in table \ref{table:cgem_perf}. In particular, the CGEM-IT is expected to improve \textit{z} determination and secondary vertex position reconstruction compared to the MDC \cite{GEM_magnetic_field}. The position reconstruction benefits from the combination of two different algorithms: 
\begin{itemize}
    \item Charge Centroid extracts the average position by weighting the signal amplitude of the firing strips.
    \item $\upmu$TPC (micro Time-Projection Chamber) uses the drift gap like a time-projection chamber. The positions of the primary ionizations in the drift gap are reconstructed by knowing the drift velocity and the arrival time of the signal at the anode. Then the points are fitted to extrapolate the position of the tracked particle in the gap center \cite{Alexopoulos2010,Riccardo2016}.    
\end{itemize}
These algorithms and the BESIII environment place special demands on the readout functions in terms of charge measurement, time resolution, and sustainable rate, requiring the development of a dedicated readout chain.\\
The two innermost layers of the detector have already been built, assembled, and are being remotely operated for cosmic data acquisition (see section \ref{setup_beijing}). The third, outermost layer, is under mechanical review and its construction will begin soon.
\section{Contribution to the project}
The PhD project presented in this thesis regarded the development of the readout chain for the CGEM-IT (see chapter \ref{cap:readout_chain}). The contribution to this project included all the readout sections:
\begin{itemize}
    \item On-detector electronics: I contributed to the calibration and tuning of the ASICs. I was also responsible for evaluating and investigating the noise  (chapter \ref{cap:noise}), and checking the operation and performance of the installed ASICs.
    \item Off-detector electronics: I worked on debugging  and optimizing the FPGA firmware (chapter \ref{cap:firmware}), performed tests and checks to verify the correct operation of the device, examined and analyzed the code to find the cause of the malfunctions, and finally proposed solutions and patches to fix such errors.
    \item Data acquisition software: I developed a complete toolkit to manage the configuration of readout electronics and data acquisition. I also developed many tools for the online monitoring of detector conditions and status reporting (chapter \ref{cap:software}).
    \item Data analysis: I wrote a set of analysis scripts to perform online analysis during data acquisition in various configurations (see appendix \ref{setups}). From these scripts I built a complete analysis tool that I used to examine the overall system status, with  particular attention to the conditions and performance of the electronics (chapters \ref{cap:agg_analysis}, \ref{cap:res_TB}).
    
\end{itemize}
\chapter{Readout chain and power supply}
\label{cap:readout_chain}
\section{System overview}  \label{sec:overview}
\begin{figure}
\centering
  \includegraphics[width=1.2\linewidth, angle=90 ]{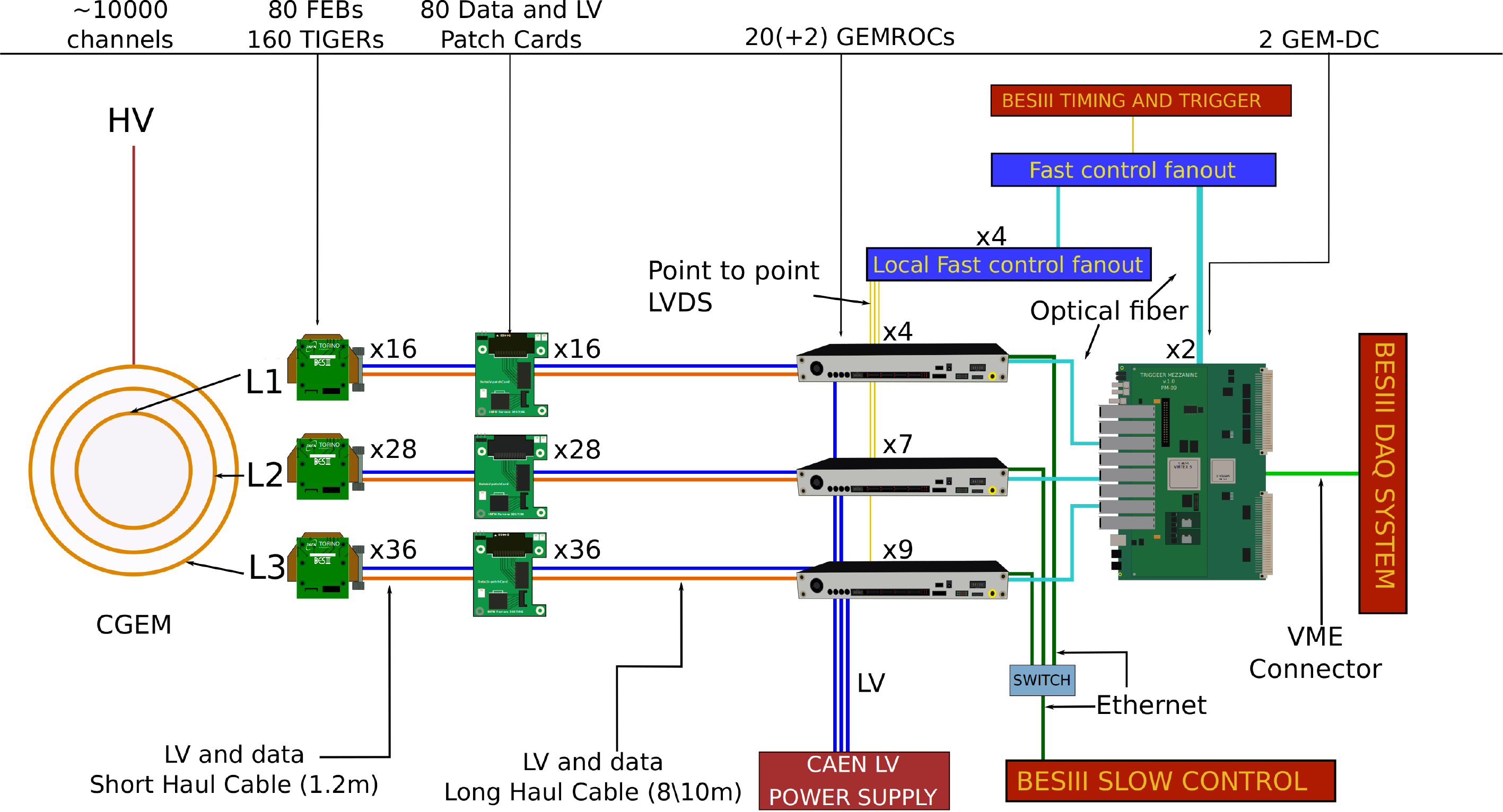}
  \caption{Scheme of the full readout chain. The blue and the orange lines represent LV and data bus respectively, cyan lines represent optical fibers and green lines represent Ethernet cables.  }
  \label{fig:chain}
\end{figure}
A custom readout chain \cite{Bortone_2021} was developed for the CGEM-IT detector. A scheme of the whole chain is shown in figure \ref{fig:chain}.\\
The signals induced on the anode strips are converted into hits by an  ASIC developed inhouse, dubbed TIGER (Torino Integrated GEM Electronics for Readout, section \ref{TIGER}). The TIGER chips are assembled in pairs on Front-End Boards (FEB) and installed on the detector. Data and ASIC Low Voltage (LV) are fed through Data Low Voltage Patch Cards (DLVPC) by the GEM Read Out Cards (GEMROC). The connection from GEMROC to DLVPC is made via long haul cables (\SI{8}{\meter} and \SI{10}{\meter} long for data and  LV, respectively), while flexible short haul cables (\SI{1.2}{\meter} long) are used inside the spectrometer between DLVPCs and FEBs. \\
The GEMROC boards receive  signals from the BESIII timing and trigger interface, communicate with the BESIII slow control via Ethernet interface and via optical fibers with the GEM-Data Concentrator (GEM-DC) cards,  which build the events and communicate with the VME-based BESIII DAQ. The GEMROC boards also manage the  power supply and configuration of the front-end boards as well as the TIGER output data acquisition
 (section \ref{backend}).\\

\subsection{Overall system requirements}
For the operations in BESIII, the whole readout chain needs to sustain a peak rate of \SI{14}{\kilo \hertz \per strip} of signal hits for the innermost layer \cite{design_2014}. This number was extrapolated considering the current rate on the MDC innermost layers. To ensure that the system had sufficient  bandwidth and rate capability headroom to accommodate signal and noise, the rate was multiplied by a safety factor of four, requiring  a system with a capability of \SI{60}{\kilo \hertz \per channel}. \\
The requirements met for the CGEM-IT project also make the electronics suitable for the readout of other innovative micro-pattern gas detectors. For this reason, the entire readout chain was developed with strong adaptability and modularity in mind.\\

\section{Front-end electronics: TIGER ASIC}\label{TIGER}
TIGER (Torino Integrated GEM Electronics for Readout) is the ASIC designed for the readout of the CGEM-IT strips \cite{RIVETTI2019181}. Each mixed-signal chip can handle the complete readout of incoming data from 64 channels and provides time and charge measurement to meet all the design requirements. The chip has been implemented in \SI{110}{\nano \meter} CMOS technology with eight metal layers, operates at \SI{1.2}{\volt}  and occupies an area of 5x5 \SI{}{\milli \meter \squared}.
\subsection{Front-end ASIC requirements}
The requirements that led to the design of the ASIC are based on the preliminary studies of the full CGEM-IT presented in the BESIII CGEM-IT Conceptual Design Report \cite{design_2014}.\\
 The innermost layer of the CGEM-IT will be exposed to a Total Ionizing Dose around \SI{10}{\kilo \rad \per yr}. The technology (\SI{110}{\nano \meter}) used has already been tested for a Total Ionizing Dose of about \SI{5}{\mega \rad} with only minor degradation tested. The printed circuit board that hosted the chips is a test-board used for TIGER prototype characterization. It was  irradiated with \SI{30}{\kilo \rad} to test the radiation hardness of passive components and voltage regulators. A deviation of \SIrange{0.7}{0.8}{\percent} was found on voltage output, which is considered perfectly acceptable for the kind of voltage regulators used on the test board.\\
The requirements for the very front-end are derived directly from the detector characteristics and are: 
\begin{itemize}
    \item Input signal in the range of \SIrange{2}{50}{\femto \coulomb}
    \item Input capacitance from a few \SI{}{\pico \farad} up to \SI{100}{\pico \farad}
    \item Noise below 2000 electrons r.m.s. for the strips with the highest capacitance
\end{itemize}
The wide range of input capacitance does not allow a shaping stage fully optimized for noise reduction.\\
Considering the desired  spatial resolution of $\sim$\SI{130}{\micro \meter}, the use of an analog readout was mandatory: for a binary readout, such a resolution would require a strip pitch of $\sim$\SI{300}{\micro \meter}, which implies a number of channels for the complete readout of the detector of about 25000. Considering the limited space available for the CGEM-IT in the BESIII spectrometer, the requirements of such a solution in terms of space, number of cables and power consumption would be unmanageable. The use of an analog readout chain allows for a manageable number of channels ($\sim$ 10000) and a larger strip pitch (\SI{650}{\micro \meter}).\\ The maximum available power consumption is about \SI{12}{\milli \watt \per channel}, given about 10000 channels.\\
To exploit the analog charge information and the time measure, the position is calculated using two different algorithms for position determination: Charge Centroid and micro-Time Projection Chamber (see \ref{design}). For efficient use of the $\upmu$-TPC a time resolution better than \SI{5}{\nano \second} is required, while the charge centroid requires a reliable hit charge estimation.
The ASIC features a trigger-less (or data-push) readout architecture: the chip generates output data without the input of an external trigger. The internal trigger logic processes the data output every time the signal is above the specified thresholds. To reconstruct the event with respect to an external trigger is thus necessary to use the timing information provided in the data outstream.
\subsection{ASIC architecture}
\label{asic_arch}
\begin{figure}
\centering
  \includegraphics[width=0.95\linewidth]{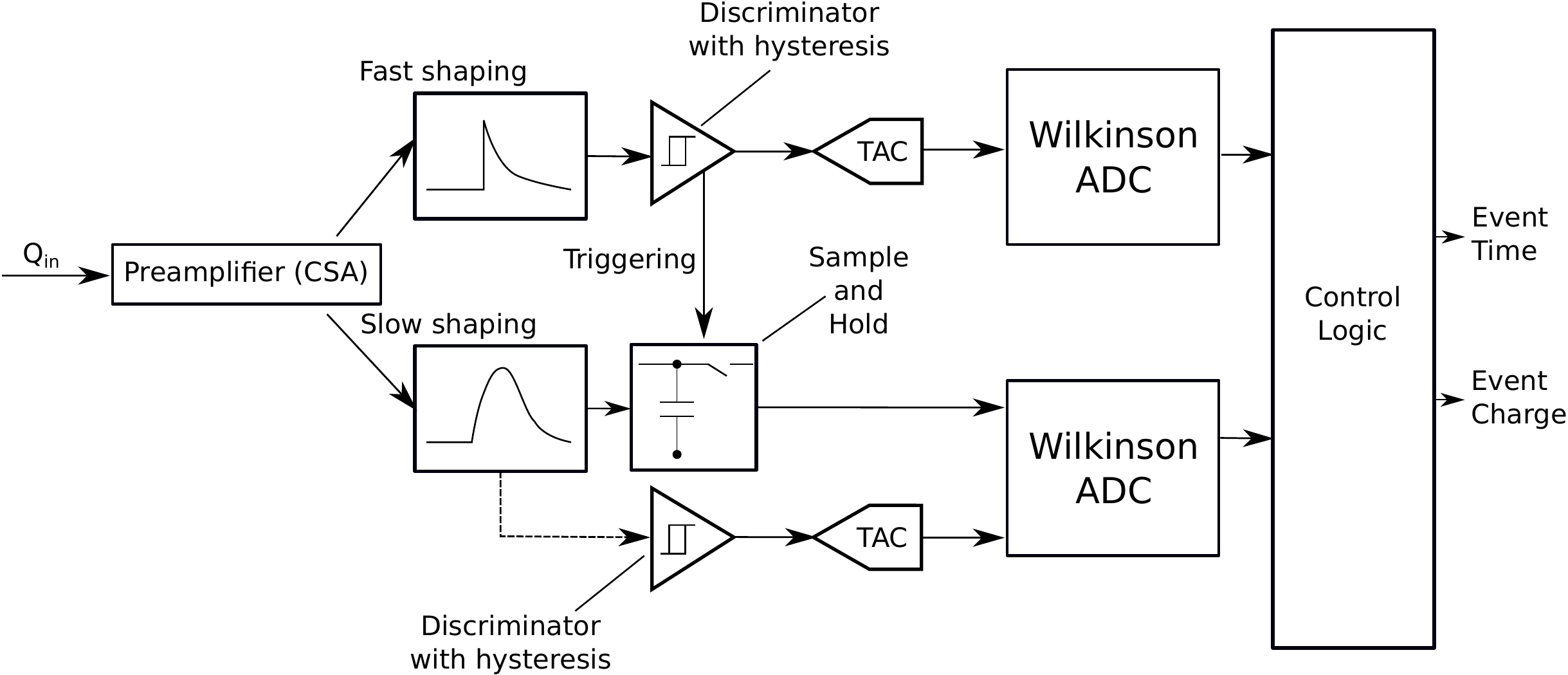}
  \caption{Overview of TIGER channel architecture.}
  \label{fig:scheme}
\end{figure}
On each channel (figure \ref{fig:scheme}), the signal is amplified and inverted by a three stage cascoded common source Charge Sensitive Amplifier \cite{Geronimo}. The signal is then duplicated to feed two different shaping branches, one optimized for time measurements (\SI{60}{\nano \second} peaking time) and the other optimized for charge measurements (\SI{170}{\nano \second} peaking time). The fast rising time on the time-optimized branch, which is equal to the expected charge collection time, provides  timing measurements with low jitter (hence it is called  the T branch). The flatter peak on the charge-optimized branch enables to sample the peak voltage with optimized Equivalent Noise Charge (ENC) for accurate charge measurements (E branch).\\ 
Fine timing  can be performed on both branches using a time to amplitude converter in combination with a Wilkinson ADC, pushing the time resolution below the clock period (\SI{6}{\nano \second} for BESIII). The jitter on the T branch dominates the final resolution (see table \ref{table:perf}), while the resolution of the TDC is better than \SI{50}{\pico \second}.\\ 
TIGER does not require an external trigger for operation, since it has one discriminator on both branches. 
The thresholds on the branch discriminators can be set independently for each branch and each channel with 6-bit DACs, while the great flexibility of the digital part of the ASIC allows many combinations of discriminator triggers to be used to validate the signals.\\ The output of the discriminators is fed to the channel controller, a digital logic unit which generates the signals that control the mixed signal back-end operations to extract the time and charge information.
Charge can be measured in two different modes: Time-over-Threshold and Sample-and-Hold mode. In Time-over-Threshold mode, the TDCs of both branches on each channel are used to measure the rising and falling edges of the signal, to obtain the signal amplitude. The time information can be acquired at the output of both shapers with different configurations.\\
In Sample-and-Hold mode, the signal on the E Branch is stored on a capacitor at a settable time after crossing the threshold,  to sample its peak value. Then, the stored signal is digitized using the Wilkinson ADC shared with the TDC, providing a linear relationship between the digital output and the amplitude of the input signal. This mode was chosen for the CGEM-IT readout since it is more practical to use and the detector signals differ in shape and duration.
To maximize signal integrity and minimize interface noise, all relevant signal processing is performed on-chip.
The ASIC outputs three types of data words: hit words, which contain the hit information in 54 bits; counter words, which are used for debugging to count the hits on a channel; and frame words, which are sent every $2^{15}$ clock cycles, as a timing reference.\\
The ASIC digital back-end and the fully-digital interface were derived from the TOFPETv2 ASIC \cite{DIFRANCESCO2016194}. Four TX LVDS links can transmit data in 8b/10b encoding at up to \SI{200}{\mega \hertz} in both Single Data Rate and Double Data Rate, while ASIC internal registers  are programmed via a \SI{10}{\mega \hertz} SPI-like configuration link. In TIGER, triple redundancy of digital registers has been added to provide Single Event Upset protection for operation in a high radiation environment.\\
On-chip calibration circuitry allows injection of a programmable amplitude test pulse at the channel input to test and characterize the channel response in the full charge input range. A digital test pulse can be sent directly to the channel controller, bypassing the front-end, to test and calibrate the TDCs and digital back-end.\\
The ASIC performance is summarized in table \ref{table:perf} \cite{fabio_tesi}.\\
\begin{table}
\centering
\begin{tabular}{|c|c|}\hline
Input dynamic range            &  2-\SI{50}{\femto \coulomb } \\ 
Gain (E branch)             &  \SI{11.8}{\milli \volt \per \femto \coulomb} \\ 
Noise (E branch)            &  < 1800 $e^-$ ENC (\SI{0.29}{\femto \coulomb}) \\ 
Jitter (T branch)           &  < \SI{4}{\nano \second}\\  
Sample-and-Hold residual non linearity &  < 1\% in the whole dynamic range \\ \hline
\end{tabular}
\caption{TIGER key performance measured on silicon with $C_{in}=$\SI{100}{\pico \farad} and $Q_{in}=$\SI{10}{\femto\coulomb}.}
\label{table:perf}
\end{table}
\subsection{Front-end ancillaries}
\begin{figure}
\centering
\begin{subfigure}{.5\textwidth}
  \centering
  \includegraphics[width=1\linewidth]{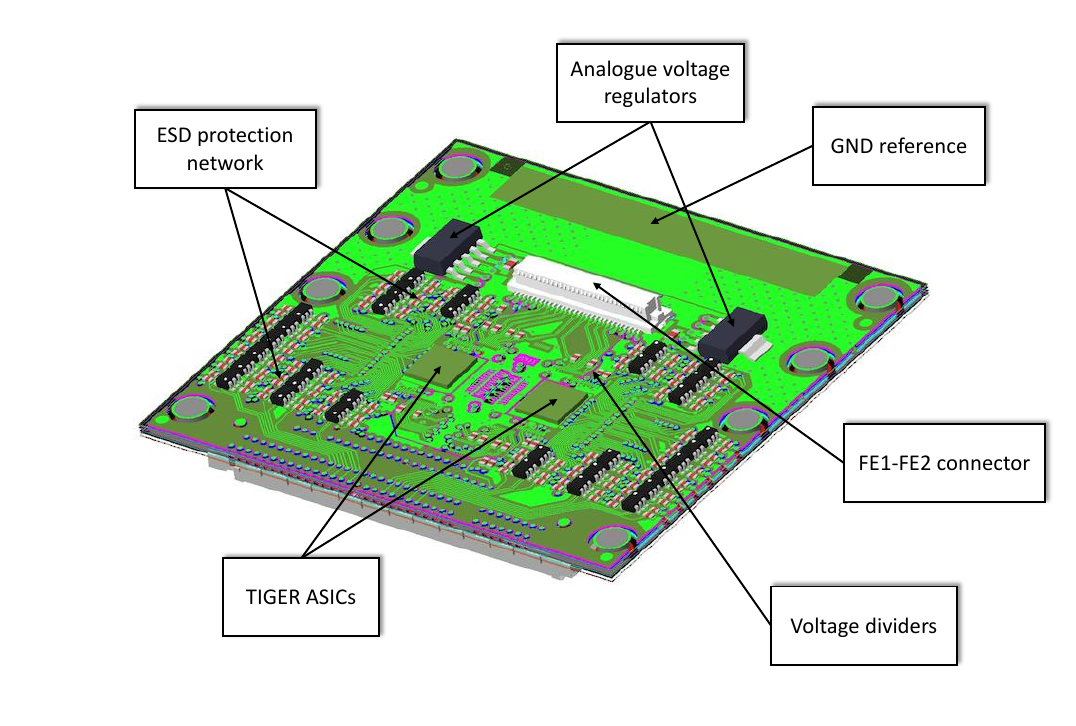}
\end{subfigure}%
\begin{subfigure}{.5\textwidth}
  \centering
  \includegraphics[width=1\linewidth]{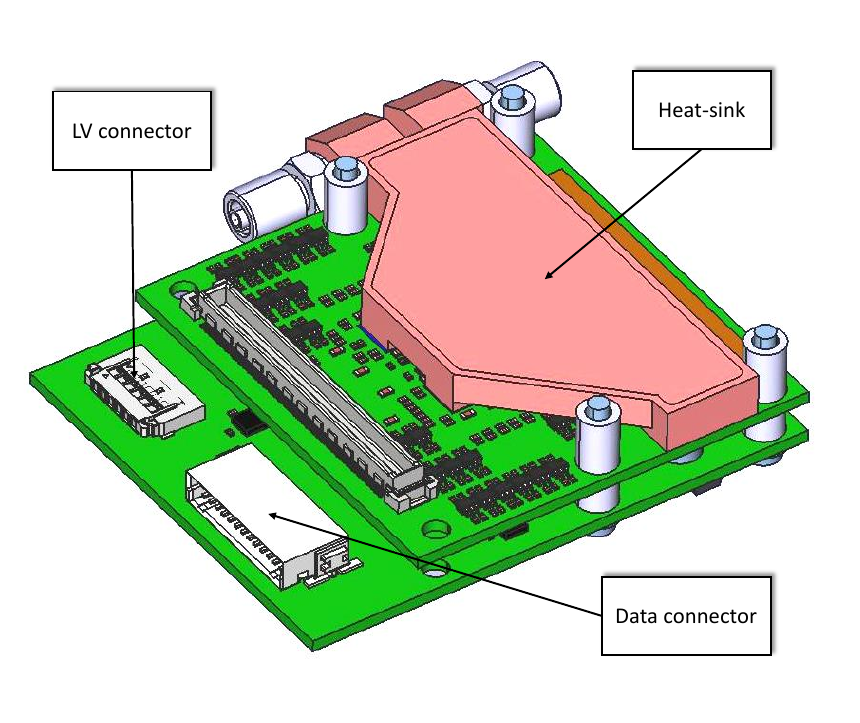}
\end{subfigure}%
  \caption{Front-End Board design for the CGEM-IT layer 1 and 2: top-side routing on FE1 (left) and FE1/FE2 assembly with liquid cooling heat heatsink (right).}
\label{fig:FEBs}
\end{figure}

\begin{figure}
\centering
\begin{subfigure}{.5\textwidth}
  \centering
  \includegraphics[width=1\linewidth]{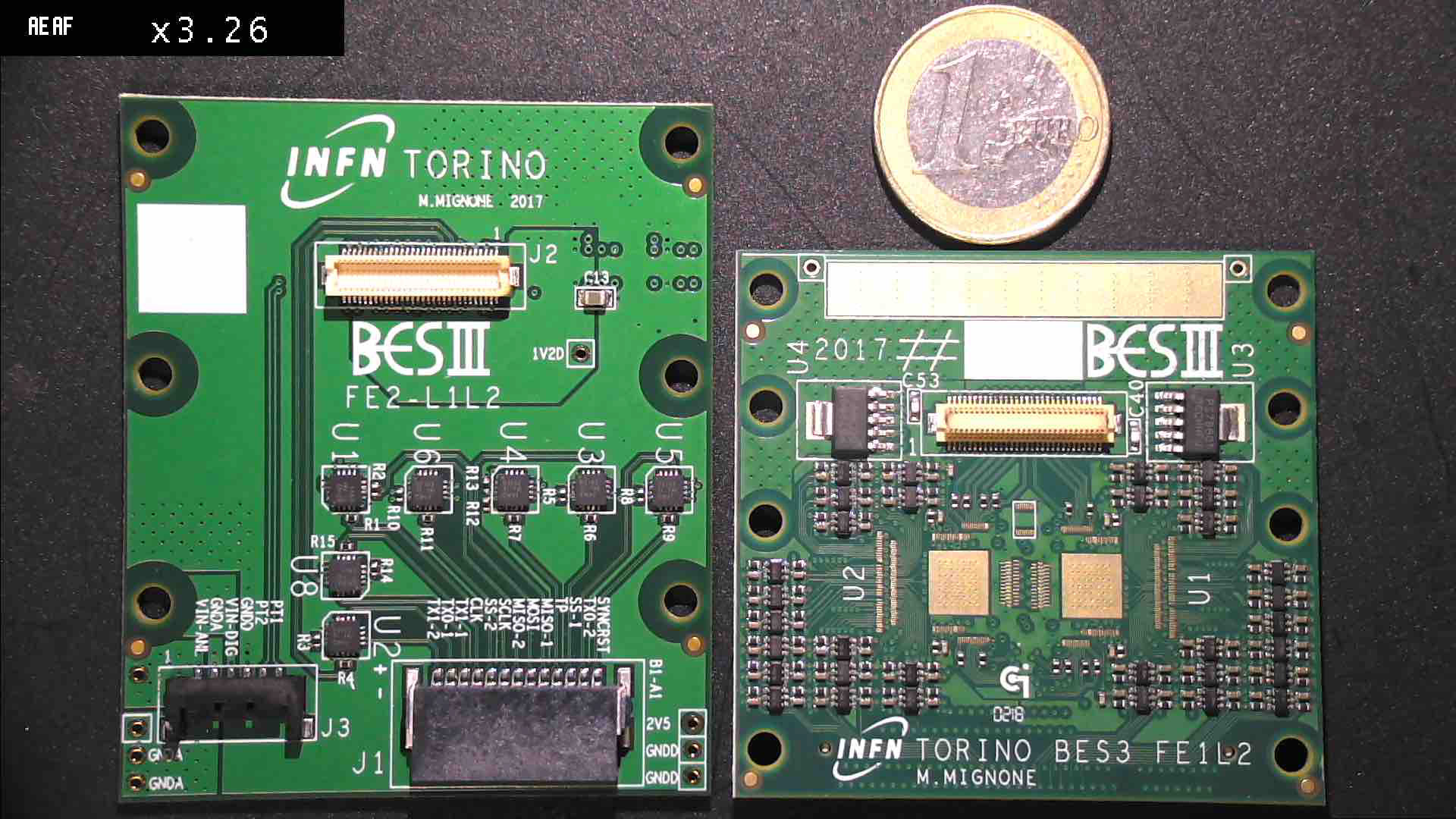}
\end{subfigure}%
\begin{subfigure}{.5\textwidth}
  \centering
  \includegraphics[width=1\linewidth]{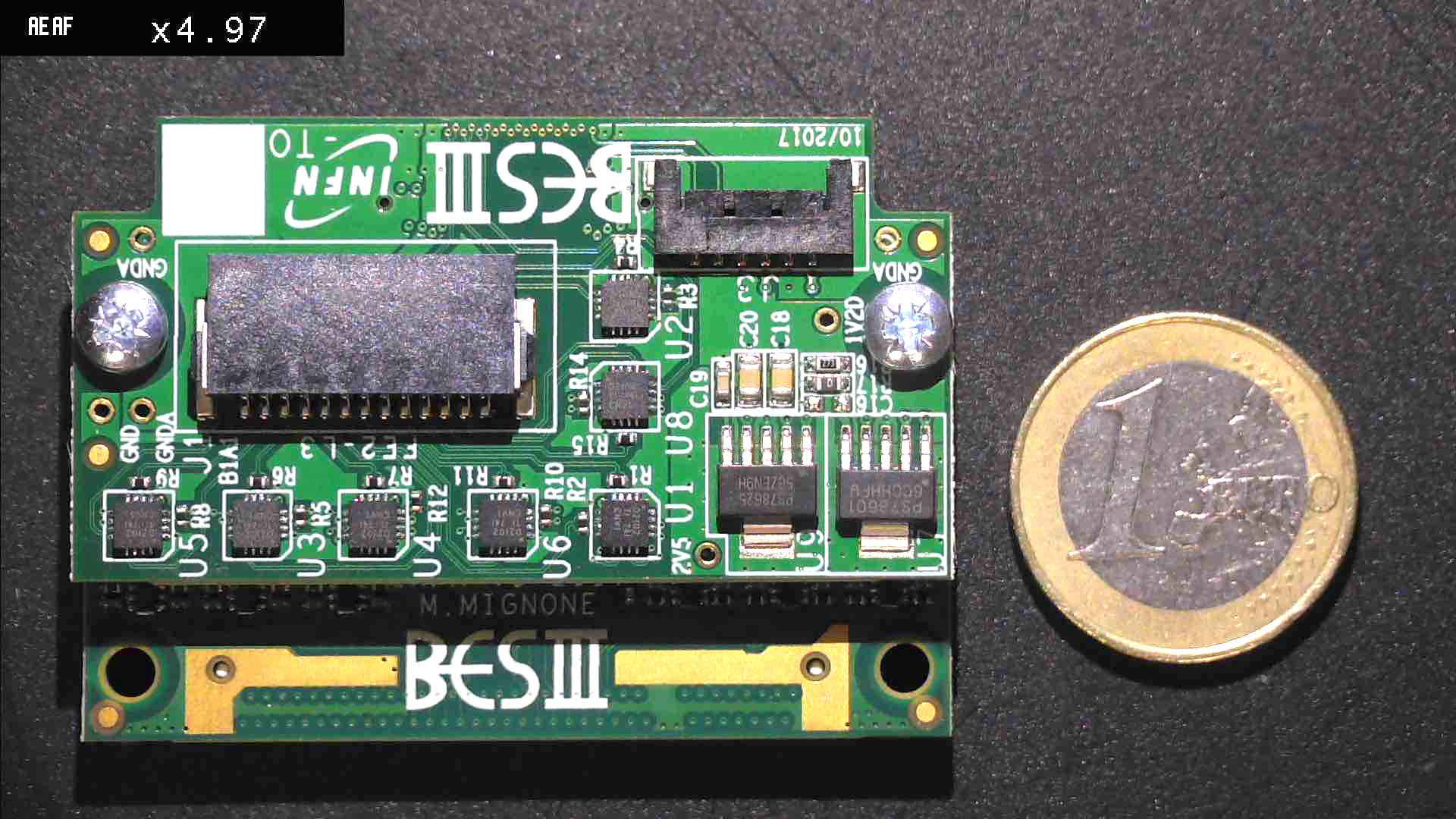}
\end{subfigure}%
  \caption{Picture of the FEBs. Left: the layer 1 and layer 2 design. Right: the layer 3, more compact, design.}
\label{fig:FEB_foto}
\end{figure}

The ASICs are mounted in pairs on Front-End Boards (FEBs)\footnote{Designed by Marco Mignone (INFN-Torino)}. A compact board design was necessary due to the limited space in the BESIII spectrometer. The boards shown in figure \ref{fig:FEBs} consist of two different PCBs (called FE1 and FE2), that are connected to each other via a HIROSE DF12-60-DS-0.5V connector. The two ASICs are bonded to the FE1 board, where power regulators and SMD voltage dividers bias the chips. This board also houses the ESP protection network, which protects each input channel of the ASIC from damages due to detector discharge. A Hirose FX10A144P connector is used to interface to the detector strips.\\
The FE2 board provides the interface with the off-detector electronics: it houses the data connector (ERNI SMC-B26), the power connector (MOLEX 504050-0691) and the LVDS buffers.\\
The different geometry of the three detector layers of the CGEM-IT required a different and even more compact design of the layer 3 FEBs. A picture of the FEBs for the different layers can be seen in figure \ref{fig:FEB_foto}.
A cooling system was designed to stabilize the temperature of the FEBs, ensure operation of the ASIC at a near constant temperature, and limit heat dissipation to the outer MDC.  A specially designed copper heat sink is mounted on each FEB. The inlet and the outlet are provided by radiation-resistant polyurethane tubes. The heat sinks are grouped to limit the pressure drop per layer. In BESIII, the  CGEM-IT cooling system will be connected to that of the EMC, using a booster pump and a patch panel with three lines in and three lines out per detector side. The cooling system maintains the temperature of the FEBs between \SI{25}{ \celsius} and \SI{30}{ \celsius} by circulating chilled water at \SI{20}{ \celsius}, \SI{2.5}{\litre / \minute} and \SI{2}{\bar}. The thermal load is about \SI{3.2}{\watt} per FEB (half for the ASICs and half for the voltage regulators), for a total of \SI{256}{\watt} for the whole system.\\
To fully equip the detector, a total of 80 FEBs (160 TIGERs) are required: 16 FEBs for layer 1, 28 FEBs for layer 2 and 36 FEBs for layer 3.\\
All necessary FEBs have been validated and calibrated.\\
\section{Off-detector electronics}\label{backend}
\begin{figure}
\centering
\begin{subfigure}{.5\textwidth}
  \centering
  \includegraphics[width=1\linewidth]{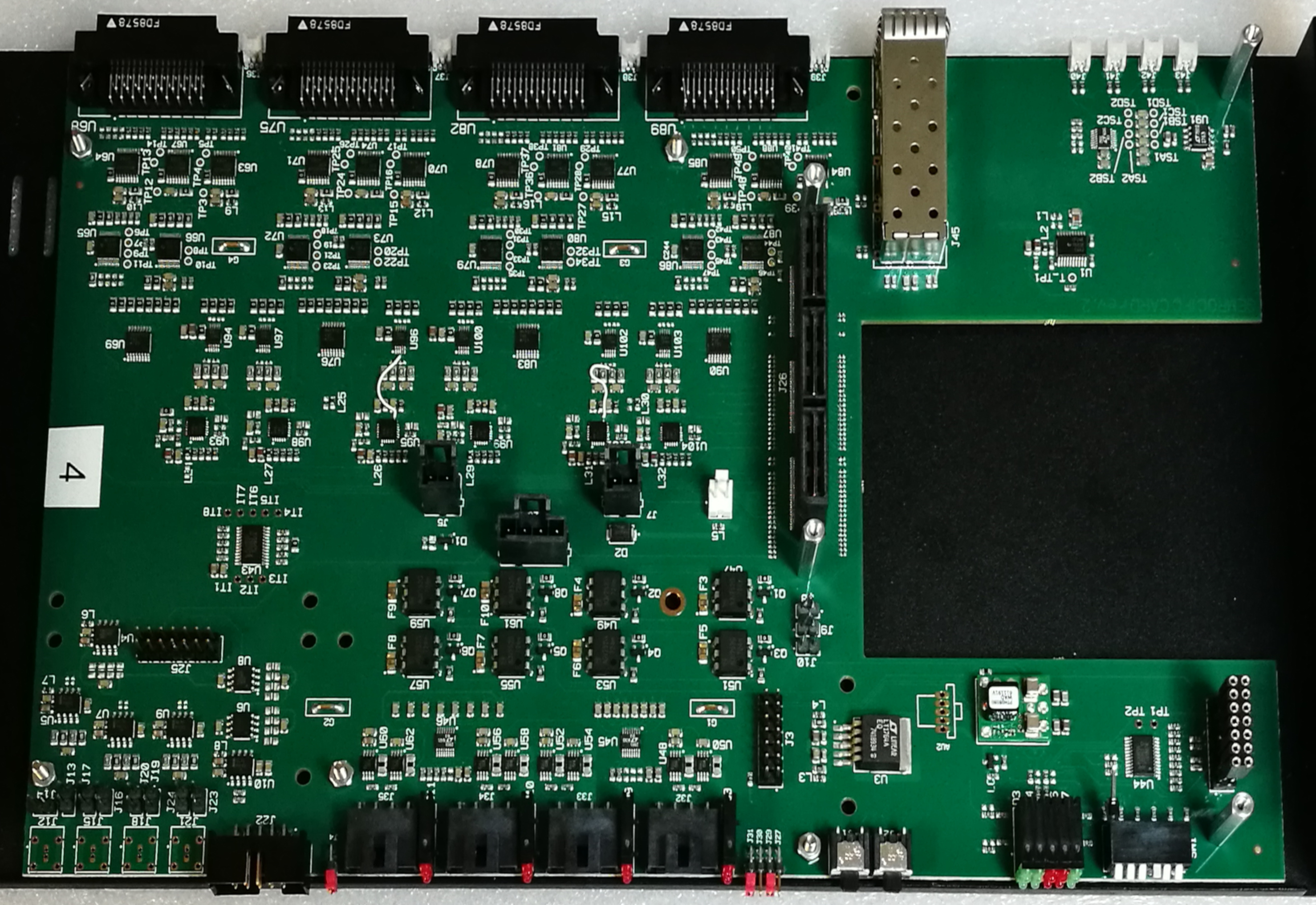}
\end{subfigure}%
\begin{subfigure}{.5\textwidth}
  \centering
  \includegraphics[width=1\linewidth]{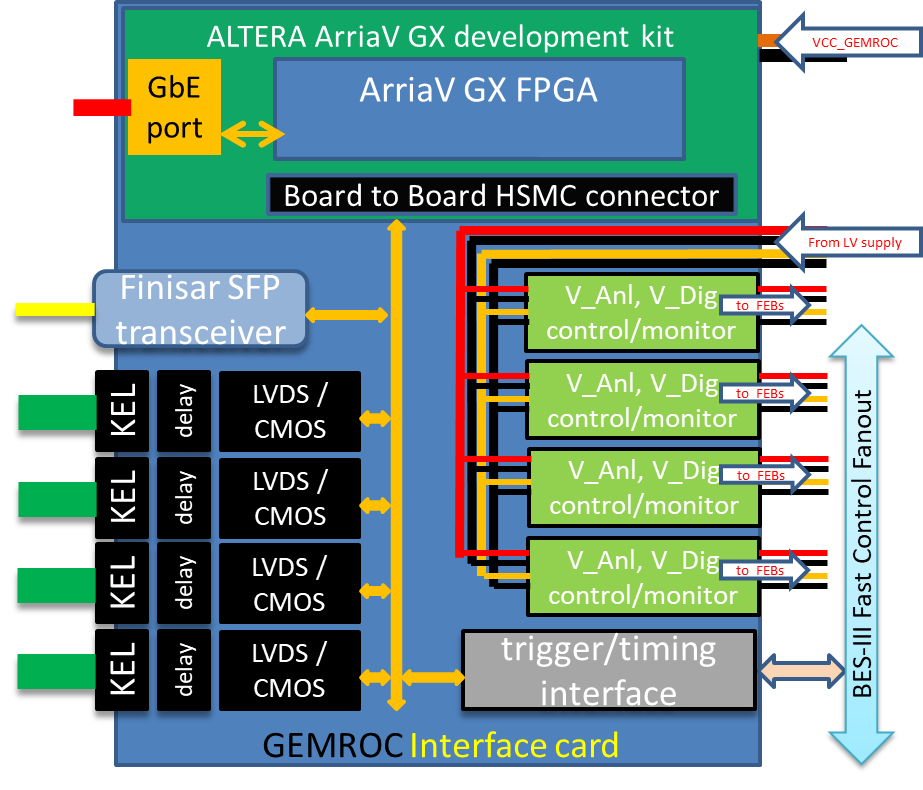}
\end{subfigure}%
  \caption{The GEMROC interface card (left) and the block diagram of the GEMROC hardware and connectivity (right).}
\label{fig:gemroc_interface}
\end{figure}

\begin{figure}
    \centering
    \includegraphics[width=1\linewidth]{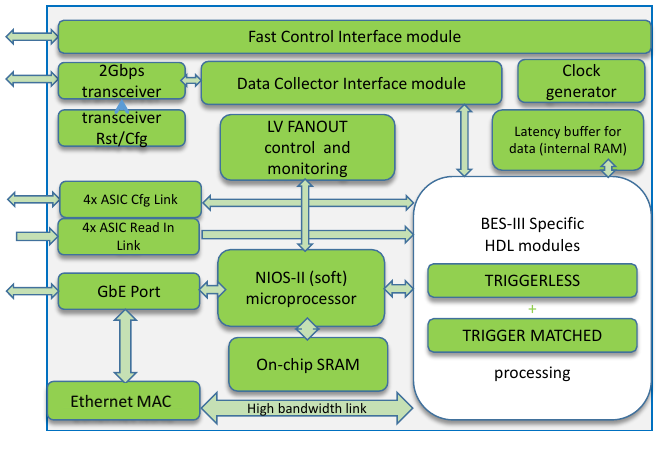}
    \caption{GEMROC firmware block structure.}
    \label{fig:gemroc_firmware_blocks}
\end{figure}

The off-detector electronics consists of two data processing units and two signal propagation modules. The data processing units are the GEM Read-Out Cards (GEMROC)\footnote{Designed by Angelo Cotta Ramusino (INFN-Ferrara)} and the GEM Data-Collector\footnote{Designed by Pawel Marciniewski (Uppsala University)} (GEM-DC), while the Fast Control Signal (FCS) propagation modules are the System FCS Fanout module and the Local FCS Fanout Module.\\
The GEMROCs are the modules designed for configuring and reading TIGERs. Figure \ref{fig:gemroc_foto} shows some pictures of the modules.
\begin{figure}
\centering
\begin{subfigure}{.5\textwidth}
  \centering
  \includegraphics[width=1\linewidth]{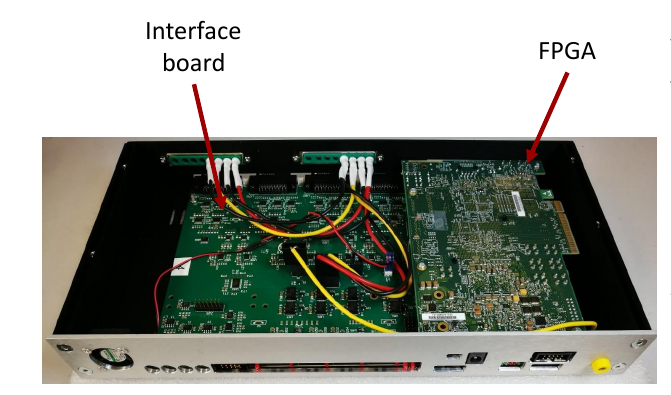}
  \label{fig:gemroc_foto}
  \end{subfigure}%
\begin{subfigure}{.5\textwidth}
  \centering
  \includegraphics[width=1\linewidth]{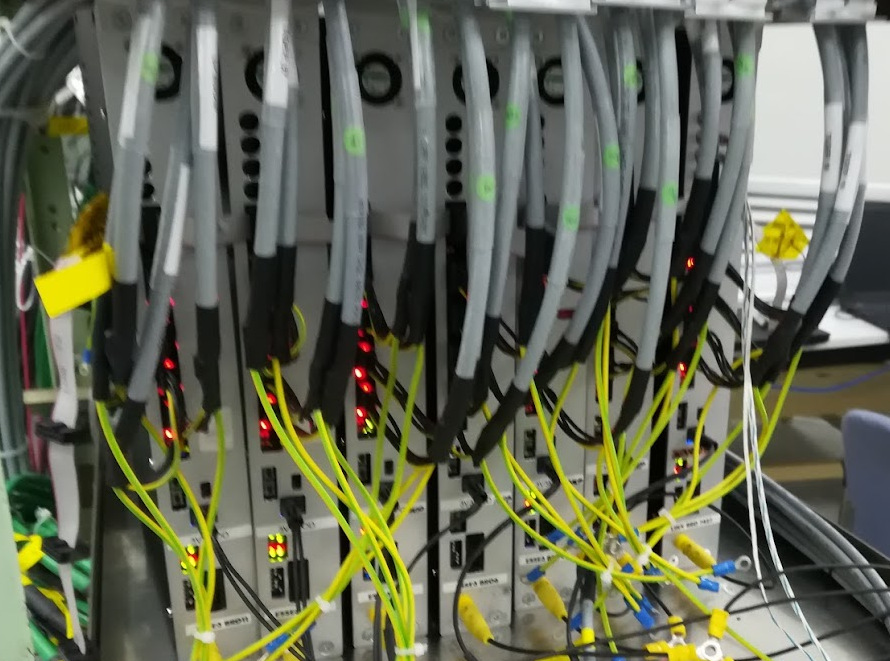}
  \end{subfigure}%
  \caption{Left: GEMROC picture showing the interface card and the FPGA; right: an array of GEMROCs during the layer 1 and layer 2 tests.}
  \label{fig:gemroc_foto}
\end{figure}
Each GEMROC handles four FEBs, for a total of eight TIGERs. The core of each GEMROC is a development kit based on an  FPGA of the Intel/ALTERA ARRIA V GX family  \cite{ARRIVA}, connected to an interface card developed for the BESIII experiment. The interface cards provide connectivity to the FEBs,  external signals, and the FEB power supply (figure \ref{fig:gemroc_interface}).\\
The GEMROC modules distribute digital and analog supply voltages to the FEBs, monitoring their supply currents and operating temperatures, via thermistors installed on the FEB, to ensure safe operation of the on-detector electronics. The entire system can be read using 20 modules, but, to ensure a symmetrical distribution on the two sides of each layer, 22 will be used.\\
The GEMROC modules receive via a suitable distribution system the timing signals which form the BESIII Fast Control System:
\begin{itemize}
    \item \textit{Clock}: BESIII distributes a \SI{41.65}{\mega \hertz} clock, equal to the radio-frequency of the BEPCII storage ring divided by 12. The GEMROC derives all the time references from this signal. The GEMROC FPGA uses internal PLLs to derive signals that toggle at a frequency of 166.6 MHz, four times the BESIII clock rate, to drive the clock inputs of TIGER.
    \item \textit{L1 trigger}: the BESIII Level 1 (L1) trigger is used by BESIII DAQ and by the CGEM-IT readout system to flag event data to be saved. It has a rejection rate of about $1:10^4$, a latency, with respect to the event, fixed at \SI{8.6}{\micro \second}, an acceptance window of \SI{1.6}{\micro \second}, an average frequency of \SI{4}{\kilo \hertz}  and a dead time of \SI{3}{\micro \second}. The trigger signal lasts for eight BESIII system clock cycles \cite{trigger}.
    \item \textit{Check}: every 256 L1 triggers, a \textit{check} signal is sent to verify the synchronization of the subsystems. This line can also be used in standalone mode by the GEMROC modules for debugging purposes (\textit{e.g.} to synchronize test pulse generation).
    \item \textit{Full}: this signal is used by all BESIII subsystems to notify that the buffers which hold the event data, pending transmission to the DAQ, are filling up. When the BESIII Fast Control System receives the FULL signal from one of the sub-detectors, it stops sending any more L1 triggers.
    \end{itemize}
The FPGAs are programmed to control the system data acquisition. The firmware blocks are shown in figure \ref{fig:gemroc_firmware_blocks}. \\
The GEMROC modules can process the data received from the FEBs in two different ways. In the first, called trigger-less (TL), which is used in the standalone configuration for debugging purposes, the data received from the enabled TIGERs are merged and transmitted over the Ethernet output port using the UDP protocol. A UDP packet, limited to the default size of \SI{1500}{B}, is transmitted either when it contains all the data collected in eight TIGER time frames ($2^{15}$ TIGER clock periods), or when the number of TIGER data words collected  reaches 180, with the remaining portion of the TIGER data transmitted in subsequent UDP data packets. \\
In contrast, in trigger-matched (TM) mode, a finite state machine selects the hits to be sent over as follows. The incoming data from each TIGER pair are stored in a \textit{latency buffer} circular memory, organized in to pages of 32 locations each. This memory is intended to buffer the incoming TIGER data, until the L1 selection trigger occurs. A latency buffer page, also called a bucket, contains all the data recorded by the two TIGERs of a FEB in a time interval corresponding to $2^8$ TIGER clock cycles (\SI{1.53}{\micro \second}). The bucket memory is circular, so the data are overwritten at each address rollover (\SI{24.6}{\micro \second}). The paged organization of the latency buffer is intended to speed up the search, which is started when the L1 trigger signal is received.\\
When the BESIII trigger arrives, the FPGA logs the trigger time of arrival timestamp, waits for a programmable delay to account for the stochastic transmission latency of data over the TIGER output serial links, and then reads the buckets determined by the L1 trigger time of arrival timestamp to search the TIGER data with coarse event timestamps that fall within the BESIII trigger window (by default \SI{1,7}{\micro \second}). The hits from all the FEBs connected to a GEMROC and enabled to data taking are merged to prepare a trigger-matched data packet to which a header and a trailer are added. The L1 trigger to which the data have been matched can be correctly identified by the header and trailer. The latter also contains diagnostic information about the GEMROC status.\\
The trigger-matched data packet is then sent over the \SI{2}{\giga \hertz}  fiber optic link to the GEM-DC modules and also sent as a UDP packet over the Ethernet port.\\
This last function is used in standalone operation of the CGEM-IT setup and enables trigger-matched data to be also collected by an alternative Ethernet-based data acquisition system.\\
The GEM-DC are VME 6U cards used to collect the trigger-matched data packets transmitted by the GEMROC modules over optical links and assemble them into sub-detector events identified by the common trigger number.
A VME interrupt is then generated by the GEM-DC to prompt the VME crate CPU to read the trigger-matched events stored in the GEM-DC buffers. The GEM-DC boards inherit the hardware design and most of the firmware from the Read-Out Driver (ROD) modules used by the KLOE-2 inner tracker \cite{DAQ_KLOE}. \\
Each FEB is connected to a GEMROC via short-haul and long-haul  multiple twisted-pair (MTP) shielded cables that are interconnected via the DLVPC boards. 
\subsection{GEMROC firmware structure}
\label{subsec:firm}
In this section the GEMROC firmware structure is presented.\\
\subsubsection* {Top-module} The design top-module defines the I/O between the FPGA, the I/O pins and the other components in the GEMROC, such as the NIOS processor. It also interconnects the various blocks of the GEMROC, generating clocks and routing the signals.
\subsubsection*{BESIII FC signal simulator} This module handles the generation of the Fast Control signals when the GEMROC is used in standalone mode.
\subsubsection* {Reset manager}This module is responsible for generating the various resets, which can be generated automatically on the first trigger, or manually as needed.
\subsubsection* {L1 distributor}The L1 distributor receives the L1 trigger signal from outside, via the top-module, and handles the calculation of the trigger window that is transmitted to the trigger-matching logic.
\subsubsection*{TIGER Test Assembly} The TIGER Test Assembly handles direct communication with the ASIC, both for the data lines and for the configuration lines. For each TIGER, a test assembly must be initialized. This module has a counter that logs the number of 8b/10b errors to monitor the communication lines and optimize their delays. It can also count the number of hits per TIGER.
\subsubsection*{FEB merger} This module consumes the data from a pair of TIGER (one FEB) and stores them in a FIFO. In trigger-less mode the data are steamed into the "Four FEB merger", while in trigger-match mode the data are stored into a Bucket DPRAM memory, waiting for a trigger. When the trigger arrives, the data are then matched to the trigger signal and forwarded. 
\subsubsection*{Four FEB merger TL} This module collects the data from the four FEBs connected to the GEMROC and assembles the UDP packets to be sent. The module sends the data when the maximum size is reached or every four frame words.
\subsubsection*{Four FEB merger TM} This module assembles the incoming data  from four FEB mergers and build the header and trailer packet information. It consists of two finite state machines that collect the data from the first and  second FEB pairs, respectively, and put them into FIFOs. Another state machine reads the data from these FIFOs and assembles the final packet. 
\subsubsection*{Diagnostic DPRAM} Many diagnostic variables are stored in this DPRAM. The NIOS processor can then access them automatically or by operator call. The DPRAM contains flags to mark whether some key buffers in the data analysis pipe are full, and to check the locking status of the clock and optical links PLLs.
\subsubsection*{Ethernet link manager} This MAC (Media Access Control) module\footnote{Designed by Stefano Chiozzi (INFN Ferrara)} was originally developed for the NA62 experiment \cite{chiozzi}. UDP communication is handled at a very low level, which allows for high performance and flexibility.
\subsubsection*{Optical link manager} This module manages communication over an optical link with the GEM-DC module. The communication is two-way, so in principle this line can be used for complete control of the GEMROC module.
\subsection{Fast Control Signals distribution modules}
\label{sec:FCS}
The MTP cables have relatively low losses, but due to the length of the data path and the unavoidable discontinuities, the fast signals exchanged between GEMROC and FEB, and in particular the TIGER clock, have been found to be susceptible to EMI and ground noise. This causes transmission errors that are detected thanks to the error detection capabilities of the 8b/10b encoding of the serial outputs of TIGER.\\
To improve the signal integrity issues, a set of patch cards was installed on the GEMROC modules' interface card; these cards have a  pre-emphasis LVDS driver that boosts the amplitude of the TIGER clock signal received by the FEBs by about \SI{20}{\percent}, resulting in a signal level that matches well the input specifications of the clock receivers on the FEBs. \\
To improve the quality of the fast control signals (FCS) and overcome the signal integrity issues, an upgrade of the fast control signal distribution system was performed. The upgraded FCS distribution system consists of a System FCS Fanout (S-FCS-S) module driving four Local FCS Fanout (L-FCS-F) modules\footnote{Both designed by Angelo Cotta Ramusino (INFN-Ferrara)}, located near the BESIII DAQ system and near the CGEM-IT GEMROC modules respectively. The GEMROC modules are grouped in four locations around the BESIII detector. Therefore, four Local FCS Fanout modules are mounted.\\
The System FCS Fanout is programmable and can simulate the BESIII timing signals to test the CGEM-IT in standalone mode.
\begin{figure}
\centering
  \includegraphics[width=0.9\linewidth]{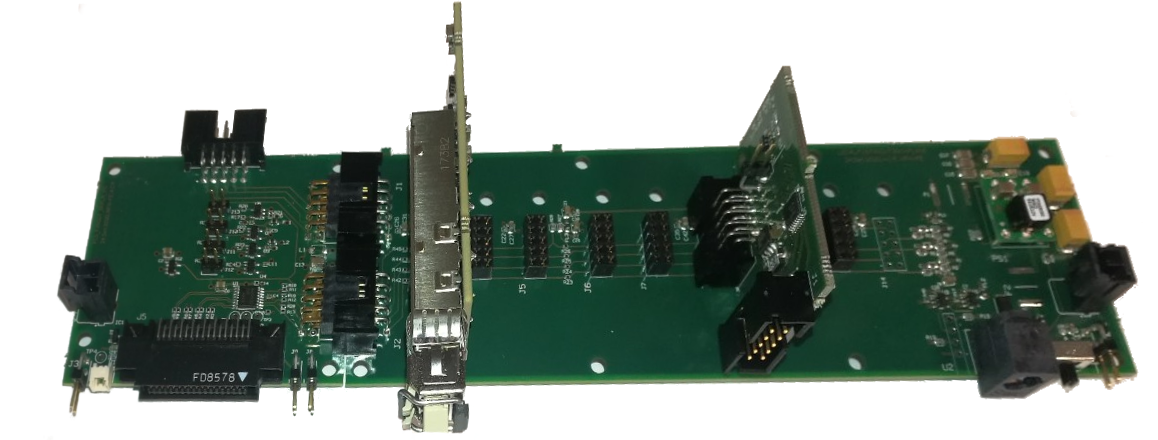}
\includegraphics[width=0.9\linewidth]{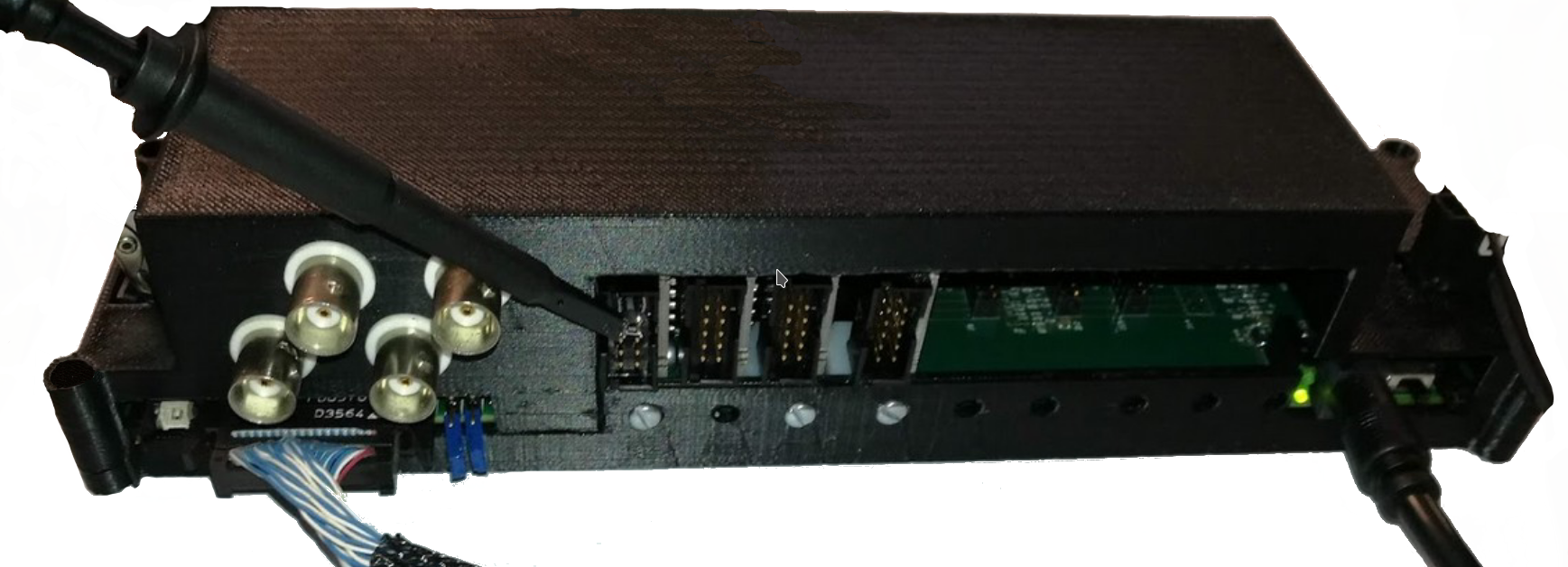}

  \caption{Local FCS Fanout module. The top picture shows the L-FCS-F PCB, with the modular slots for the connection of optical transceivers or LVDS connectors. The bottom picture shows the L-FCS-F enclosed in its 3D printed case. }
  \label{fig:LFCS}
\end{figure} 
The S-FCS-S is  directly interfaced to the BESIII FCS System and it is connected to the L-FCS-F by bidirectional optical links operating at a frequency of DC to \SI{50}{\mega \hertz}. The L-FCS-F modules (figure \ref{fig:LFCS}) use transceiver daughter cards to drive the received electrical signals to a multi-drop backplane with single-ended signaling. Up to six Flat-Cable Port (FC-Port) daughter cards are also installed on the L-FCS-F backplane. The FC-Port cards have a 10-pin connector for the flat cable segments onto which the output LVDS signals are delivered to the target GEMROC module.\\ 
\section{Low Voltage distribution}
The power distribution design is based on the following FEB bias requirements: \SI{1.08}{\ampere} at \SI{1.4}{\volt} for the ASIC analog part, \SI{0.32}{\ampere} at \SI{2.5}{\volt} for the ASIC digital circuitry. Another demand for the LV power supply system is to provide the GEMROC modules with a bias voltage of \SI{15}{\volt} at an average current consumption of about \SI{1.1}{\ampere}.\\
The power supply system uses the following commercial modules produced by CAEN S.p.A.:
\begin{itemize}
    \item One SY4527LC mainframe \cite{SY4527LC} (max \SI{600}{\watt}) and three A2517 boards \cite{A2517} (24 channels, max \SI{5}{\ampere/ channel} and \SI{50}{\watt/channel}) to power the FEBs;
    \item One SY5527 BASIC  mainframe \cite{SY5527} (max \SI{600}{\watt}) and four A2519 boards \cite{A2519} (32 channels, max \SI{5}{\ampere/ channel} and \SI{50}{\watt/channel}) to power the GEMROC modules.
\end{itemize}
A pair (analog and digital supplies) of the output channels of the CAEN A2517 is connected in parallel to the FEB power input of two GEMROC boards, which in turn supply and monitor the analog and digital power supplies for eight FEBs.\\
This modularity of eight was chosen because it fits well with the dimensions and complexity of the GEMROC modules, and makes it possible to have a separate ground reference for each detector layer when 22 GEMROC modules are used.\\
 All power supply cables are shielded and fitted with connectors to reference the shields to the grounding network. The length and current capacity of the cables used for LV power distribution are listed below:
\begin{itemize}
    \item from the mainframe rack to the GEMROC:
    \begin{itemize}
        \item GEMROC supply cable: \SI{17}{\meter}, carrying (maximum) \SI{2}{\ampere};
        \item FEB supply: \SI{17}{\meter}, carrying (maximum) \SI{9}{\ampere} (analog+digital);
    \end{itemize} 
    \item from the GEMROCs to the DLVPCs:  \SI {10}{\meter}, carrying (maximum) \SI{1.1}{\ampere};
    \item from the DLVPCs to the FEBs: \SI{1.2}{\meter}, carrying (maximum) \SI{1.1}{\ampere}.
\end{itemize}
Throughout the whole LV distribution system, the IR drop is a concern but the hardest constraints apply to the cables connected between the CAEN A2517 boards and the GEMROC modules; for these cables the four-wire cross-section is \SI{4}{\milli \meter^2}.\\
The power dissipation, taking into account the optimization of the TIGERs operating conditions, is expected to be \SI{330}{\watt} for the total number of FEBs and \SI{484}{\watt} for the total of 22 GEMROC modules.
It will be increased by the power of the FCS Fanout system, which is estimated to be \SI{33}{\watt} for the FCS System  Fanout and \SI{22}{\watt} for the FCS Local Fanout.\\
To leave headroom to the power supply system, an A4532 Power Booster Unit will be installed in the SY5527 CAEN mainframe to increase its output power by \SI{600}{\watt}.\\
The LV power distribution system is equipped with hardware and firmware features that allow remote control at the CAEN power supply mainframe level and at the GEMROC level.\\
The CAEN boards control and monitor their individual output channels that provide analog and digital power to the groups of eight FEBs, while the GEMROCs distribute, control, and monitor the power to the individual FEBs. The LV power distribution system is managed in standalone mode by custom software at the FEB level (see chapter \ref{cap:software}) and by the CAEN GECO system at the power supply mainframe level. Once the CGEM-IT is installed inside the BESIII spectrometer, the entire system will be managed by the Detector Control System (DCS) \cite{Ablikim2010}.
\section{High Voltage distribution}\label{HV}
\begin{figure}
\centering
  \includegraphics[width=0.5\linewidth]{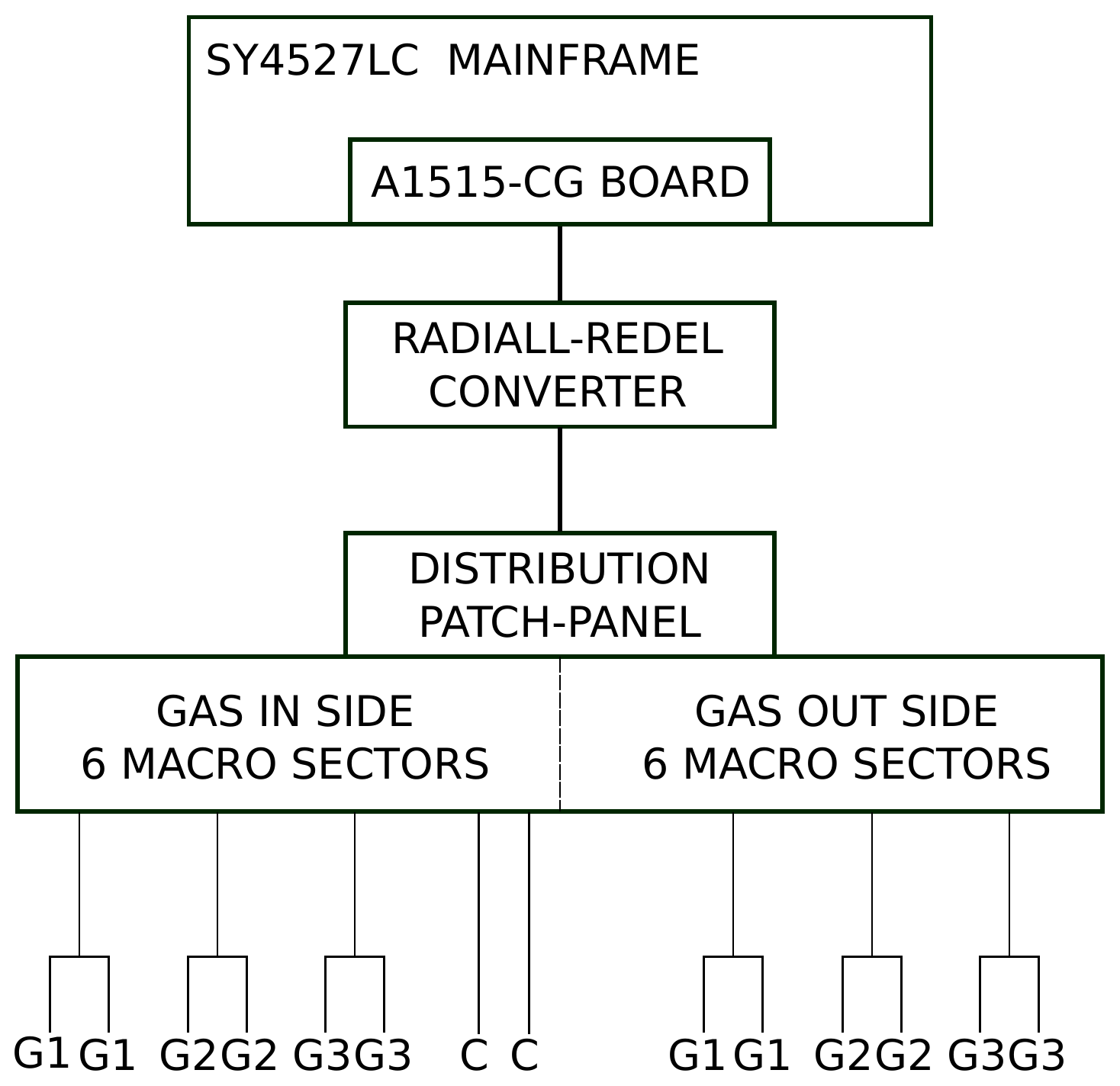}
  \caption{Block diagram of the layer 1 HV distribution system. Each multi-core cable departing from the distribution patch-panel comprises two macro-sectors cables, the 20 corresponding micro-sector cables and six reference ground cables. The power line splits in the two macro-sector cables for the last meter before the detector. }
  \label{fig:HV_scheme}
\end{figure}

The description of the High Voltage chain is included to complete the picture of the power distribution systems and their interconnection.
For each triple-GEM, seven different electrodes need to be biased (the two copper planes of each GEM and the cathode). One side of each GEM foil is segmented in macro-sectors, the other in micro-sectors. Each macro-sector is matched by 10 micro-sectors. The cathode electrode is not segmented. The subdivision of the electrodes allows to reduce the energy in case of GEM discharge and to disconnect a part of the detector in case of a short-circuit between GEM foils. The number of sectors depends on the layer size: for each GEM, there are 4 macro-sectors for layer 1, 8 for layer 2 and 12 for layer 3. The electrodes of layer 1 are etched on a single sheet, while those of layer 2 and layer 3 on two sheets, glued together.\\
The HV distribution for layer 1 is shown in figure \ref{fig:HV_scheme}. The electrodes are supplied from both sides of the cylinder ("gas in" and "gas out"), to optimize the space.\\
CAEN A1515CG boards \cite{A1515} are used as the main generators, housed in a SY4527LC  mainframe \cite{SY4527LC}. CAEN A1515CG are boards used to power triple-GEM detector, and provide seven floating outputs on the corresponding detector electrodes. The HV reference ground is connected via a \SI{10}{\kilo \ohm} resistor to the ground plane of each layer.
The boards are interfaced through a Radiall-REDEL converter to a custom passive distribution patch-panel that splits the seven inputs into all the levels required to power the detector. Each output connector supplies two macro-sectors and the related micro-sectors. A complete panel supplies up to 18 macro-sectors and the related micro-sectors. The panels allow to disconnect single sectors for diagnostic purposes or to isolate a shorted micro-sector.\\
The power supplies, the Radiall-REDEL converter and the distribution panel will be placed on the platform on top of the spectrometer. 
Braided and halogen free cables, rated up to \SI{4.5}{\kilo \volt}, are used for the off-dectector routing to bring the power from the patch panels to the HV connectors cards inside the spectrometer. Appropriate shielding of these \SI{18}{\meter} long cables is essential to lower the HV induced pick-up noise of the system. The \SI{2}{\meter} long cable for on-detector interconnections has to be lightweight, multi-core and flexible to allow routing in confined spaces. Custom cables with small cross-section and low-weight  were assembled using extruded FEP insulated cables. Custom connectors for off-detector/on-detector connections were designed for the same reasons of size and weight. The connectors consist of two boards which are plugged into each other. Electrical insulation is provided by a protective coating and 3D-printed plastic housings.\\
The connectors to the CGEM-IT are similar to those of interconnections. Since the tight space does not allow to install any housing, the electrical insulation is ensured by the board clearance and by the insulating coating. They mount protection resistors (\SI{1}{\mega \ohm} for macro-sectors and \SI{10}{\mega \ohm} for micro-sectors). The coupling between the protection resistors and the macro/micro-sectors capacitance ensures RC filtering on the HV lines. \\
All power supplies are completely floating and each supply has its own return path. The entire detector is contained within a Faraday cage. The detector module is connected to earth at only one point (star structure).
\chapter{GEMROC firmware optimization}
\label{cap:firmware}
In order to achieve the performance goal required by the CGEM-IT project and to be able to acquire stably
under different conditions, the GEMROC firmware (described in section \ref{subsec:firm}) had to be optimised and patched. This chapter describes some of the issues discovered during testing and acquisition and the corresponding solutions.
\section{Packets shift}
During acquisition, a fairly quite common bug in the trigger-matched UDP packets assembly was discovered early on. The GEMROCs firmware builds the UDP event packets in the \textit{Four FEB merger TM} module. This module receives the hits selected by the trigger-matching logic for each FEB and merges them, first in two pairs and then all together. It also appends the header and trailer information to the data, which are then downstreamed for transmission.\\
The header and trailer information have already been partially filled in the previous phase and the \textit{Four FEB merger TM} module joins them together by selecting the header from the data of the first FEB pair and the trailer from the data of the second FEB pair.\\
After some time of acquisition, especially with sustained data input, the state machine handling the process used to lose synchronization between the first and second FEB pairs. It built packets with the header and the data of the first FEB pair, corresponding to the nth trigger, while the trailer and data of the second FEB pair corresponded to the (n-1)th trigger.\\
The problem was due to an asymmetry in the state machine that manages the merging of the four FEBs data: if the first two FEBs data were ready two clock cycles before the end of the assembly process of the last two FEBs, the data packet was incomplete and the data of the last two FEBs were sent in the next packet. This shift could also extend to more than one trigger.\\
Prior to the firmware fix, the problem was corrected via software that detected the shift in the data stream and reassigned the hits to the correct event. However, this patch was not able to correct all shifts and sets limits on online data quality control.\\
Two patches were developed to address this bug. The first patch was to make the state machine fully symmetric, with flags for completion of the process mandatory for both the FEB pairs to complete the packet process. \\
We also decided to synchronize the whole process by requiring that a trigger be fully processed (by raising a flag from \textit{Four FEB merger TM} module) before starting the next trigger procedure, holding the trigger information in a buffer in the \textit{L1 distributor} module.\\
The effects of the patches were checked  at the acquisition facility for planar detectors in Ferrara (see \ref{setup_fe}) in similar runs of $\sim$ 8000 triggers. We went from $\sim$ \SI{50}{\percent} of mispacked data corrected by the software patch and \SI{1.2}{\percent} of uncorrectable data to \SI{0.6}{\percent} mispacked data and \SI{0}{\percent} uncorrectable data. \\The remaining mispacked data are due to other bugs that will be addressed in the future.\\
\section{Trigger-matching bugs}
The efficiency of trigger-matching was tested under different conditions. To perform a controlled test, but with a signal source independent from the GEMROC clock, a different FPGA was used to generate normally distributed signals. These signals were fed into the GEMROC and used to generate test pulses on four TIGER channels, and triggers  that were timed to extract those test pulses. The other TIGER channels were disabled and the threshold of the pulse-driven channel was high, so only the generated signal was collected by the GEMROC.  When the data were acquired in triggerless mode, all the test signals were present, but only about \SI{98.5}{\percent} were present in the trigger-matching mode.\\
To determine the causes of this inefficiency, a dedicated signal was added to the firmware that rises each time there are only three hits in the packet being sent. By triggering Quartus Signal-Tap logic analyzer (see appendix \ref{cap:third_party}) it was possible to read the contents of the FPGA registers at the time of the error.\\
The problem was in the timing of the read operations from DPRAM memory. The time required to read was underestimated by one clock pulse, so that when the read addresses were scanned, the data read always referred to the previous memory position.\\
When reading two pages of 32-bit memory, this resulted in last memory position of the second pages never being read. Since the hit position was random within the time-addressed pages, one hit per 64 was lost: about the \SI{1.6}{\percent}.\\
During these tests, another bug was discovered. Each time the trigger arrived with a timestamp of exactly zero, the hit extraction time frameword was miscalculated. Since the trigger timestamp counter has 32 bits, the impact of this bug on efficiency was very small. Nevertheless, it was fixed with a simple code adjustment.\\
After these patches, the efficiency of the trigger-matching in the test described above was of \SI{100}{\percent}, tested with one million triggers.\\
Testing the algorithm with some noise on the channels (as tested during the test beam and described in section \ref{sec:tb_thr_setting}) showed a small loss of efficiency, of up to \SI{3}{\percent}, although the test conditions were well below the maximum buffer capacity of the system. Further studies will be conducted to fully understand and optimize the process.
\section{Trigger glitch sensitivity}
\begin{figure}[h]
\centering
  \includegraphics[width=1\linewidth]{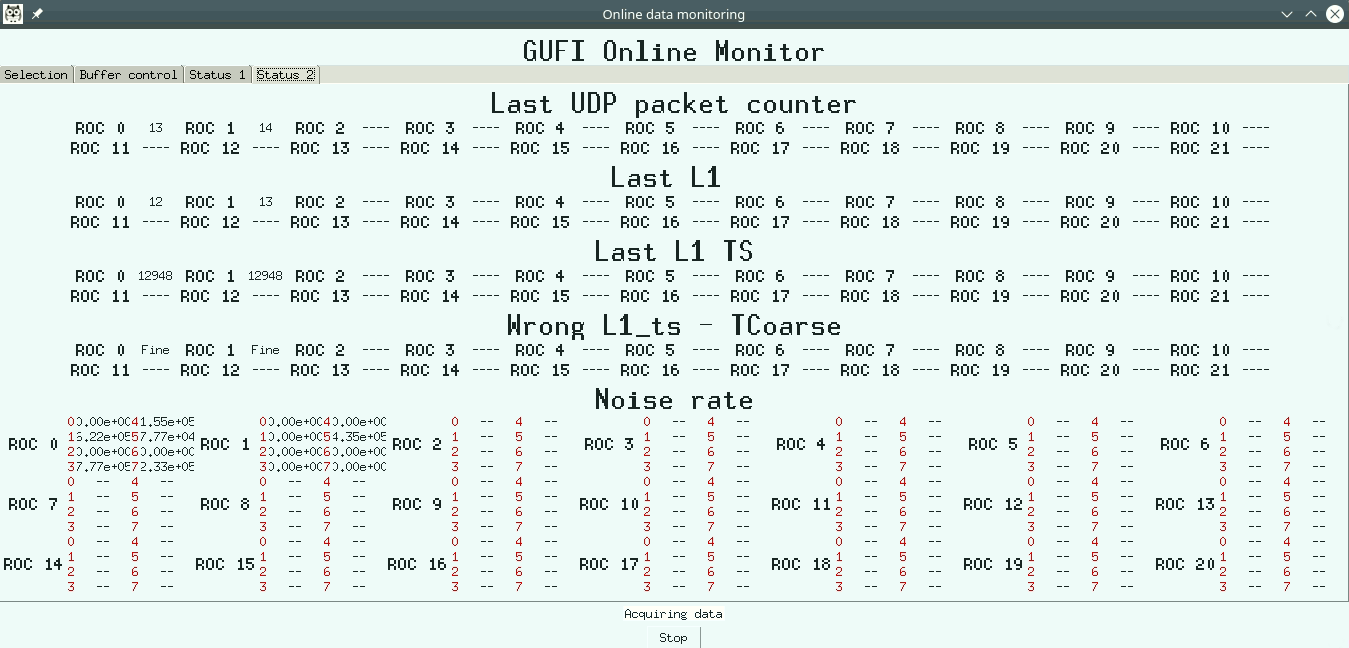}
  \caption{Screenshot of the GUFI online monitoring page created to investigate the incorrect trigger alignment. During this acquisition, both GEMROC marked the last trigger with the same timestamp (12948), but with different trigger numbers (12 and 13). }
  \label{fig:trig_bug}
\end{figure}

The tests with the planar detectors (\ref{setup_fe}) were performed with two GEMROCs reading eight FEBs. During these tests,  the triggers detected by the two GEMROCs sometimes did not match. To investigate this problem, a GUFI online monitoring tool (see chapter \ref{cap:software}) was developed (figure \ref{fig:trig_bug}).
One of the GEMROC seemed to count one or more additional triggers compared to the other. \\
Assuming that this was due to glitches in the transmission line of the trigger signal, a control for the length of the trigger signal was added.\\
According to the specifications, BESIII trigger signal lasts eight BESIII clock cycles, so the firmware was updated to require a minimum length of seven clock cycles in order to accept and process a trigger. This filter eliminates the problem of incorrect trigger alignment.
\chapter{Data Acquisition Software}
\label{cap:software}
Control software was developed to characterize, debug and test the system before the installation. This software (called Graphical User Frontend Interface - GUFI), written in Python, handles electronics operations and acquisitions. The software has been developed over the years according to the needs of the working group.\\
Third-party software and libraries mentioned in this chapter are discussed in more detail in appendix \ref{cap:third_party}.
\section{GUFI}
\begin{figure}
\centering
  \includegraphics[width=0.9\linewidth]{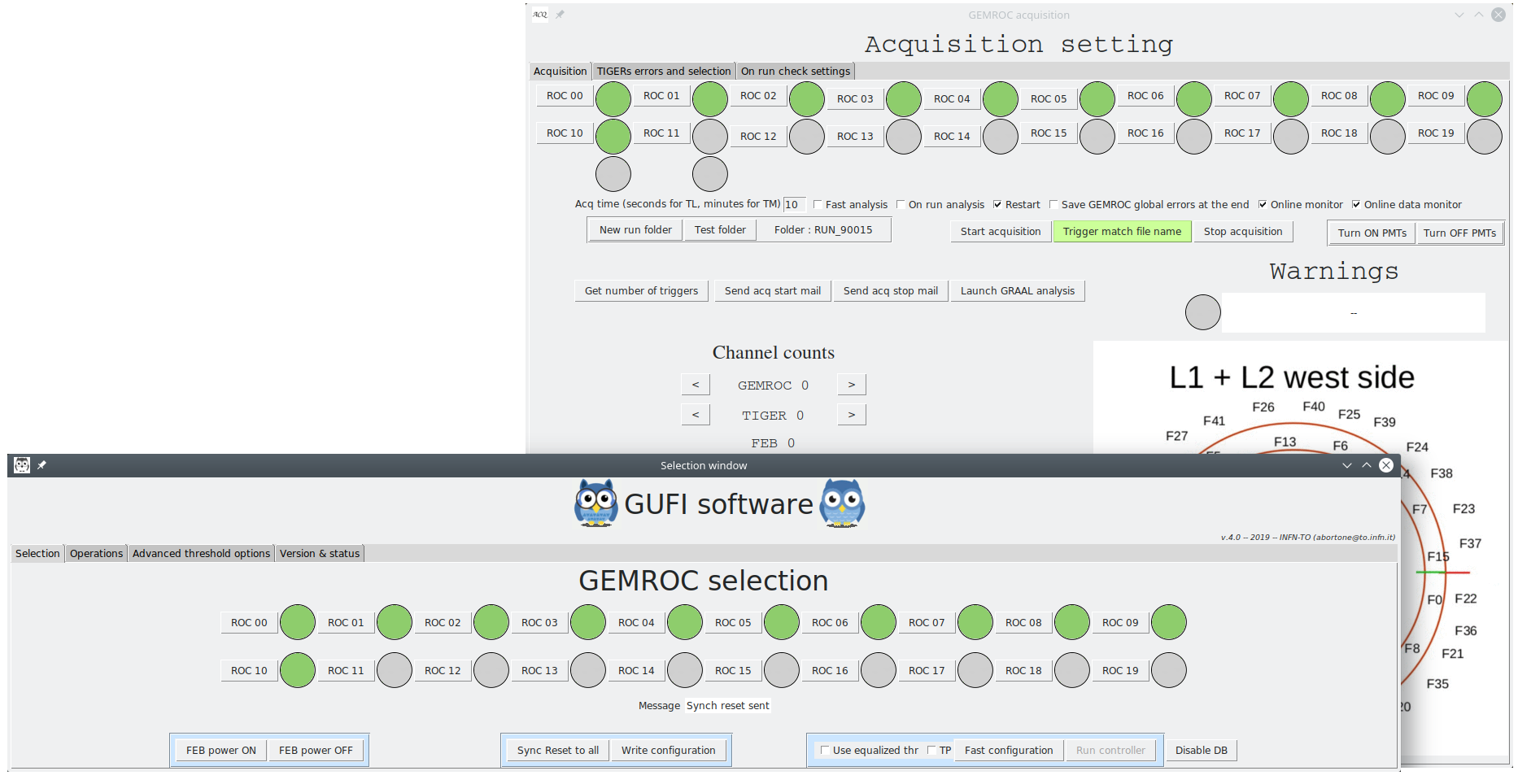}
  \caption{The front-page and the acquisition page of the GUFI software.}
  \label{fig:gufi_page}
\end{figure}

The software is written with an object-oriented approach and structured in classes  to be easily scalable and extensible.
The core of the software consists of the classes contained in the files \textit{GEM\_CONF\_classes.py} and \linebreak \textit{GEM\_COM\_classes.py}.\\
The \textit{GEM\_CONF\_classes.py} contains the classes for configuring the entire system: two classes for the GEMROC (one for DAQ configuration and one for power settings) and two classes for TIGER (one for TIGER global configuration and one for TIGER channel configuration).\\
The \textit{GEM\_COM\_classes.py} contains two classes: a handler that identifies the physical GEMROC and the \textit{GEM\_CONF\_classes.py} instances related to that GEMROC and the corresponding TIGERs, and the communication class that manages the UDP settings and provides the functions to write the different configurations to the GEMROCs and to the TIGERs.\\
Other high-level classes are contained in the files \textit{GEM\_ACQ\_classes.py} and \\ \textit{GEM\_ANALYSIS\_classes.py}. Data acquisition is managed by the classes contained in the \textit{GEM\_ACQ\_classes.py}, for both the trig\-ger-match\-ed and the trig\-ger\--\-less modes. To manage large bitrates during data acquisition, the number of write operations on disk is reduced by a RAM buffer of customizable length. Data are saved in folders (runs), and divided into smaller units called sub-runs to ease the data management and the analysis. The length of the runs and sub-runs can be determined by both the acquisition time and/or  the number of triggers acquired.\\
The file \textit{GEM\_ANALYSIS\_classes.py} mainly contains the functions needed to perform the threshold scan, estimate the noise, and set the thresholds.\\
On top of these modules, various user interfaces can be initialized to ease the user experience and perform advanced operations. All graphical interfaces are based on the \textit{tkinter} Python library.\\
The main Graphical User Interface is contained in the \textit{conf\_GUI.py} file (figure \ref{fig:gufi_page}). This interface can handle communication with up to 20 GEMROCs and is designed to keep the configuration of a large number of TIGER (up to 160) and channels (up to 10240) quickly and easily. Configuration can also be scripted to make it easy for a user to load them and operate the system.\\
To manage the acquisition, another GUI, \textit{acq\_GUI.py}, can be run standalone or from the main GUI. This interface manages the acquisition and the sub-run division by the \textit{GEM\_ACQ\_classes.py} instances and, when started from the main GUI, can also monitor TIGER and GEMROC parameters, reconfigure the system, and handle acquisitions independently.\\
The classes in the file \textit{DB\_classes.py} are not necessary for GUFI operation, but can be enabled and used  to log some system metrics to a database for monitoring and analysis purposes (see section \ref{sec:on_monitoring}).
A library of functions was extracted from GUFI and converted into C++ for integration into the BESIII DAQ software. The library is currently in the debugging phase and will be soon used for the CGEM-IT interface within the main DAQ software\footnote{In development by ZENG Tingxuan (IHEP) }.\\
\section{Online monitoring}
\label{sec:on_monitoring}
\begin{figure}
\centering
  \includegraphics[width=0.8\linewidth]{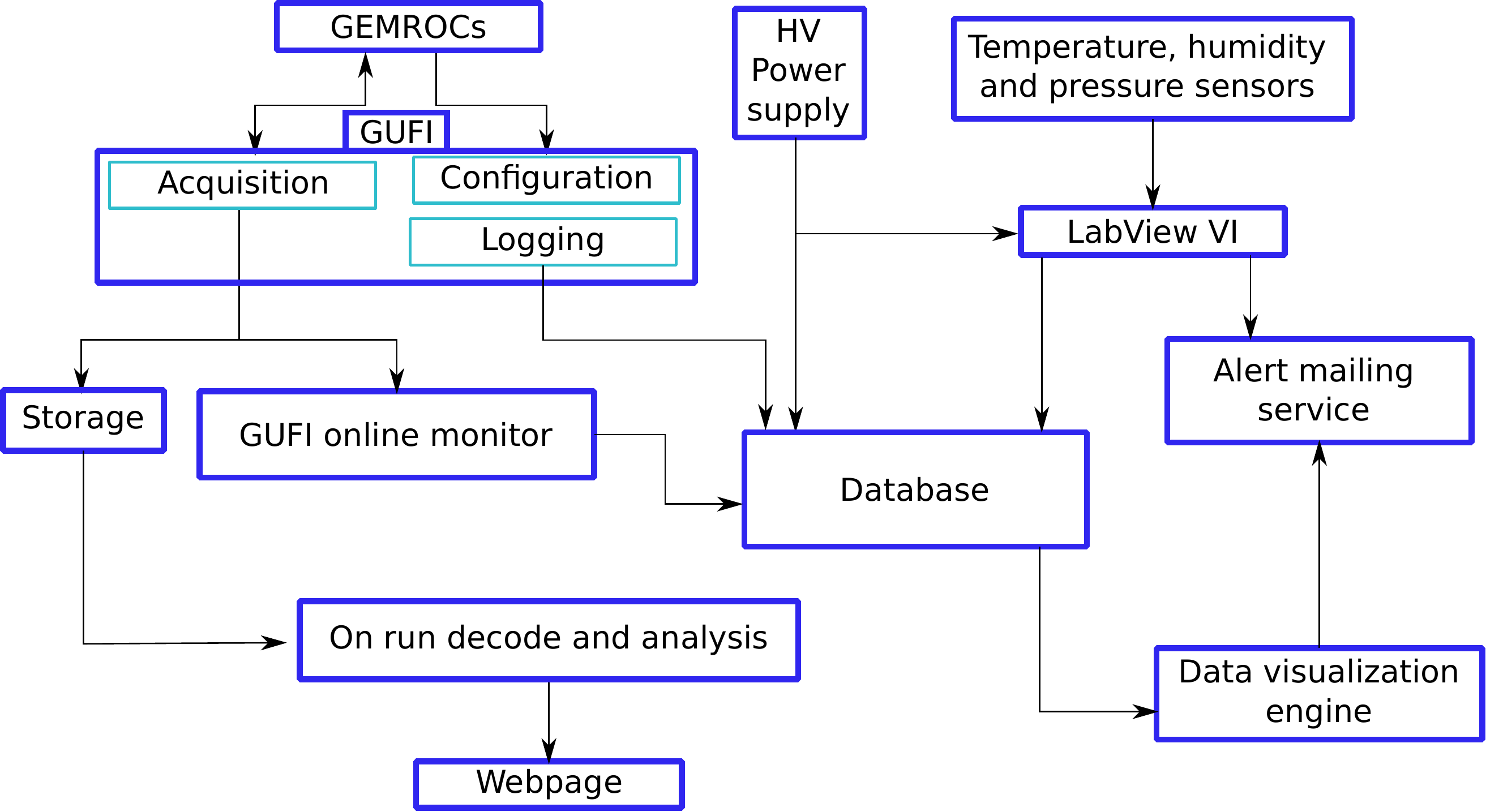}
  \caption{The acquisition and test control software infrastructure. Many parts of the software can be installed and run in different computers, since the communication takes place via UDP or TCP protocol.}
  \label{fig:gufi_scheme}
\end{figure}

To ensure security and data quality over long acquisition periods, an online monitoring system was implemented. This infrastructure was even more important during the pandemic outbreak: even if it was not possible to physically reach the detectors in China, we were able to keep the detector on and collect data remotely while ensuring the safety of the operations.\\
The described system was implemented for the Beijing cosmic stand, consisting of the two innermost layers of CGEM-IT. Due to its high scalability and modularity, the whole system or part of it can  be easily implemented for other setups.\\
The overall scheme of the acquisition and monitoring software can be seen in figure \ref{fig:gufi_scheme}.
The core of the system consists of an InfluxDB database.\\ 
The database can run on any PC that is accessible to the acquisition system, and can be populated by any other program via http requests. Many data sources feed the database:
\begin{itemize}
    \item GUFI can provide a stream of data on voltage, current and  temperature of TIGERs and GEMROCs, both during the acquisition or in idle state.
    \item GUFI can also send a copy of each data packet via UDP communication to another software, called GUFI Online Monitor. This software monitors the data stream and provides entries in the database about the bit rate and the number of incoming packets from each GEMROC. It can also perform a fast decode of the packets, to log the rate of incoming signal and noise from each TIGER and run hit-level analysis to produce output plots of detector performance.
    \item Three sensors, controlled by an Arduino board, monitor temperature and humidity in the room. A LabVIEW program reads the sensors and logs the information to the database.
    \item A custom LabVIEW program was developed to manage the HV power supply to the CGEM layers. This software can also log the monitored current information to the database. A simple Python script is also used to extract the complete HV information from the LabVIEW text logs and log it to the database.
\end{itemize}
The data stored in the database can be queried using SQL-like query strings sent over the http protocol to the running instance of the database to be aggregated with other data or to perform analysis (see chapter \ref{cap:agg_analysis}).\\
The online alerting system and data visualization are implemented using Grafana software. 
The Grafana user interface is organized into dashboards that contain related metrics and provides an automatic alert system when  values exceed set thresholds. Alerts have been set to be sent to experts via email and via Telegram (where possible). The Grafana interface was also configured to display the result of the GUFI Online Monitor fast analysis.\\
During the last year, a shift system has been implemented by the BESIII Italian Collaboration in order to run the detector in China. For each shift, a DAQ expert, a detector expert and a user are requested. While the experts are responsible for switching the system on and off and intervening in case of malfunctions, the user is responsible for checking the online monitor system and alerting the experts when necessary.\\

\section{CIVETTA}
\label{civetta}
CIVETTA (Complete Interactive VErsatile Test Tool Analysis) is a fast-analysis software designed to provide a quick feedback on data collection. Written in Python, integrates all the steps to obtain complete metrics about the detector performance.  The software is fully parallelized at the subrun level,  to use all the CPU on the machine and maximize performance. The software was initially written to operate with a setup made of four planar GEM detectors, tested with cosmics rays and during the CERN test beam (see appendix \ref{setups}), but can also be adapted for the cylindrical geometry of the CGEM-IT .\\
The analysis is divided in different steps, each can be run autonomously. \\
\subsubsection*{Decoding}
\begin{figure}
    \centering
    \includegraphics[width=1.7\linewidth, angle=90]{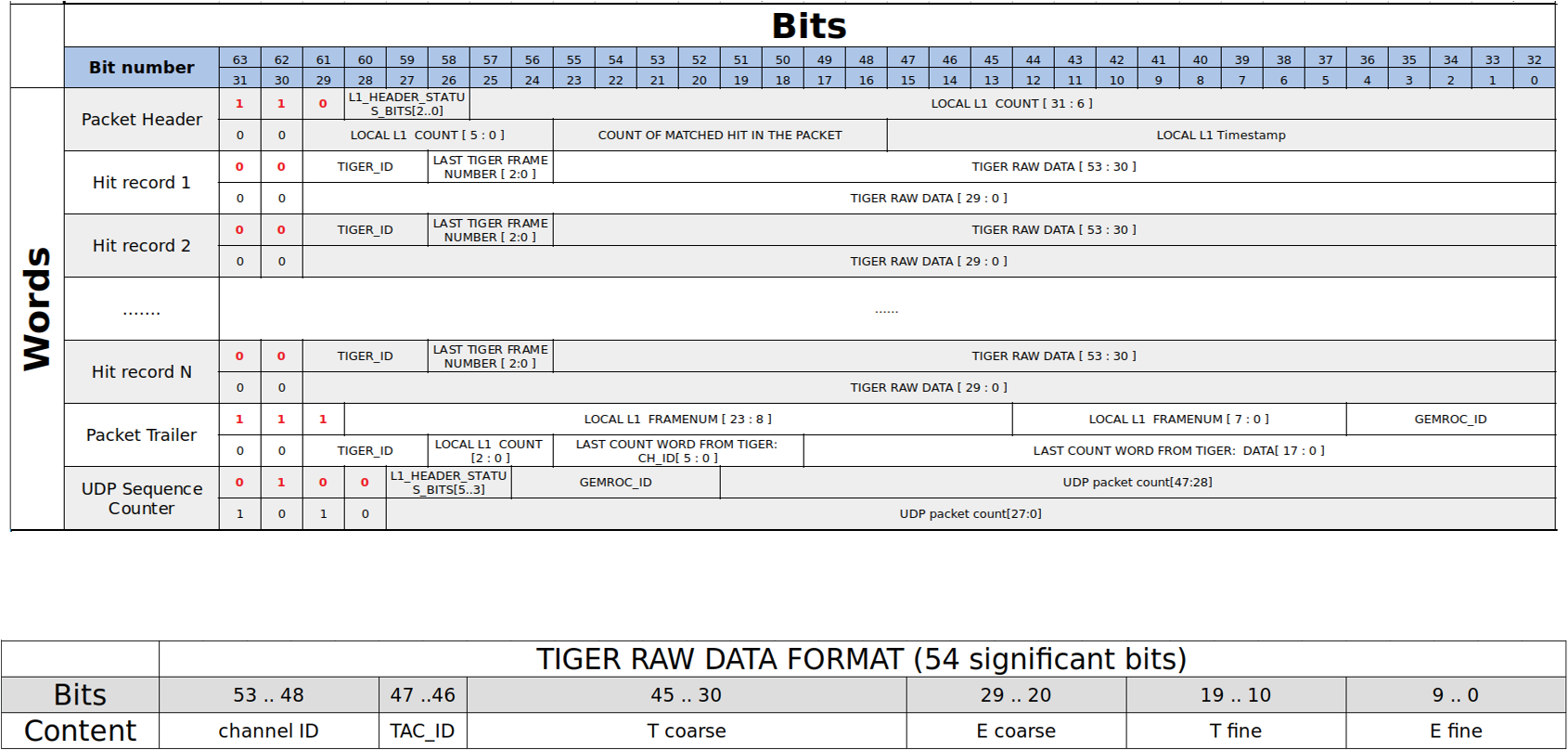} 
    \caption{Trigger-matched packet data format and TIGER hit word data format}
    \label{fig:data_format_TM}
\end{figure}

The data are stored on disk in binary format, as they arrive in the GEMROC UDP packets. CIVETTA is designed to work with the data packets in trigger-matched mode (see section \ref{backend}). The first operation to be performed is to parse the binary data. The result of this operation is a pickle compressed file (appendix \ref{cap:third_party}), containing a Pandas dataframe in which all the hits are stored. To decode the data, the binary file is read eight by eight byte. Each byte is inverted to match the endianness of the data. A group of eight byte data forms a word. The data format is shown in figure \ref{fig:data_format_TM}.\\
Each data file (one for each GEMROC and sub-run) is decoded independently and then merged with the files of the same sub-run.\\

\subsubsection*{Mapping and Calibration}
A pickle mapping file is prepared prior to operations. This file contains the complete description of the channel-strip mapping and the identification number of each mounted FEB.\\
Using the mapping file, the strip position and the detector plane information is added to the decoded file. The hit charge is calculated using the calibration files and added to the dataframe. The software supports only the Sample-and-Hold mode, but the Time-over-Threshold mode will be implemented soon.\\
To speed up the process, the online data analysis during the test beam applies the mapping and calibration by vectorization, instead of iterating on each hit. This is a kind of approach, allows to save computation time on large data sets. For a test run, consisting of 1388 files (two GEMROCs for 694 subruns),  the processing time on a notebook for the decoding and calibration operations using iteration over ROOT files is $\sim$\SI{14}{\minute}, while using vectorization on the Pandas dataframe is $\sim$\SI{5}{\minute}. For a server-like system, with a higher number of cores, the difference is even larger.\\
After this operation the data are merged into a single Pandas dataframe containing all event hits.\\
\subsubsection*{Clusterization}
To perform the clustering, the single hit data are first selected by a time cut in the signal domain. The signal is in a time window whose size depends on the size of the drift gap size and the drift velocity of the electrons, and whose position depends on the trigger latency.\\
The software then builds the clusters on each view event by event. The clusterization algorithm is the following:
\begin{enumerate}
    \item If  there is at least one hit, the algorithm assumes the existence of a cluster (number of clusters $K=1$)
    \item Using a pre-built implementation of the K-means algorithm (included in the SciKit-Learn package), the position of the cluster is estimated.
    \item If all hits are close enough to their relative cluster center, the process terminates and the clusters found. The maximum distance can be adjusted to be more or less tolerant of missing strips in the cluster. The default maximum distance is $\frac{N_{Hits}}{2} +1$ where $N_{Hits}$ is the number of hits in the cluster.
    \item If one or more hits are too far from the center of the cluster, the number of clusters $K$ is increased by 1 and the process starts again at point 2.
\end{enumerate}
After clusterization, the position of the clusters is determined using the charge centroid algorithm.\\
To form the 2-D clusters, the cluster with the higher charge in both views is selected. This selection, although coarse, is a good discrimination between signal and noise.
\subsubsection*{Tracking}
For each event, if there are at least three clusters on different detector planes in a view, the software tries to fit them with a line. If there is more than one cluster on a detector view, the cluster with the highest charge is selected. This fit is used to make a rough estimation of the residual distribution and to better locate the signal clusters, which are selected if they are less than \SI{2}{\milli \meter} from the fitted track. The output of this procedure is a nearly clean signal cluster distribution, since the presence of a noise cluster in the selection usually invalidates the fit and thus produces no entry.\\
From the tracking information, a simple and fast estimate of the efficiency can be calculated for each detector plane and view, taking into account the tracks up to a certain tolerance from the residual peak and calculating the efficiency as $\epsilon=\frac{N_{view}}{N_{total}}$, where $N_{total}$ is the number of accepted tracks and $N_{view}$ is the number of tracks with a point on the examined view. To perform a better estimate of efficiency, a more complex and computationally intensive analysis must be performed (see \ref{perf_meas}).\\
\subsection{Data visualization}
\begin{figure}[h!]
\centering
  \centering
  \includegraphics[width=0.8\linewidth]{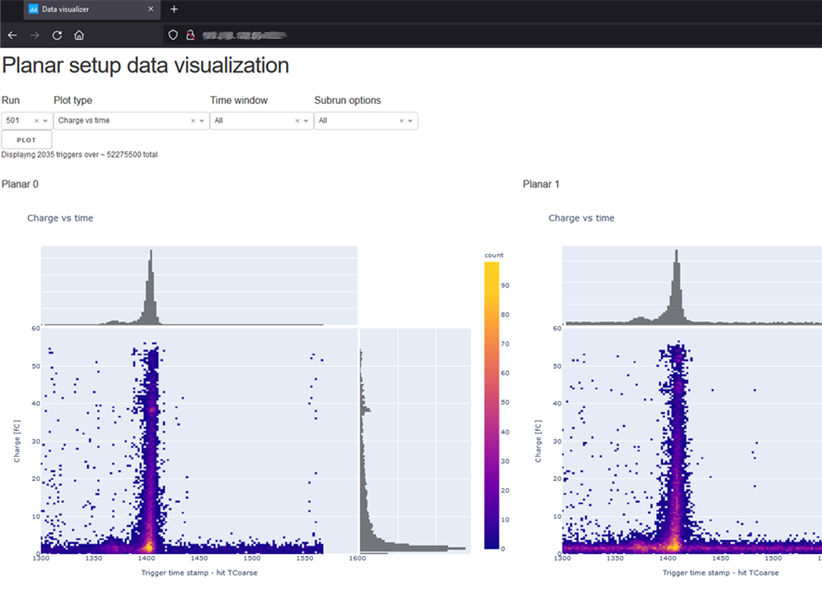}
  \centering
  \includegraphics[width=0.8\linewidth]{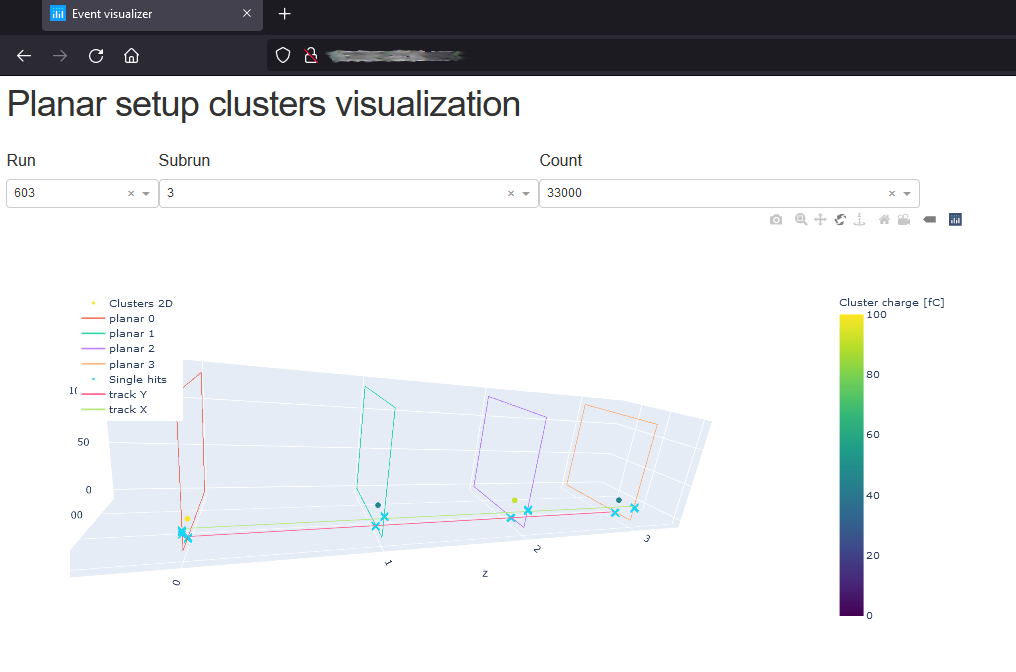}
 \caption{Example of the \textit{Data visualizer} (left) and \textit{Event visualizer} (right) dashboard pages.}
\label{fig:dash}
\end{figure}

The data generated by CIVETTA can be visualized by users a few minutes after the end of each subrun.\\
The open source \textit{Dash} platform is used to visualize  data stored in compressed Pandas dataframes. The application is deployed on a local machine running a simple Gunicorn server, a Python WSGI HTTP Server for UNIX.
This prompt online data visualization is important for detecting and fixing detector or electronics problems during a fast paced test campaign, such as the July 2021  test beam at CERN.\\
The dashboards are organized into three pages:
\begin{itemize}
    \item The \textit{Data visualizer} page contains cumulative plots of runs. The metrics available are: hit charge w.r.t. time, strip occupancy, distribution of charge and size for both 2D and 1D clusters, 2D cluster heat-map, track residuals and coarse efficiency assessment.
    \item The \textit{Event visualizer} page is used to display single events, showing single hits, 2D clusters and track fits. This page is particularly useful for examining the performance of the analysis code and studying the characteristics of inefficient events.
    \item The \textit{Trigger statistics} page contains information about data acquisition, such as the number of triggers acquired for each run and the time of each acquisition.
\end{itemize}
For an example of the \textit{Data visualizer} and \textit{Event visualizer} pages, see figure \ref{fig:dash}.

\subsection{Performance measurement}
\label{perf_meas}
To obtain a more accurate estimate of performance, CIVETTA has slower and more computationally intensive offline routines.\\ In particular, the software aims at providing  a good estimate of the efficiency and  resolution of the planar detectors, taking into account almost all related parameters.
\subsubsection{Alignment}
\begin{figure}[h!]
\centering
  \centering
  \includegraphics[width=0.9\linewidth]{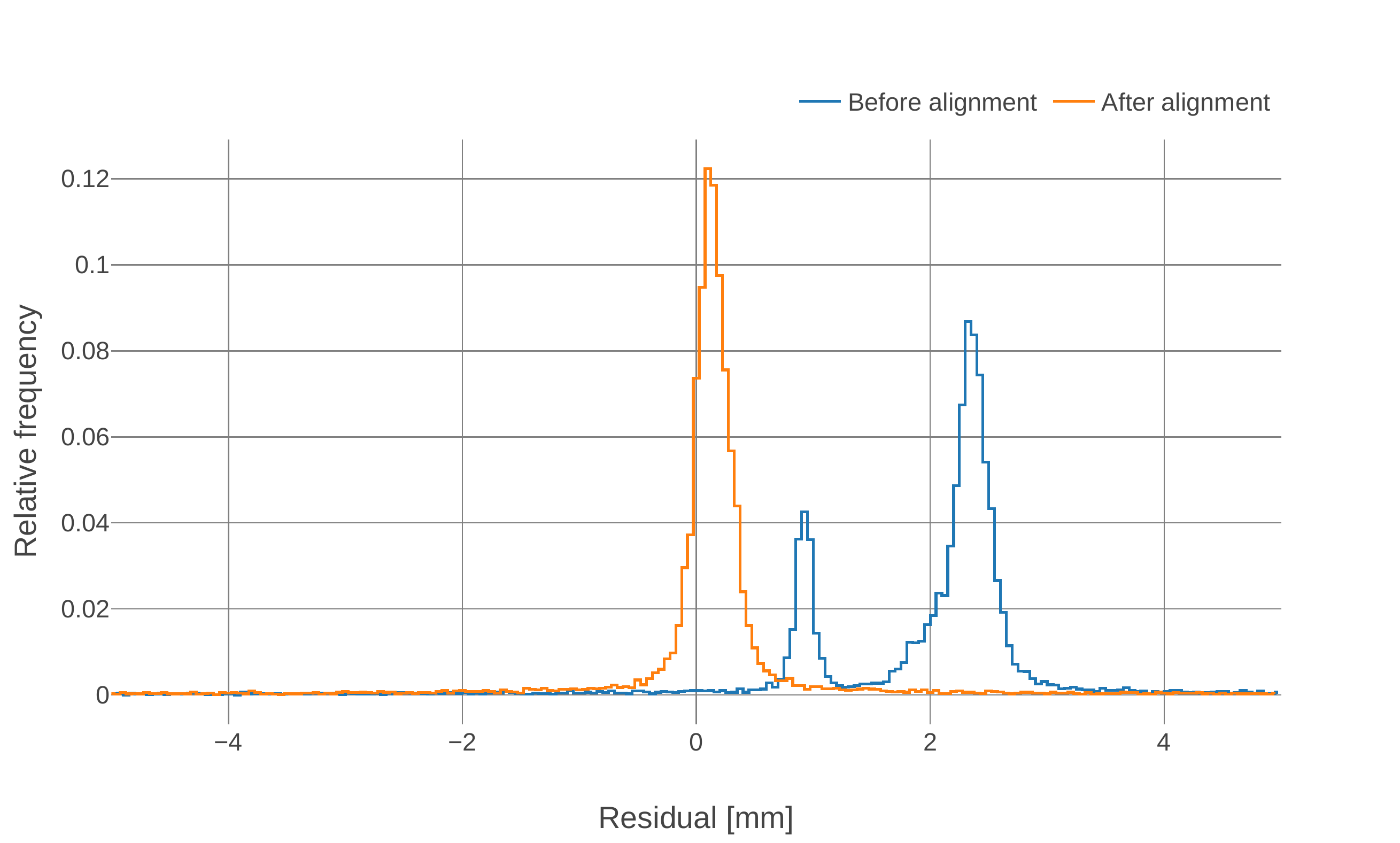}
\caption{Distribution of residuals before and after the alignment procedure. Histogram integrals are normalized to one.}
\label{fig:res}
\end{figure}

Even when mounted on a rigid mechanical structure, detectors are often slightly misaligned. Unfortunately, both efficiency and resolution are affected by this misalignment. Therefore, it is necessary to minimize the impact on performance by properly correcting the position of the strips with respect to each other. \\ For simplicity, since $z$ is the direction perpendicular to the detectors and $x$,$y$ are the directions parallel to the sides of the detectors, only the displacement in the $xy$ plane is corrected, which is considered to be the main factor.\\
An iterative process is used to align the detectors. For each view and plane, one view at a time, a fit of tracks is performed. The residual distribution of these tracks is calculated. Then the position of the strips is shifted so that the peak of the residual distribution moves to zero. The process is repeated twice  for each view and plane in the same way.
At the end of the process, the residuals are better centred at zero and the standard deviation of the residual distribution is slightly reduced (figure \ref{fig:res}).\\

\subsubsection{Efficiency estimation}
The procedure for estimating  the efficiency is described below and has to  be performed from the beginning for each view and plane of the detector after  alignment. For four detection planes, the plane for which we want to estimate  efficiency is called plane under test, while  the other three planes are defined as trackers. 
\begin{itemize}
    \item A track fit is performed for the most charged clusters on the trackers.
    \item The residual distribution of all the tracks for a given run is formed. The mean ($\upmu_T$) and the standard deviation ($\upsigma_T$) of the residual distribution of the entire run are extracted.
    \item For each event, if all trackers have residuals below $2.5 \upsigma_T$ from $\upmu_T$, the track is considered good. These steps ensure us to consider only good tracks.
    \item By considering only the good tracks, the projected position on the plane under test is calculated from the fit.
    \item The non-inclusive residuals between the projected position and the nearest cluster position on the plane are calculated.  The mean ($\upmu_U$) and the standard deviation ($\upsigma_U$) of the non-inclusive residual distribution are extracted.
    \item If the projected position is outside the edges of the plane under test, or within $5 \upsigma_U$ from such edges, the event is discarded. Other conditions may be imposed on the projected position. If the projection meets these requirements, the event is included in the denominator of the efficiency formula and counted. 
    \item If there is at least one cluster within $5 \upsigma_U$ from the projected position, the event is considered \textit{efficient} and the numerator of the formula is increased by one.
    \item The information about \textit{efficient} and \textit{inefficient} events is stored for further analysis.
\end{itemize}

\chapter{Noise measurements and optimization}
\label{cap:noise}
The noise measurement and its reduction is critical for optimising the detector performance. \\
The CGEM-IT poses special challenges for noise optimization: the detector layers are quite large compared to other GEM detector setups, so the strip and inter-strip capacitances are larger. In addition, the readout on both sides is quite far apart, so it is more difficult to define  a solid ground level for the whole readout.\\

\section{Noise amplitude}
\begin{figure}
\centering
  \includegraphics[width=0.6\linewidth]{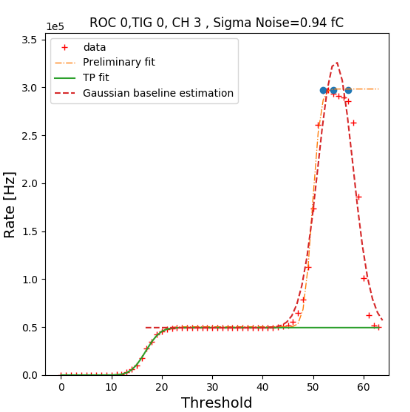}
  \caption{Threshold scan with a periodic test pulse at $\sim$ \SI{50}{\kilo \hertz}, as output by GUFI. The plot shows the acquired data, a preliminary fit to estimate the test pulse and the baseline zone, the test pulse fit to estimate the noise, and the Gaussian fit to estimate the baseline. The TIGER threshold DAC operates in inverse logic, so higher values correspond to lower thresholds.}
\label{fig:TP_scan}
\end{figure}

\begin{figure}
\centering
  \includegraphics[width=0.9\linewidth]{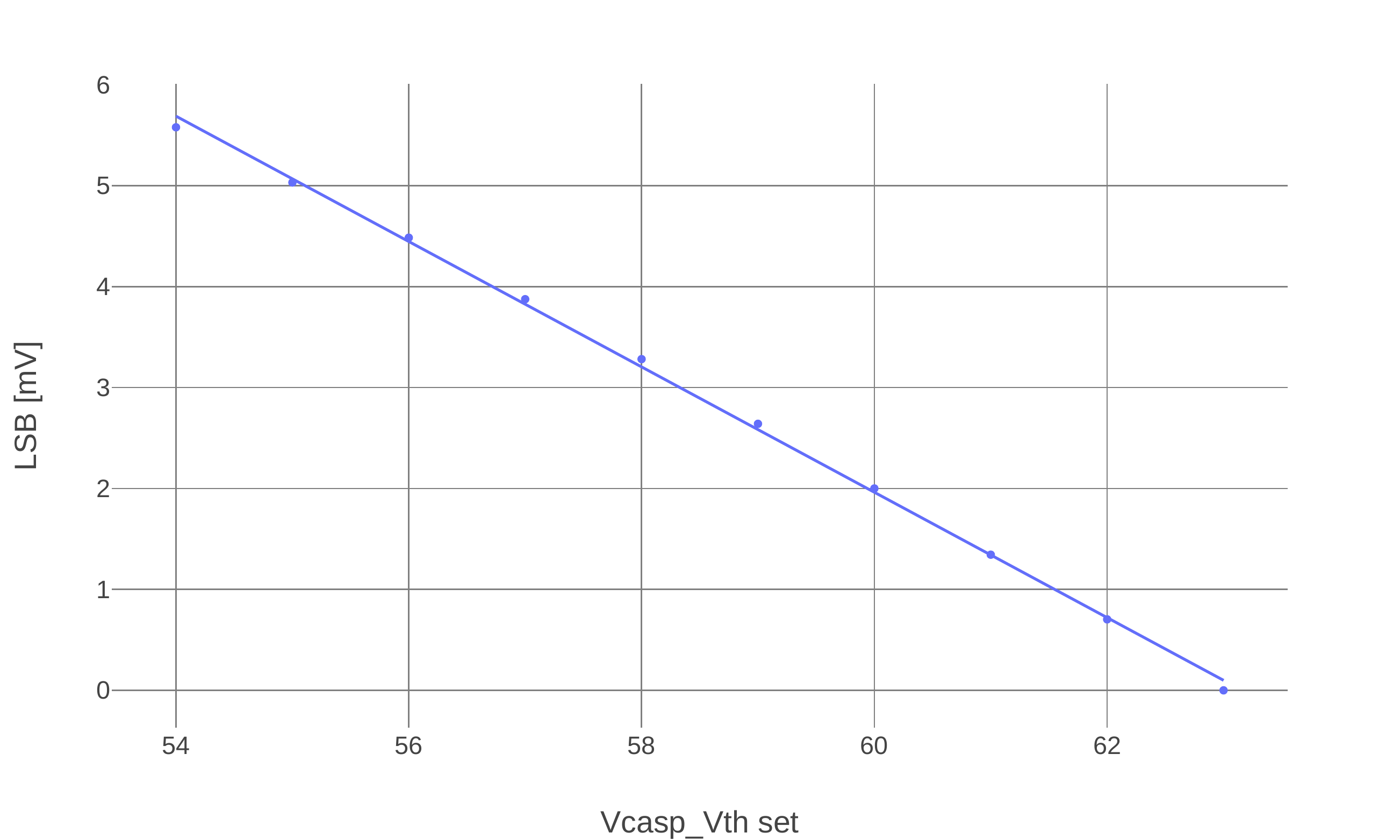}
  \caption{LSB value, in mV, with respect to the \textit{VCasp\_Vth} setting of TIGER. The measurement was performed in a range of \textit{VCasp\_Vth} values useful for chip operations. }
    \label{fig:vcasp_lsb}
\end{figure}

\begin{figure}
\centering
  \includegraphics[width=0.9\linewidth]{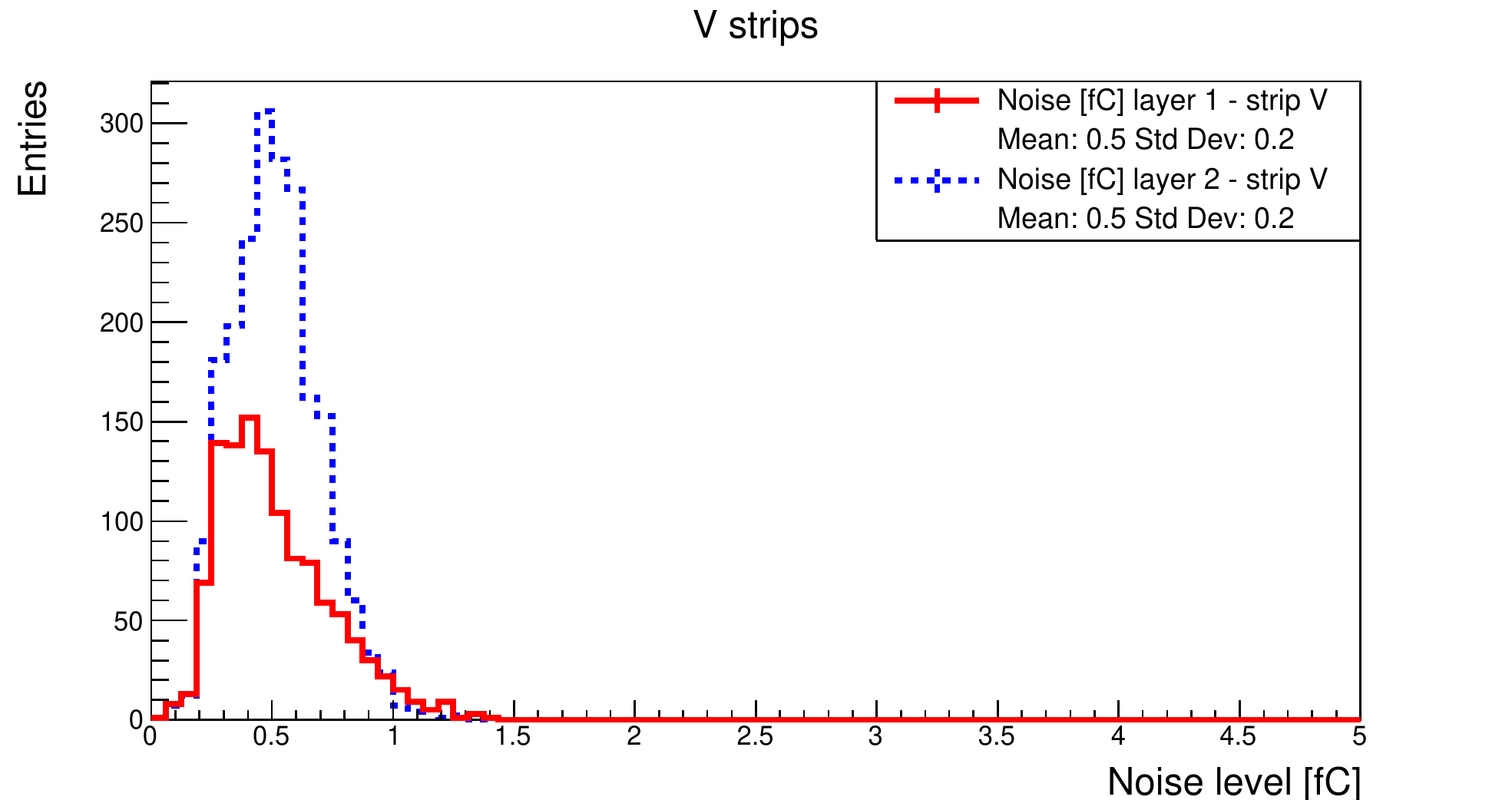}
\centering
  \includegraphics[width=0.9\linewidth]{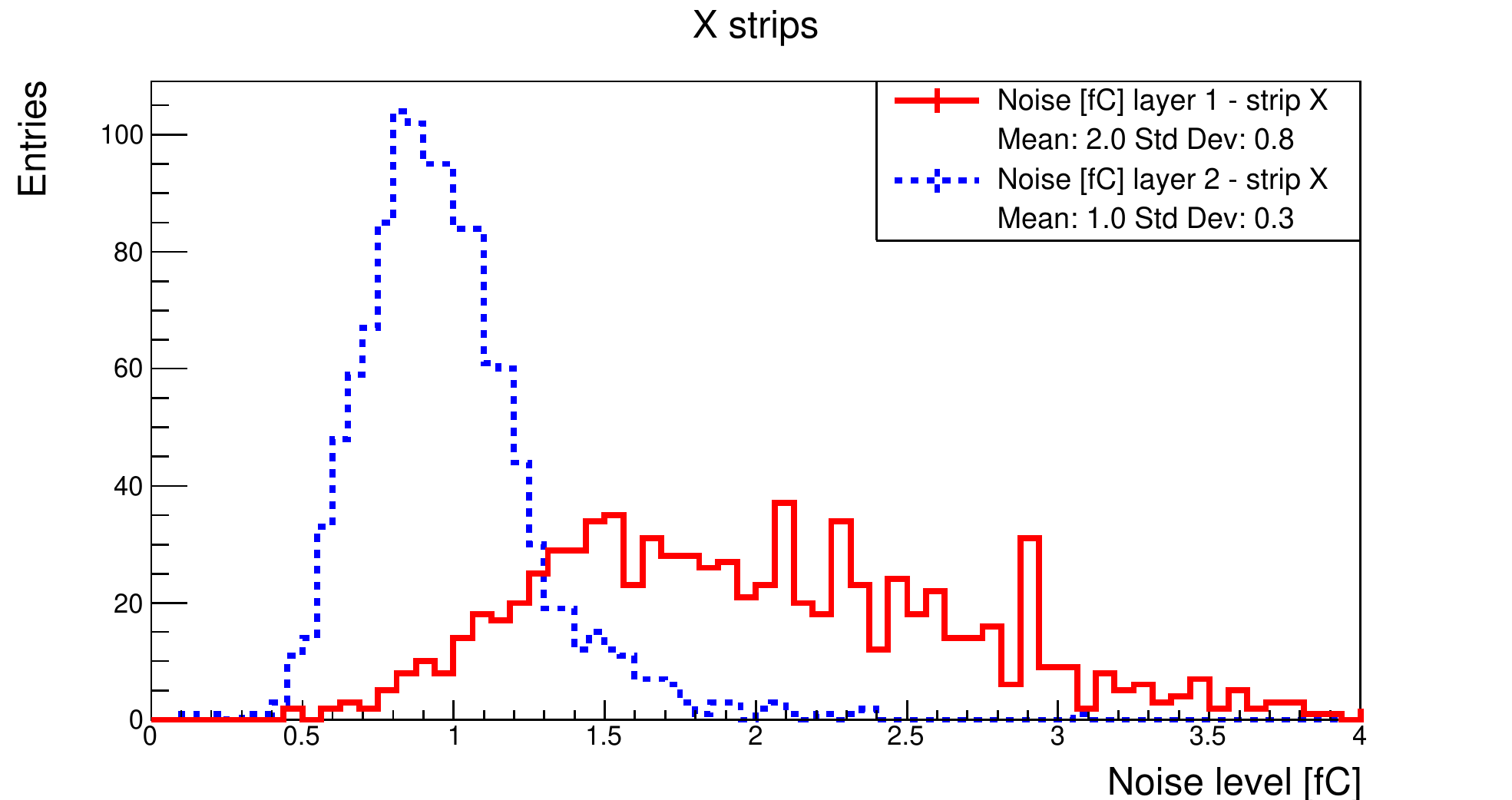}
  \caption{Noise level measured on layer 1 and layer 2, both for X and V strips, using the cosmic ray test setup in Beijing.}
    \label{fig:noise_x}
\end{figure}

\begin{figure}
\centering
  \includegraphics[width=1\linewidth]{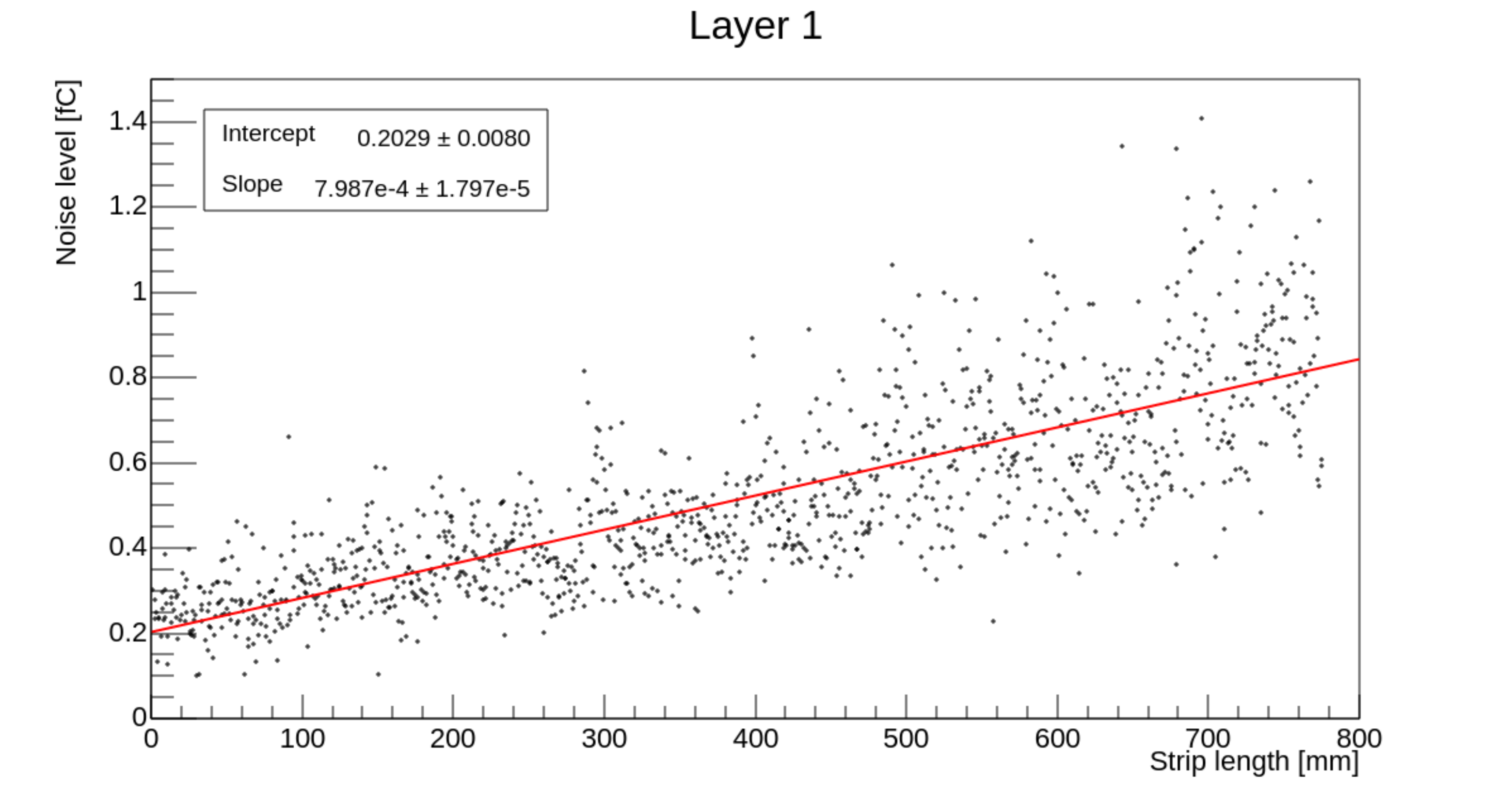}
\centering
  \includegraphics[width=1\linewidth]{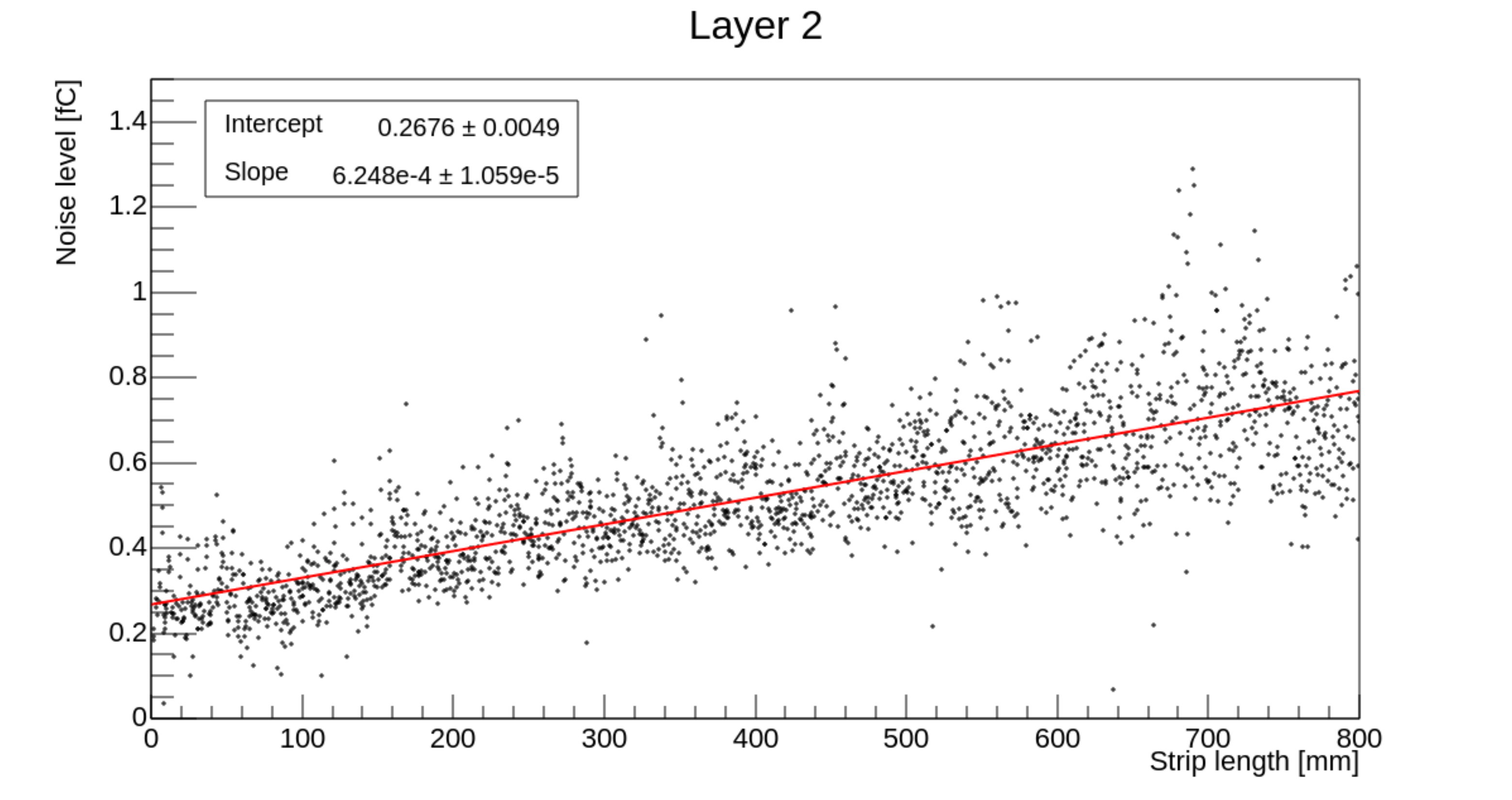}
  \caption{The trend of noise versus  V strip length is due to the increasing capacitance of longer strips. The mechanical differences between the two layers determine the differences in slope and intercept. The parameters of the linear fit are given in the legend.}
    \label{fig:noise_length}
\end{figure}

The GUFI software (see chapter \ref{cap:software}) contains the tools to measure both the noise amplitude and the noise spectrum. To measure the noise amplitude, one test pulse at a time is sent to each TIGER channel. While the test pulse is sent at a fixed rate, the threshold is swept across all possible DAC values. Assuming that the noise amplitude is Gaussian, the rate curve can be  automatically fitted to give an estimate of the standard deviation for the noise distribution.
The resulting rate vs threshold plot is shown in figure \ref{fig:TP_scan}.\\
Given $\eta$ the detection efficiency of the test pulse, defined as $\eta=\frac{\text{Rate detected}}{\text{Rate injected}}$, and assuming that the noise amplitude distribution is Gaussian, we can estimate the standard deviation of the noise distribution by fitting the threshold scan data with an error function \cite{Rivettifront}:
$$
\eta(T) =\left[ erf \left( \frac{T-\mu}{\sqrt{2}\sigma} \right) + \frac{1}{2} \right]
$$
where $T$ is the threshold in digits, $\mu$ is the test pulse amplitude and $\sigma$ is the estimate of the Gaussian noise standard deviation. The fit gives the standard deviation of the noise in units of the LSB value of the threshold, which can be converted to charge units as:
$$
\sigma_c =\sigma \cdot \text{LSBV}/ G  
$$
where LSBV is the LSB measured in mV and $G$ is the branch gain. The branch gain can be approximated to \SI{12.25}{\milli \volt \per \femto \coulomb} for both branches \cite{fabio_tesi}, while LSBV depends on the settings of TIGER. Two TIGER configuration fields control the LSB value for the branch T and E, respectively \textit{VCasp\_Vth} and \textit{VCasp\_Vth2}. The LSB versus \textit{VCasp\_Vth} trend was measured using an access point present on the FEB, which allows the threshold set on the channel 63 to be measured directly with a multimeter (figure \ref{fig:vcasp_lsb}). A linear fit was performed which resulted in a slope of -0.62 mV/digit and an intercept of 39.2 mV. These values were used for both branches. \\
Using this procedure, the noise condition of the entire setup in Beijing (see section \ref{setup_beijing}) can be assessed (figure \ref{fig:noise_x}). The difference in the noise levels of X strips for layer 1 and layer 2 can be explained by taking into account the different mechanical structure \cite{Balossino_2020}: layer 1 uses a carbon fiber anode support structure, which increases the capacitive coupling of adjacent X strips. Note that the noise level of the V strips shows a trend with the strip length, as expected (figure \ref{fig:noise_length}).\\ These scans are also a diagnostic tool, useful for checking the status of each channel and its proper operation.\\
\FloatBarrier
\section{Noise spectrum}
\begin{figure}[h]
\centering
  \includegraphics[width=0.8\linewidth]{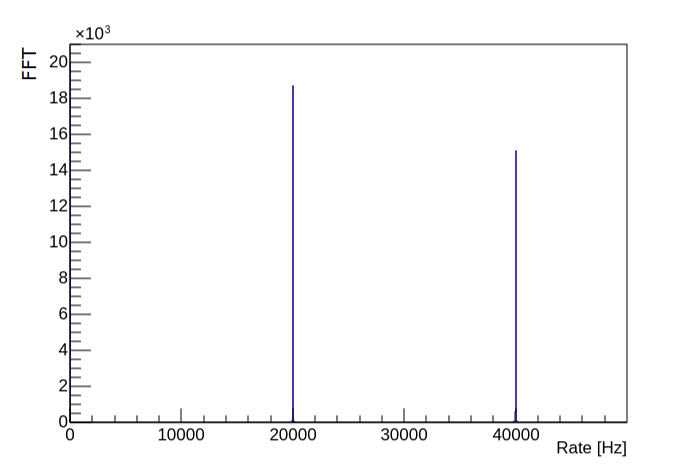}
  \caption{FFT of a 20kHz signal injected into TIGER. The main component and the first harmonic are shown.}
    \label{fig:fft_test}
\end{figure}

\begin{figure}
\centering
\begin{subfigure}{.5\textwidth}
  \centering
  \includegraphics[width=1\linewidth]{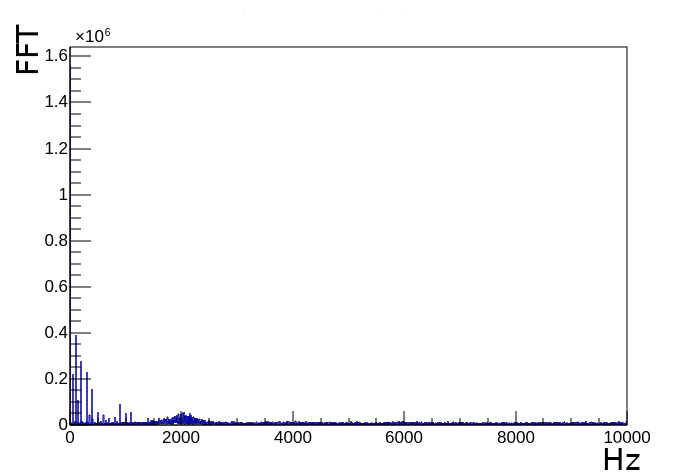}
\end{subfigure}%
\begin{subfigure}{.5\textwidth}
  \centering
  \includegraphics[width=1\linewidth]{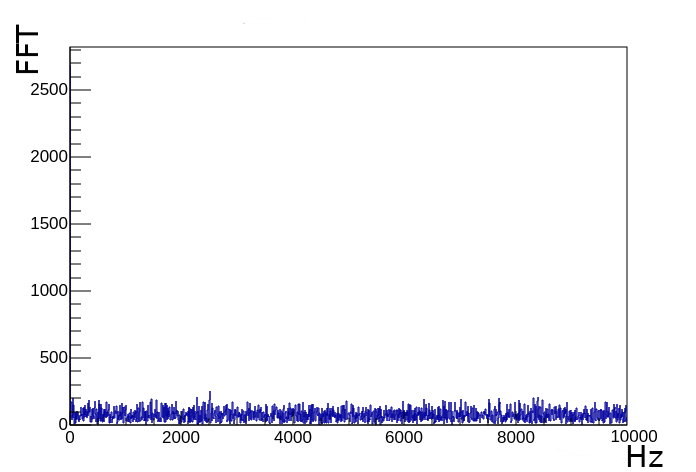}
\end{subfigure}%
  \caption{FFT of the noise acquisition, up to \SI{10}{\kilo \hertz}, with (right) and without (left) the aluminum foil as shielding. The left plot show peaks at multiples of \SI{50}{\hertz} are present, along  with a structure around \SI{2}{\kilo \hertz}. Both features were suppressed by the shielding. In both cases, the DC component dominates.}
\label{fig:shielding_low_f}
\end{figure}

\begin{figure}
\centering
  \includegraphics[width=0.8\linewidth]{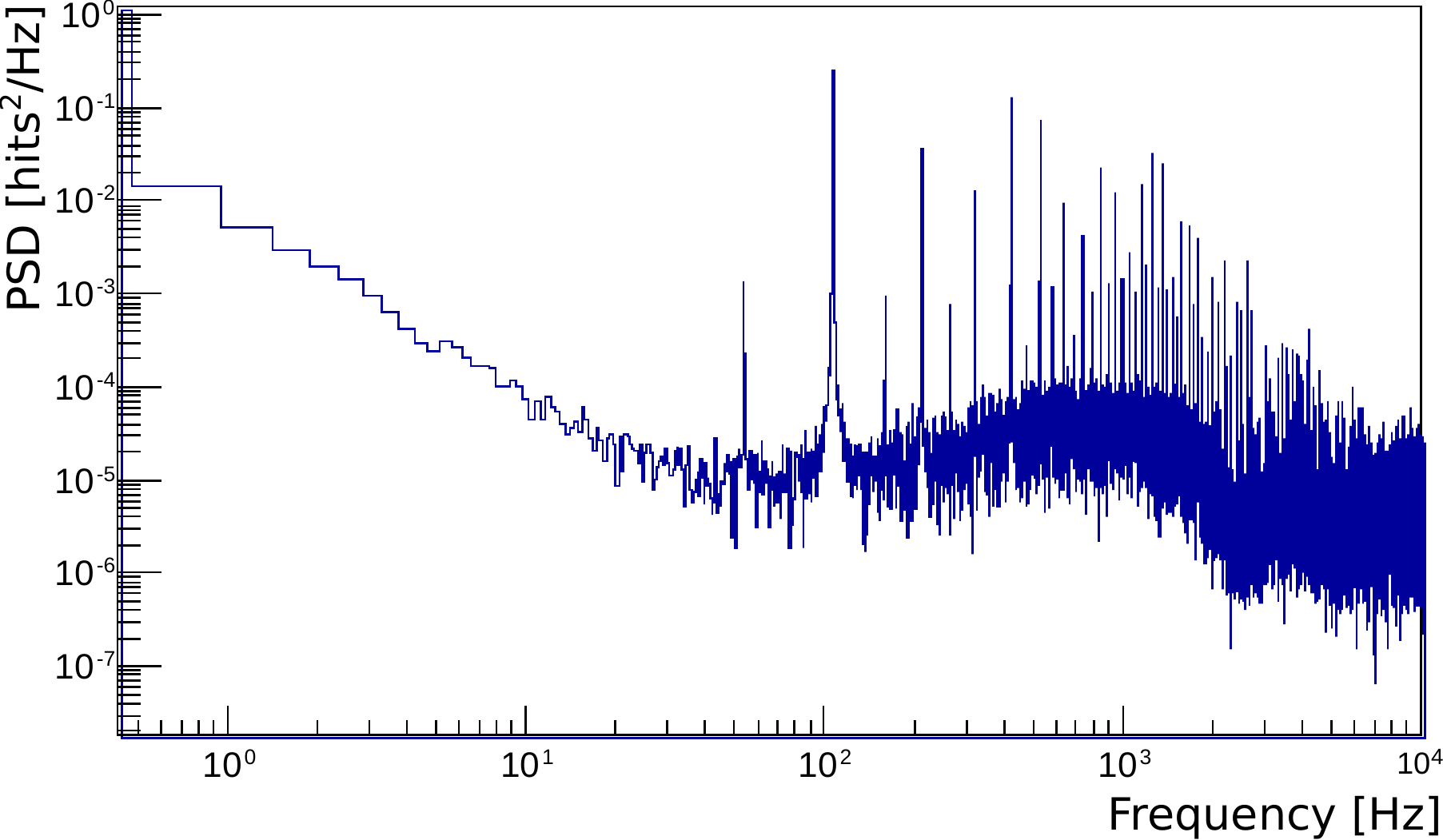}
  \centering
  \includegraphics[width=0.8\linewidth]{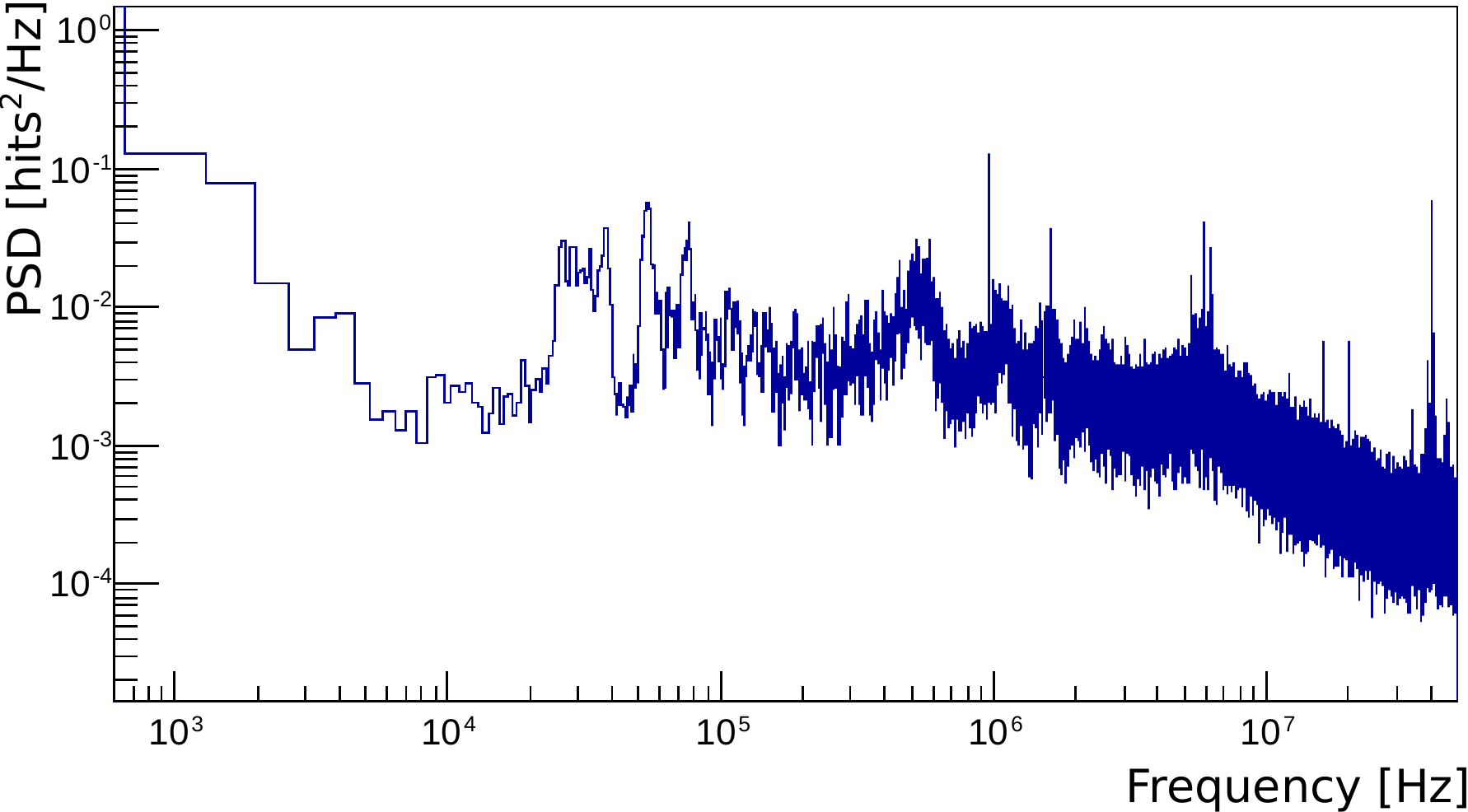}
  \caption{Power Spectral Density for the CGEM-IT layer 1 setup. Two plots are included for ease of calculation and visualization. Frequency in the top plot goes up to \SI{10}{\kilo \hertz}, in the bottom plot up to \SI{50}{\mega \hertz}.}
\label{fig:PSDcgem}
\end{figure}

To measure the noise spectrum without saturating the Ethernet bandwidth, small sections of the setup can be acquired in trigger-less mode and the data analyzed for the number of hits using Fast Fourier Transform (FFT).\\ 
To check the validity of the acquisition and analysis process, a test pulse at \SI{20}{\kilo \hertz} was injected into one channel of the TIGER and the result analyzed (figure \ref{fig:fft_test}).
Such analysis, combined with the raw noise frequency, resulted in the optimization of the noise on the planar setup (see section \ref{setup_fe}). In particular, we discovered the importance of shielding the last part of the HV cables, directly  connected with the planar detector.\\
\begin{figure}
\centering
  \includegraphics[width=0.8\linewidth]{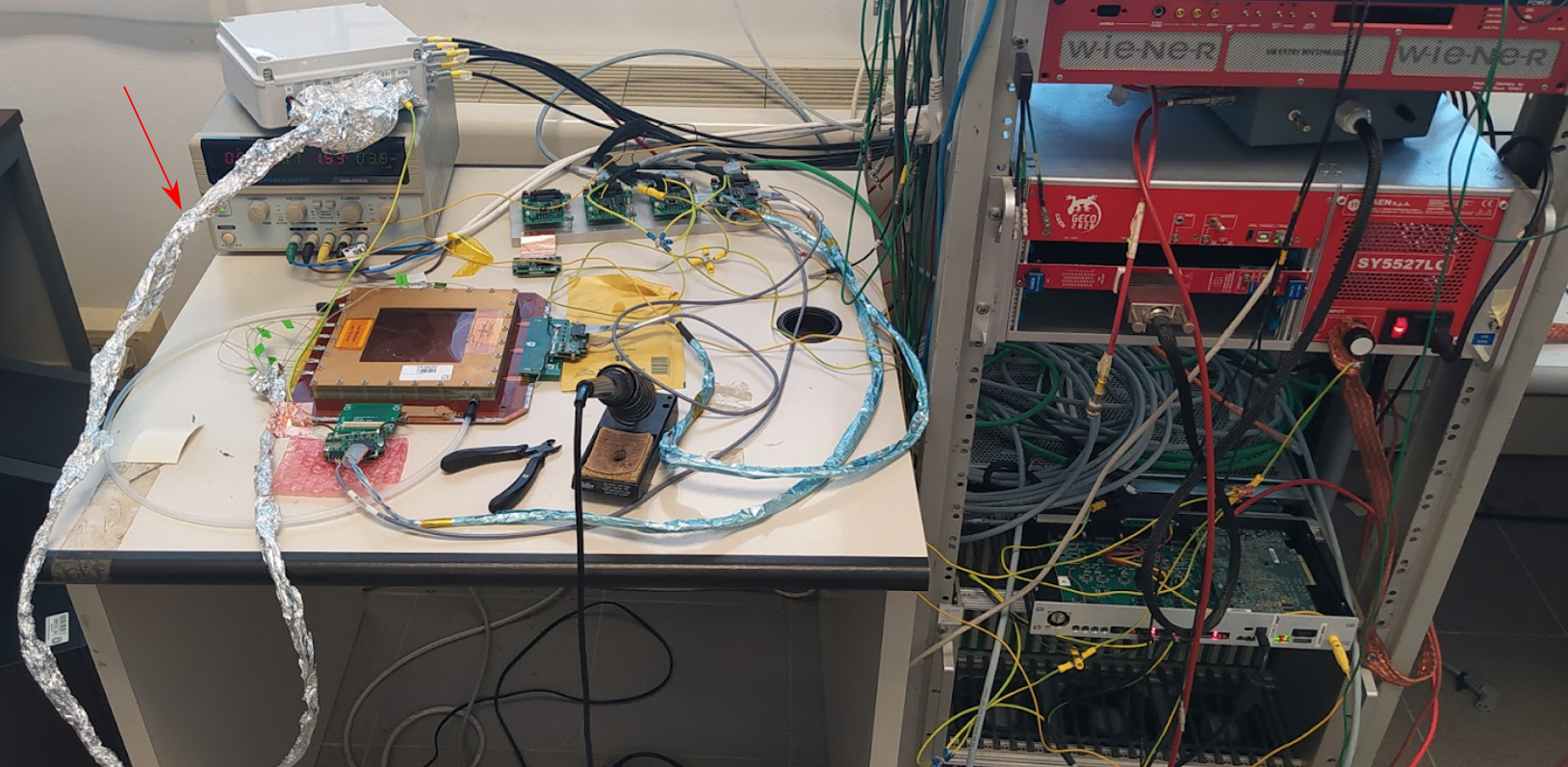}
  \caption{Small test setup used for the noise tests and optimization. The red arrow points to the HV cable under study, with an aluminum foil used for shielding.}
    \label{fig:noise_test_setup}
\end{figure}
On a small test setup (figure \ref{fig:noise_test_setup}), we tested the noise with the cable unconnected, the cable connected but unshielded, and the cable connected and shielded with an aluminum foil. We tested the noise rate and spectrum with fixed reasonable thresholds. The noise rate with the unshielded cable was few orders of magnitude higher than the case with the unconnected cable or shielded cable (from $\sim$ \SI{200}{\hertz} to tens of kHz). The noise was also strongly dependent on the cable position, indicating a strong pick-up problem.\\ Looking the spectrum (figure \ref{fig:shielding_low_f}), the shielding was able to remove most of the noise components, especially at low frequencies, making the noise almost white at the frequencies tested.\\
After these tests, shielded cables were used for HV distribution, and the Ferrara Electronics workshop developed an interconnect and filter interface, dubbed LEGO (see section \ref{setup_fe}).\\
A similar analysis was performed with the two CGEM layers assembled together in Beijing. The spectra are shown in figure \ref{fig:PSDcgem}.\\
As for the lower frequency range (upper plot), the spectrum shows a relevant peak at \SI{50}{\kilo \hertz} and multiples thereof, with the  peak of the second harmonic (\SI{100}{\kilo \hertz}) higher than the others. This may be related to some interference from the electrical power distribution of the building. At the higher frequencies, we have peaks at \SI{937}{\kilo \hertz}, at \SI{5.82}{\mega \hertz} and \SI{40}{\mega \hertz} (which is the frequency of the GEMROC's simulated BESIII clock).\\
The RC filtering on the CGEM is formed by the protection resistors at the sector connection and the capacitance of the sector itself (see section \ref{HV}). Indeed, the on-detector HV cables do not have any shielding in order to be as light as possible. We will perform additional tests for different configurations of shielding and filtering of the HV lines as the pandemic situation allows. \\
\FloatBarrier
\section{Threshold setting}
\label{thr_setting}
\begin{figure}[h]
\centering
  \includegraphics[width=0.8\linewidth]{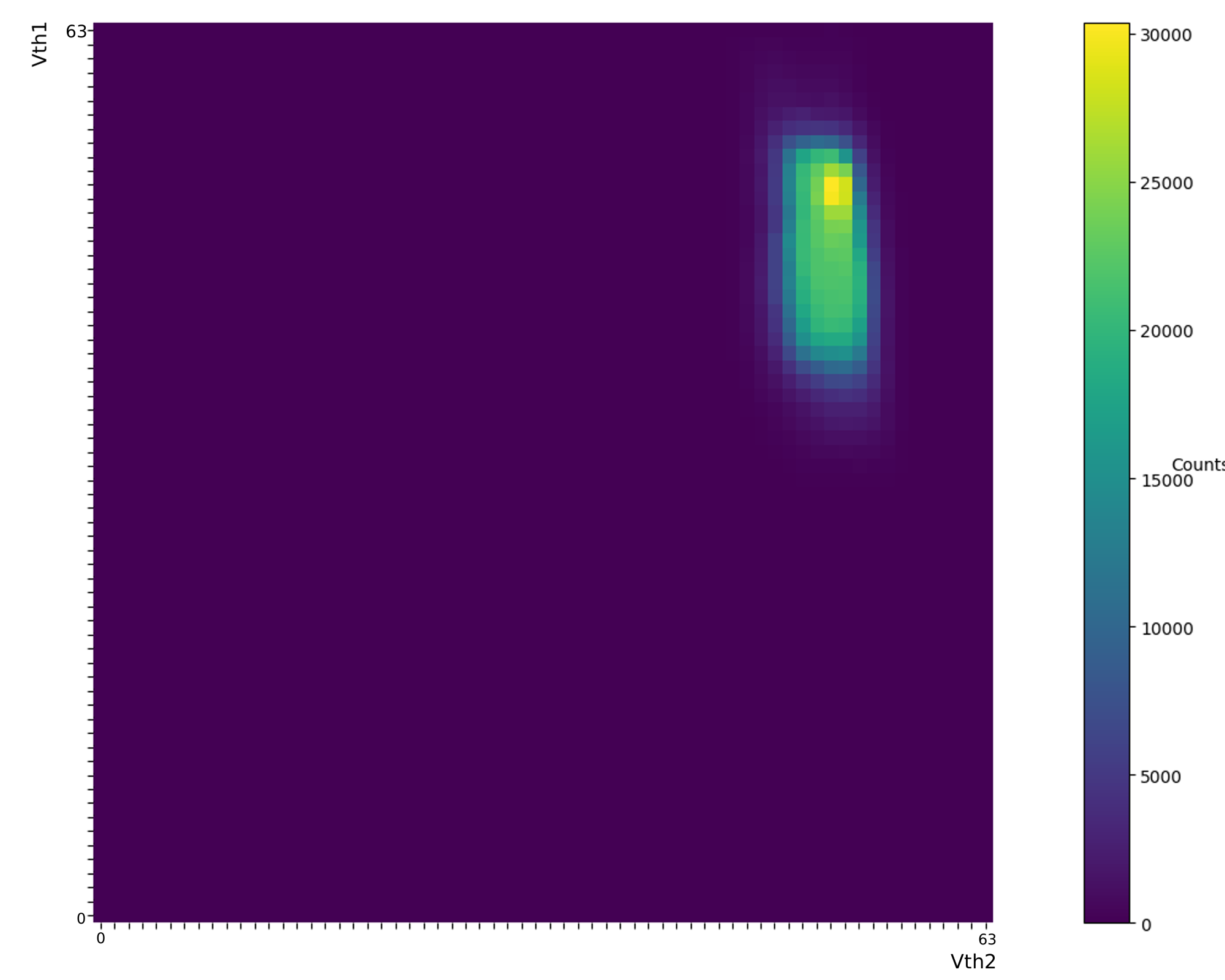}
  \caption{Threshold scan of a TIGER channel, on both the thresholds. The yellow point is the baseline while the green area is the noise area. The data are cut to \SI{30}{\kilo \hertz}. }
    \label{fig:double_scan}
\end{figure}

To optimize the performance of the detector, we chose to use thresholds on both branches to filter the signal. TIGER allows multiple configurations for validating the thresholds. In our configuration, the time and charge measurements are triggered by crossing the threshold on the T branch, while the validation and end of signal are triggered by the combination of the two threshold crossings.\\
Since the noise on the branches is not fully correlated, the use of two branches allows the effective threshold to be lowered while maintaining the same noise rate. The disadvantage of this mode is that the threshold setting space  becomes two-dimensional, and a full scan of all the channels (figure \ref{fig:double_scan}) requires too much time for many practical operations with the test setups.
To set  the thresholds, two different algorithms were developed and inserted into GUFI:
\begin{itemize}
    \item Scan a threshold at a time while the other is disabled, fit the results, and set the thresholds according to the fit. This method is fast, but since the noise correlation on the two branches is unknown, it cannot predict the noise rate/channel in advance.
    \item Starting from a point in the threshold space selected by the first method, the second algorithm will:
    \begin{itemize}
        \item Check that the noise rate of the channel is within an adjustable tolerance near the desired rate.
        \item Scan both branches in a finite number of steps around the starting point ($\pm3$ by default ).
        \item Set the threshold at which the noise rate is closest to the desired rate.
        \item Repeat the algorithm N times (2 by default).
    \end{itemize} This method requires more time, but can better equalize the noise rate/channel and optimize bandwidth usage (figure \ref{fig:THR_eq}).
    
\begin{figure}[h]
\centering
  \includegraphics[width=1\linewidth]{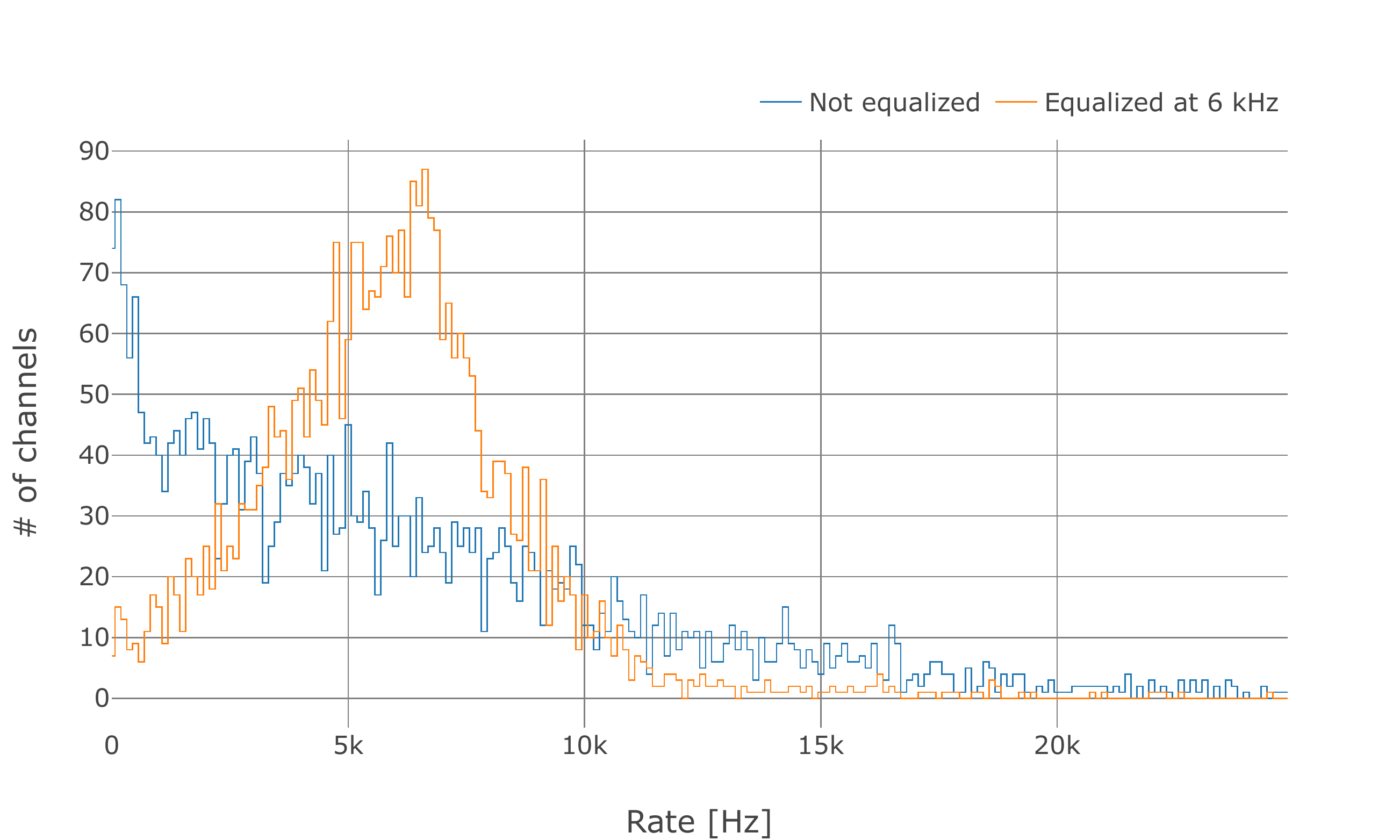}
  \caption{Noise Rate on channels of layer 2 in two different runs. In one, the thresholds were set using the first method (blu line), and in the other using the second method, with the desired rate at \SI{6}{\kilo \hertz} (red line). The noise rate is distributed around the desired rate because the thresholds have discrete steps and an optimal setting is not always possible. }
    \label{fig:THR_eq}
\end{figure}
\end{itemize}
\chapter{Aggregate data analysis}
\label{cap:agg_analysis}
Offline monitoring of the Beijing cosmic stand setup produces a large number of measurements and metrics. These data can be a useful source for cross-analysis and performance stability assessment.\\
Data on environmental parameters are of great interest as they affect the performance and stability of the detector. Indeed, over the last year, we have observed an increase in the detector current absorption associated with high ambient humidity. Since we are not able to investigate on site, it is necessary to monitor both the HV currents and the environmental parameters to ensure the safety of the detector until we can travel again.\\

\section{Data production}

\begin{figure}[h!]
\centering
  \centering
  \includegraphics[width=0.8\linewidth]{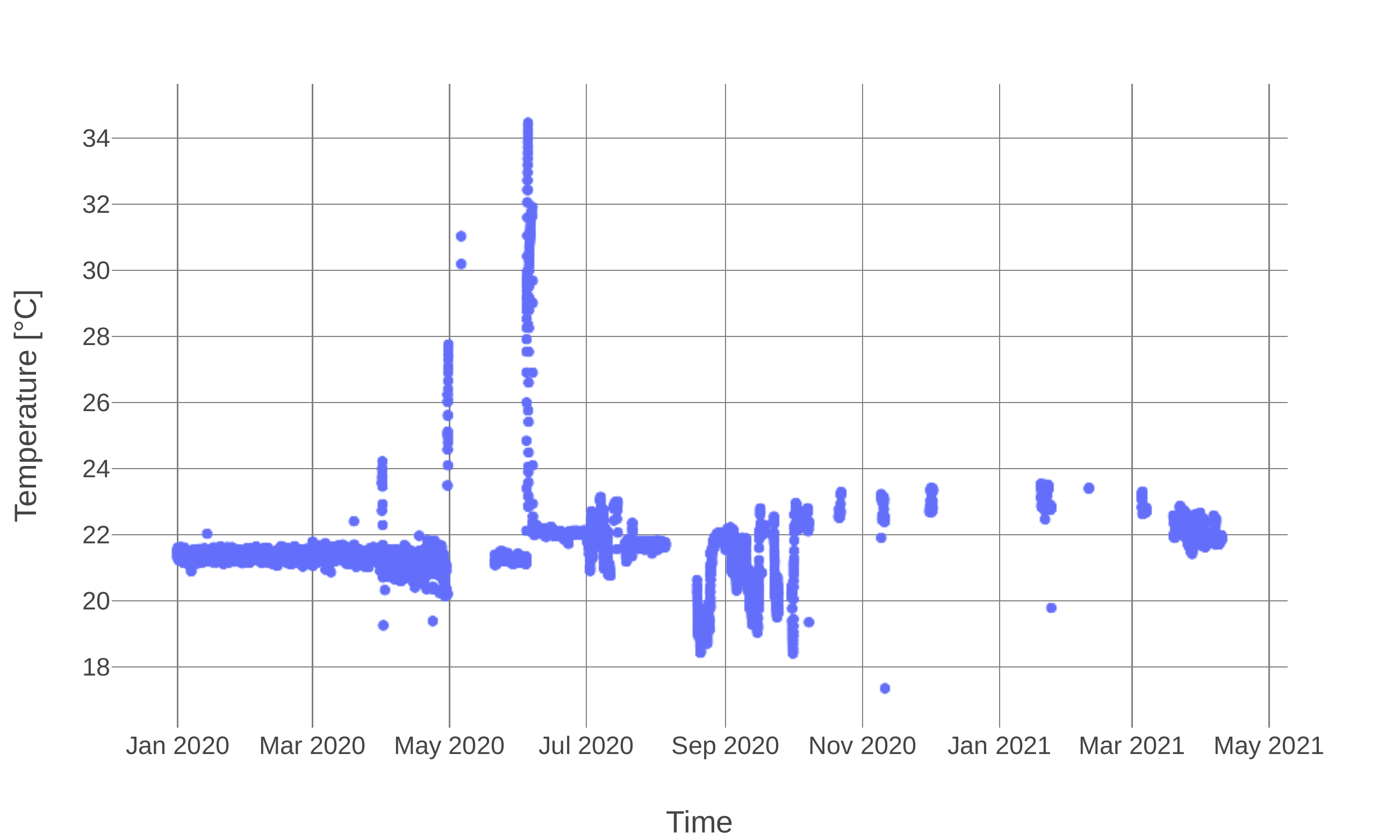}
  \centering
  \includegraphics[width=0.7\linewidth]{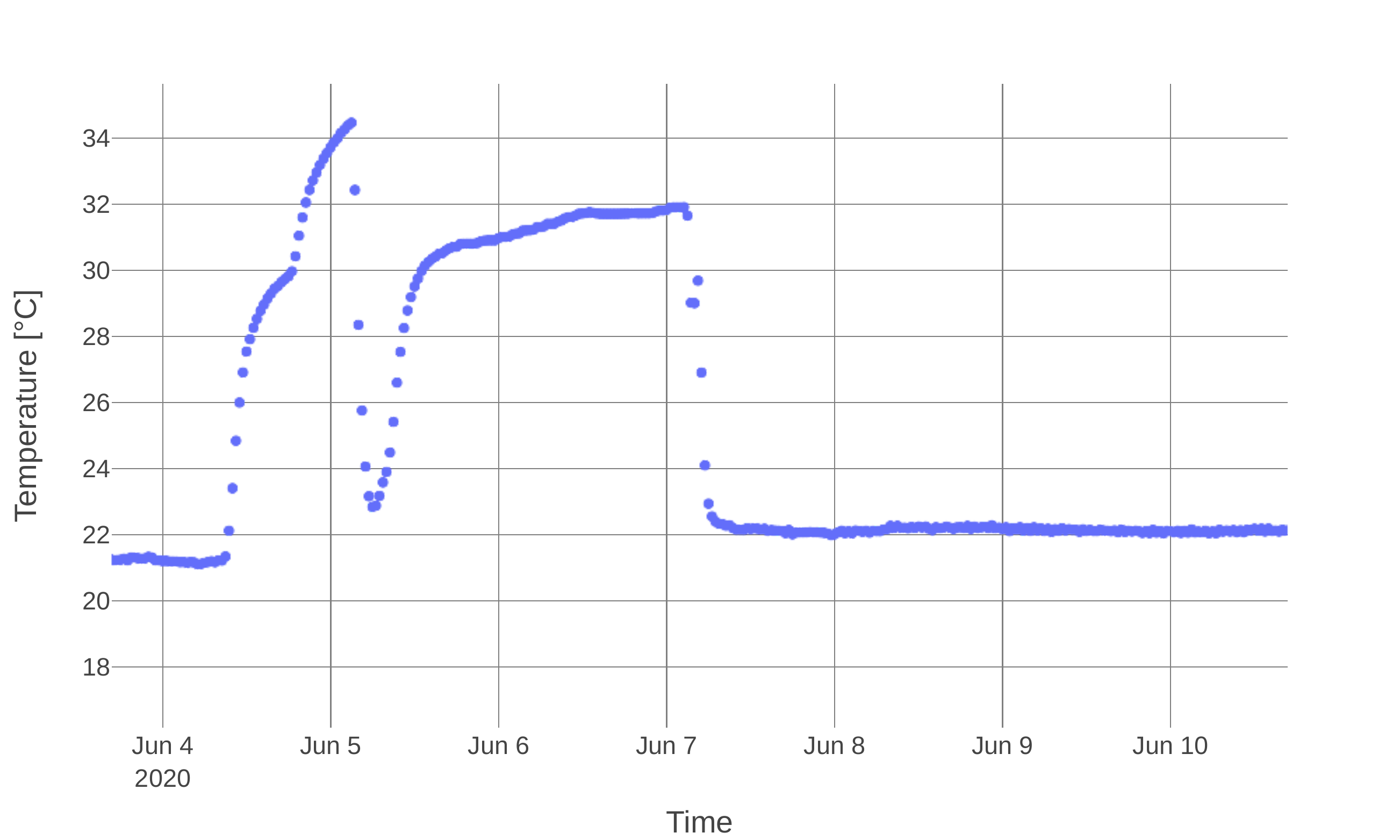}
  \caption{Top: temperature data collected during the acquisition. Bottom: a temperature rise due to a malfunction of the laboratory's air conditioning system is clearly visible.}
\label{fig:temp_data}
\end{figure}

\begin{figure}[h!]
\centering
\includegraphics[width=1\textwidth]{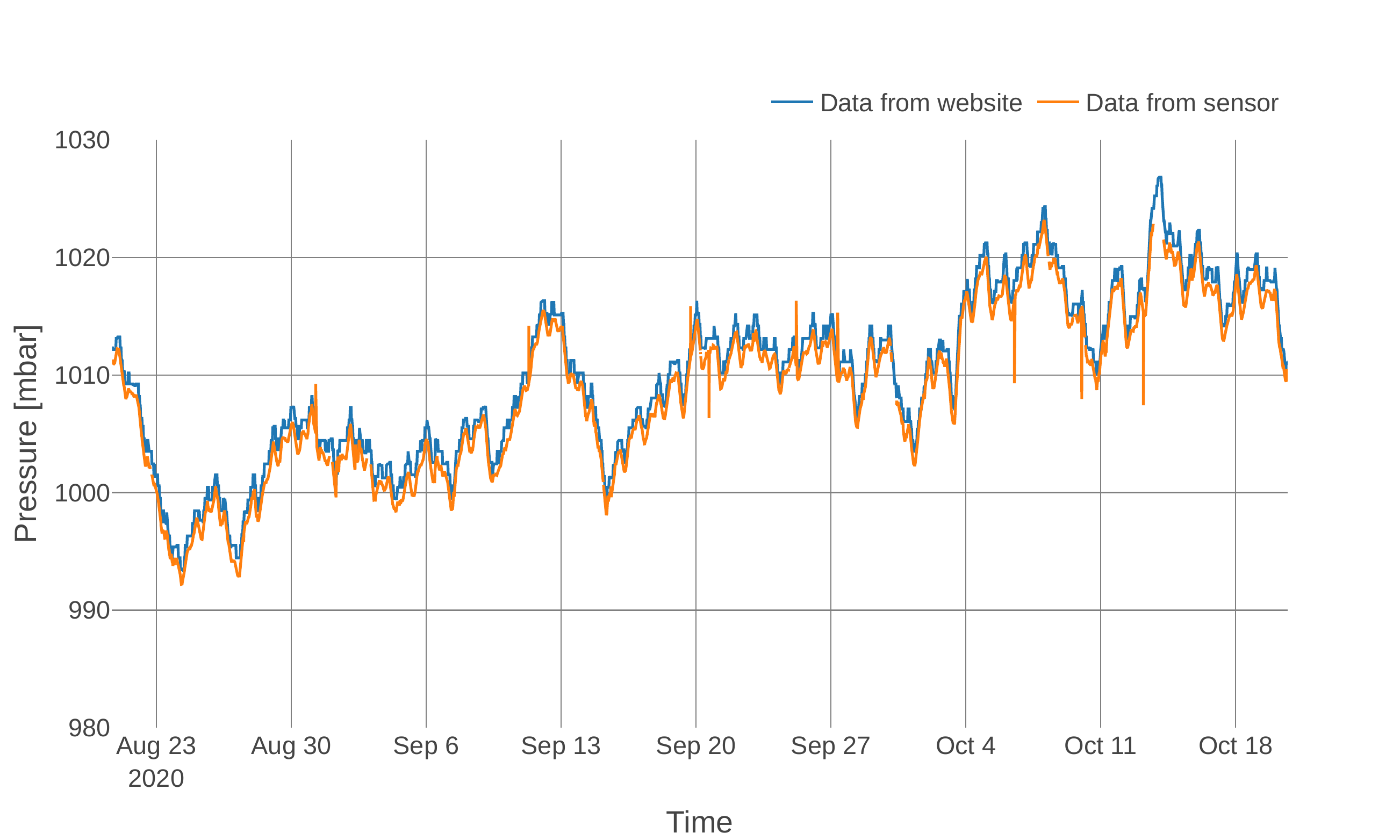}
\caption{Pressure data from Beijing airport and from the sensor in the room. The data are in good agreement, with data from the Beijing airport being more consistent and complete. }
\label{fig:pressure_Data}
\end{figure}

In order to use the data, they need to be aligned and combined. The different data sources have been treated as follows: 
\begin{itemize}
     \item \textbf{HV data}: the data are imported from the textual log file output by the LabVIEW VI to control HV. The data are stored in the text file every 30 seconds.\\
    Since the goal is to study over a long period of time under standard conditions, the data are averaged over five minutes and anomalous values, corresponding to detector discharges and detector power-on or power-off  are discarded. HV data are supplemented by the derivative of the detector electrode currents, which is calculated every 15 minutes and normalised over the hourly variation.
    \item \textbf{Humidity and temperature}: due to some operation issues of the sensors, the different datasets have been manually unified and the temporal coverage is not  yet complete (figure \ref{fig:temp_data}).

    \item \textbf{Pressure}: we have a barometer in the laboratory where the system is housed, but its measurements did not cover the entire data acquisition. To have the full range of data, we instead use the data measured at the Beijing airport, which are available online\footnote{Data from :https://rp5.ru/Weather\_archive\_in\_Beijing,\_Peking\_(airport)}. The agreement between the two datasets is very good (figure \ref{fig:pressure_Data}). An offset of  \SI{-0.7}{\milli \bar} was applied  to account for the  difference in altitude between the airport and the laboratory,  achieving a better agreement.
    \item \textbf{Cluster data}: cluster information is extracted using CIVETTA (see section \ref{civetta}). CIVETTA does not fully support the cylindrical geometry and to obtain a selection of signal clusters, without the need for alignment and fine-tuning, only the X-view is considered. The clusters are selected using an inclusive fit with three or four points. To select a cluster, it must have a residual less than \SI{1.5}{\centi \meter} and the other clusters in the track must have a residual less than \SI{3}{\centi \meter}.
\end{itemize}

\FloatBarrier
\section{Data analysis}
\begin{figure}
\centering
\begin{subfigure}{.5\textwidth}
  \centering
  \includegraphics[width=1\linewidth]{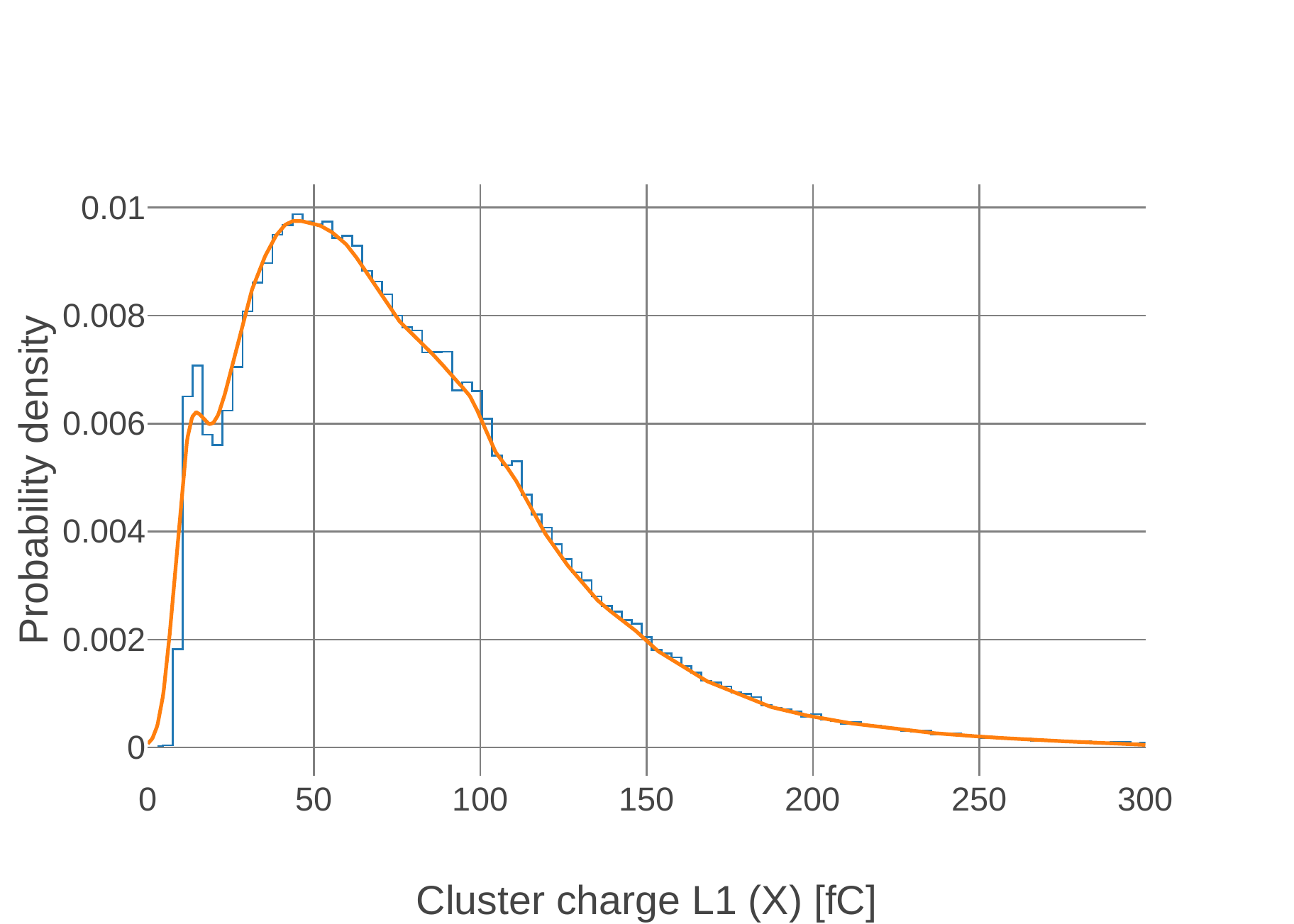}
\end{subfigure}%
\begin{subfigure}{.5\textwidth}
  \centering
  \includegraphics[width=1\linewidth]{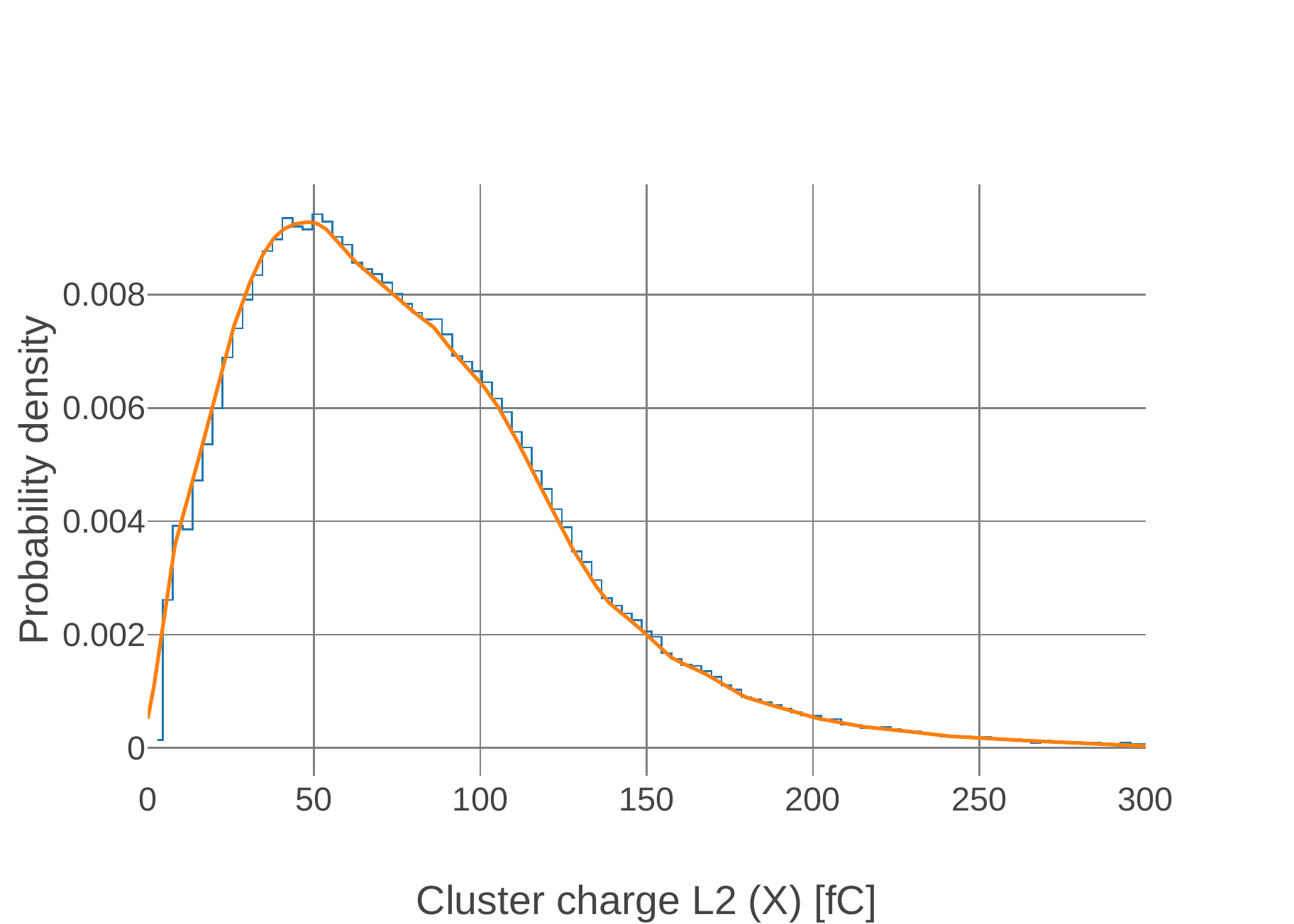}
\end{subfigure}%
 \caption{Cluster charge extracted from the data for an example run (labelled 351), for layer 1 and layer 2 X strips . The orange curve is the Kernel Density Function, extrapolated using the KDE method. The MPV is \SI{45}{\femto \coulomb} for layer 1 and \SI{48}{\femto \coulomb} for layer 2, while the average value is \SI{77}{\femto \coulomb} for layer 1 and \SI{78}{\femto \coulomb} for layer 2. Histogram integrals are normalized to one.}
\label{fig:cl_charge_351}
\end{figure}

\begin{figure}
\centering
\begin{subfigure}{.5\textwidth}
  \centering
  \includegraphics[width=1\linewidth]{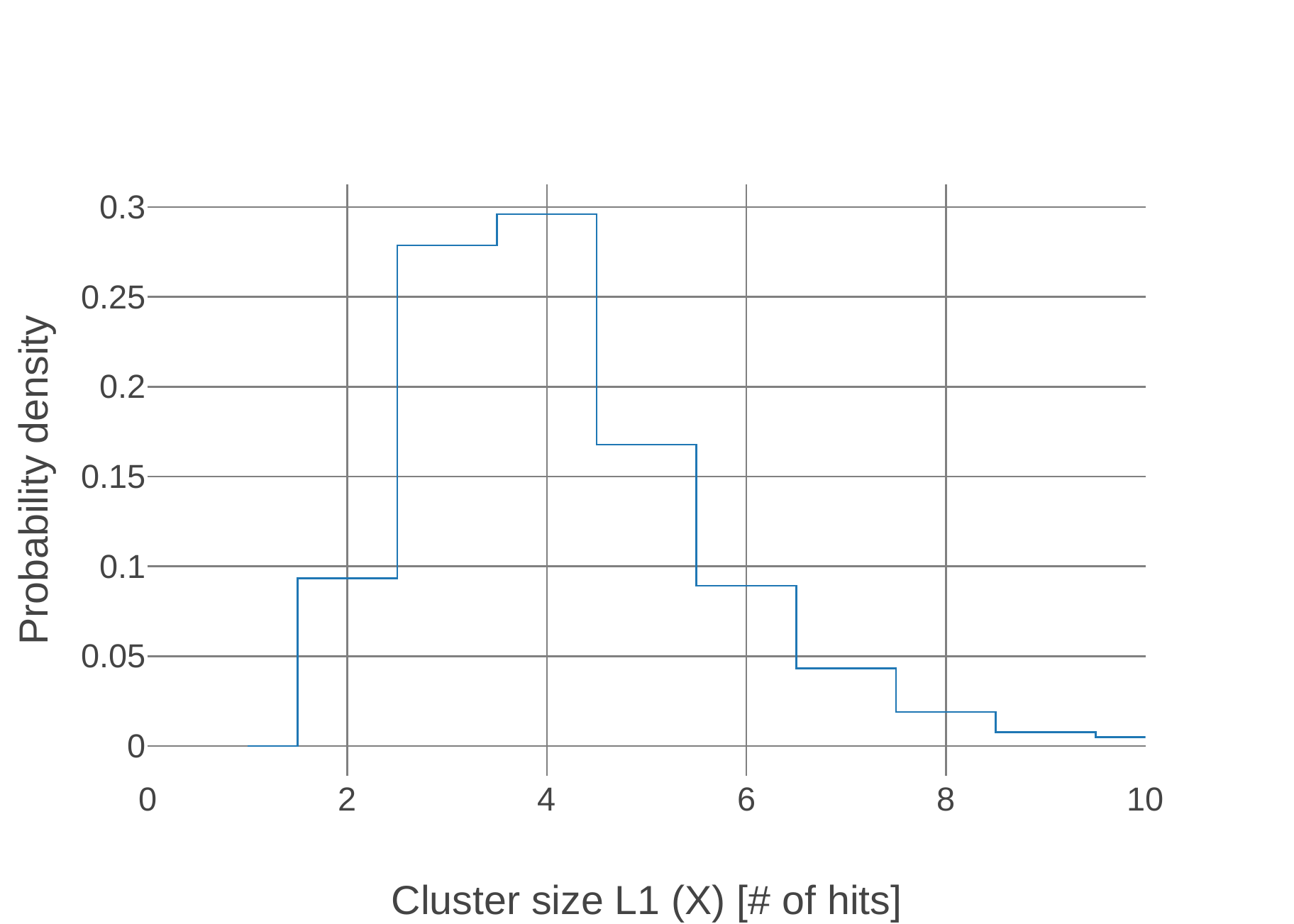}
\end{subfigure}%
\begin{subfigure}{.5\textwidth}
  \centering
  \includegraphics[width=1\linewidth]{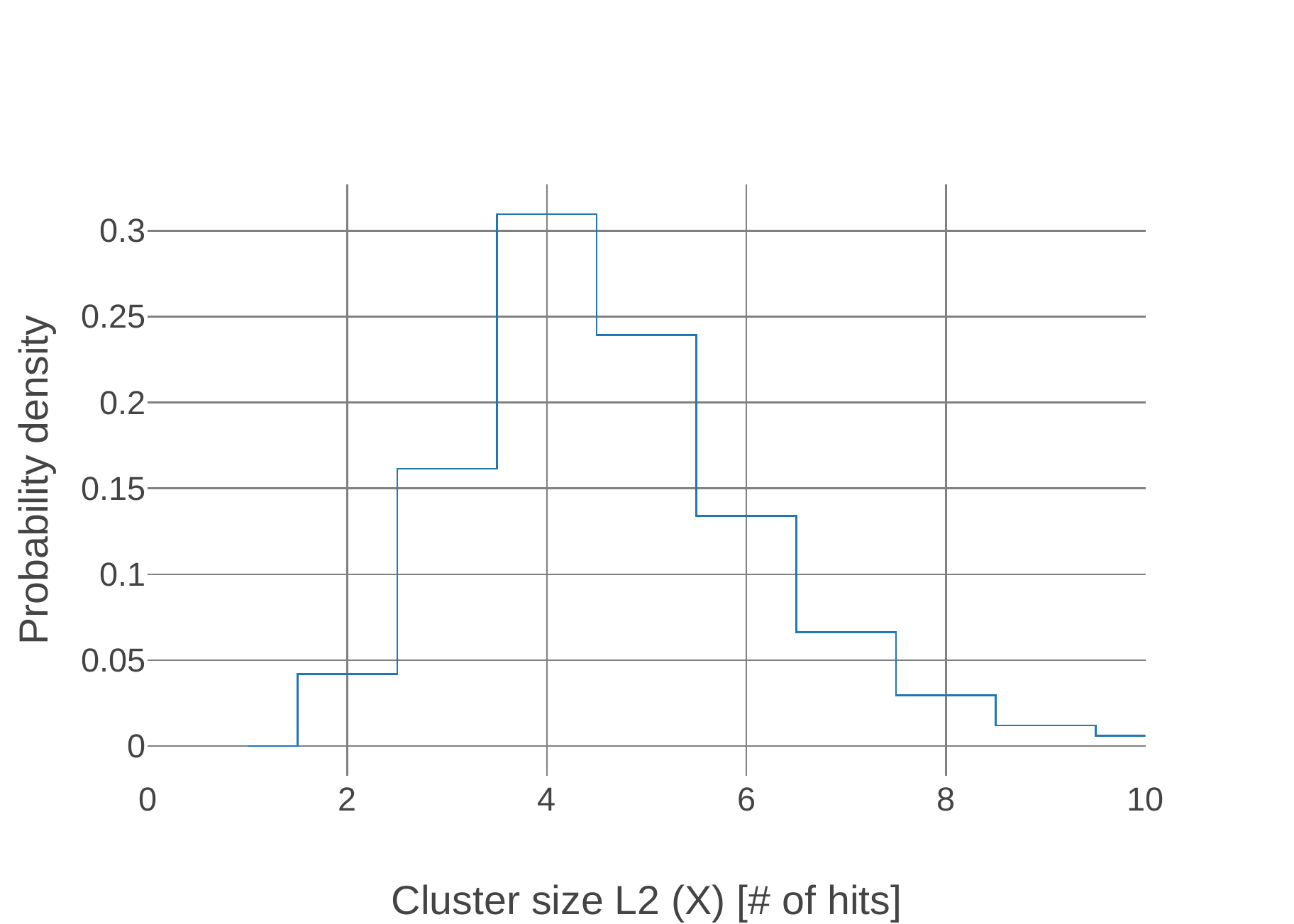}
\end{subfigure}%
 \caption{Cluster size (number of hits per cluster) extracted from the data for an example run (labelled 351), for X strips of layer 1 and layer 2 . The average cluster size for layer 1 is $3.2$, while for layer 2 is $3.7$. Histogram integrals are normalized to one.}
\label{fig:cl_size_351}
\end{figure}

\begin{figure}
\centering
  \centering
  \includegraphics[width=1\linewidth]{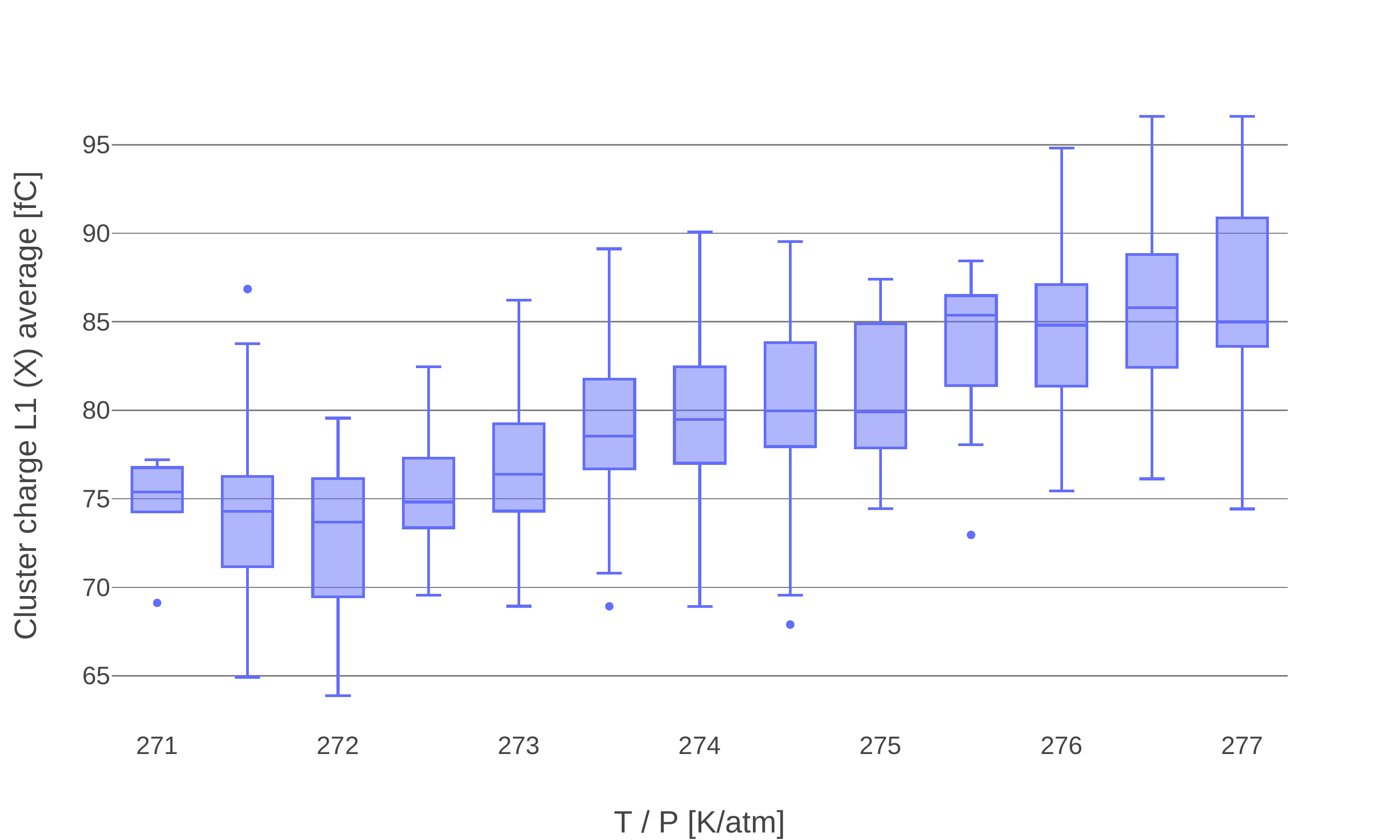}
 \caption{Cluster charge for layer 1, X strips, during a run with respect to the temperature/pressure ratio.}
\label{fig:L1_TP}
\end{figure}

These aggregate data allow to study the detector status over time under different conditions.  The distributions of cluster charge and cluster size, normalized to unity, for an example run (labeled 351 with $\sim$ 280,000 clusters) are shown in figures \ref{fig:cl_charge_351} and \ref{fig:cl_size_351}.
The average cluster charge and most probable value (MPV) can be extracted from these distributions. Compared to the average estimate, the MPV estimate has the advantage of being more robust to the saturation of single channels.\\
A Kernel Density Estimation (KDE) algorithm is  used to extract the MPV since it is difficult to fit it with a probability density function, mainly because channel saturation deforms the distribution. The KDE (also called the Parzen-Rosenblatt window method \cite{kde, kde2}) is a non-parametric method for estimating the probability density function of a random variable. To estimate the unknown density function $f$ of $n$ samples $x_i$, we estimate $f_{E}$ as:\\
$$
f_{E}(x) = \frac{1}{nh} \sum_{i=1}^{n}K\left(\frac{x-x_i}{h}\right)
$$
where $K$ is a suitable kernel function and $h$ is a scale parameter called bandwidth. To extract the PDF of the cluster charge distribution, a Gaussian kernel was chosen, where the bandwidth was calculated using the rule of thumb \cite{Bandwidth}:
$$
h = 0.9\, \min\left(\hat{\sigma}, \frac{IQR}{1.34}\right)\, n^{-\frac{1}{5}}
$$
Where $\hat{\sigma}$ is the  standard deviation and $IQR$ is the interquantile range.\\
Cluster data are available for each subrun (\SI{10}{\min}, $\sim$ 100-200 triggers). Even though the number of clusters is too small to extract the MPV, we can still perform an analysis at this level by using the average cluster charge.\\
\begin{figure}[t]
\centering
  \centering
  \includegraphics[width=1\linewidth]{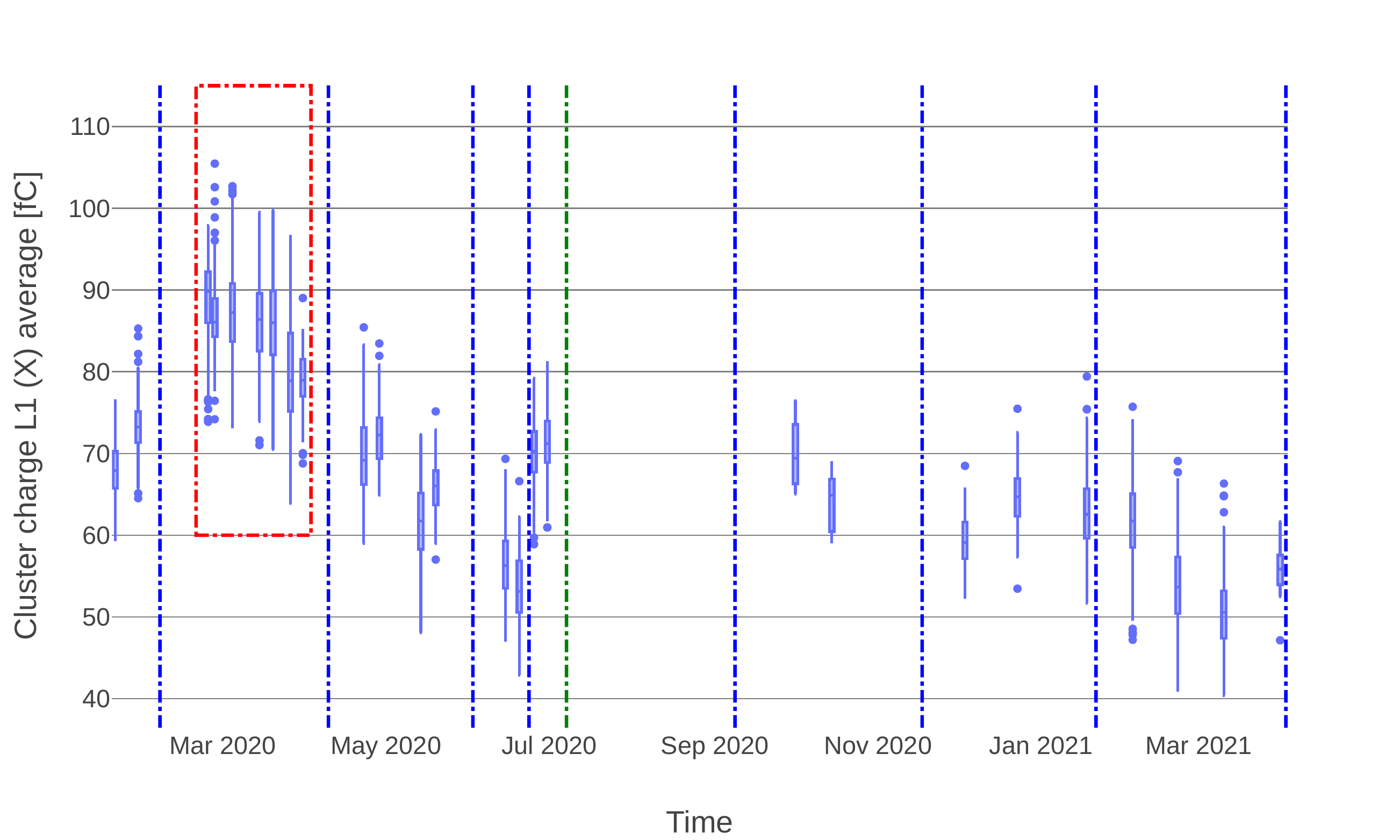}
  \centering
  \includegraphics[width=1\linewidth]{Chapters/chapter_5_files/charge_l1_time.pdf}
 \caption{Boxplot of the cluster charge for layer 1 and layer 2, X strips, throughout the acquisition campaign. }
\label{fig:cl_charge_story}
\end{figure}
Figure \ref{fig:cl_charge_story} shows the distribution of cluster charge as a function of time for the entire data collection. The data are presented with a boxplot\footnote{Each box extends from  the first quartile (Q1) to the third  quartile (Q3). The second quartile (Q2) is indicated by a line inside the box. The whiskers correspond to the edges of the box +/- 1.5 times the interquartile range (IQR: Q3-Q1), and the  points shown lie outside this range.}, with an entry for each subrun and a box for each run. The red box contains runs with higher gain, while the blue dotted lines represent argon bottle exchanges  and the green lines represents isobutane bottle exchanges.  The large variance between subruns is due to both statistical oscillation and the influence of environmental parameters. The overall trend in  cluster charge is still under study. The sudden increase in cluster charge when the argon bottle was replaced in July 2020 points to gas quality as a possible cause of the reduction in  gain. However, with this loss of gain, the detector performance does not seem to deteriorate in terms of resolution and efficiency.\\
\begin{figure}
\centering
  \centering
  \includegraphics[width=1\linewidth]{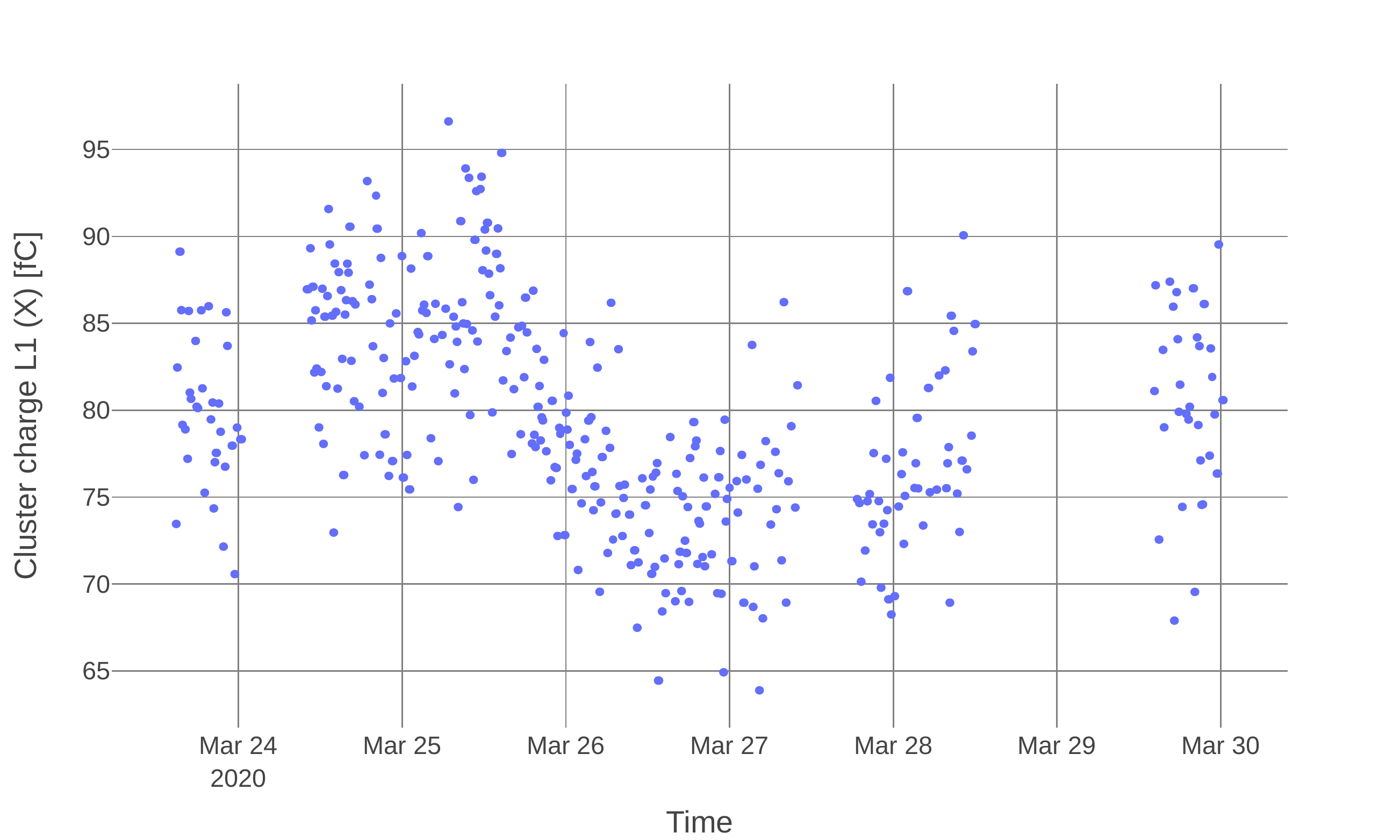}
 \caption{Cluster charge for layer 1, X strips,  during a run.}
\label{fig:cl_charge_story_377}
\end{figure}
Figure \ref{fig:cl_charge_story_377} shows the average cluster charge during a single high-gain run (run 377). The run spanned seven days, and the environmental conditions in the laboratory varied greatly due  the large  fluctuations in weather in Beijing during spring. \\
As described in the literature \cite{Altunbas:2001wd, Biswas:2013zra}, the gain of GEM detectors is affected by environmental parameters. The gain $G$ depends on the temperature $T$ and the pressure $P$ with a relation
$$
G(T/P) = A\cdot e^{B \cdot T/P}
$$
The exponential dependence is deduced by assuming inverse proportionality of the Townsend coefficient $\upalpha$ to the mass density $\uprho$, and thus $\upalpha\propto 1/\uprho \propto T/P$.
Plotting the cluster charge against the temperature (in K) over pressure (in atm) (figure \ref{fig:L1_TP}) we note an increasing trend as predicted, but due to the small range of environmental data and the low statistics of cosmic rays, we cannot produce a reliable fit to normalise the gain.\\
\FloatBarrier

\begin{figure}[h!]
\centering
  \centering
  \includegraphics[width=1\linewidth]{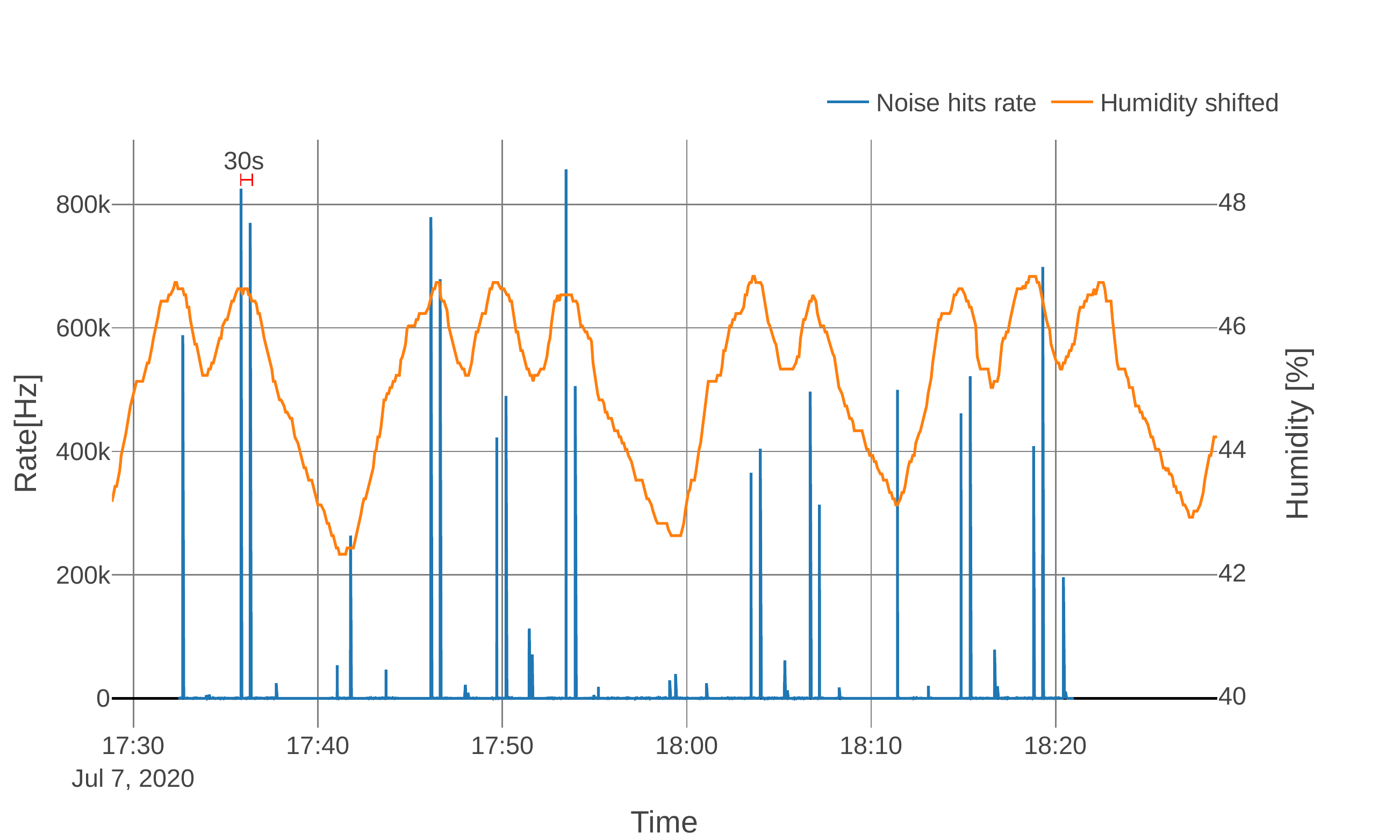}
 \caption{Noise rate and humidity, shifted by \SI{70}{\second}.}
\label{fig:TL_humidity}
\end{figure}

The data in the database can also be matched with data obtained from other sources or acquisitions. To characterize the noise and study the behaviour of the system without detector hits, we recorded a run in trigger-less mode, with the detector off, high thresholds, and only a few active TIGERs. During this acquisition, pairs of noise peaks were recorded, with exactly \SI{30}{\second} between each peak and variable time between pairs. When matching this acquisition with the humidity data (figure \ref{fig:TL_humidity}), we found a correlation with the room's air conditioning. If you shift the humidity data by \SI{70}{\second}, the noise peaks correspond to the maximum or the minimum of humidity. Therefore, the noise peaks are likely due to the air conditioner turning on and off, while the time shift is due to the thermal inertia of the room. This phenomenon does not affect the cosmic ray detection because the noise peaks are not correlated with the triggers, so their detection is very unlikely. \\
\chapter{Data analysis with CIVETTA}
\label{cap:res_TB}
\begin{figure}[h!]
\centering
\includegraphics[width=1\textwidth]{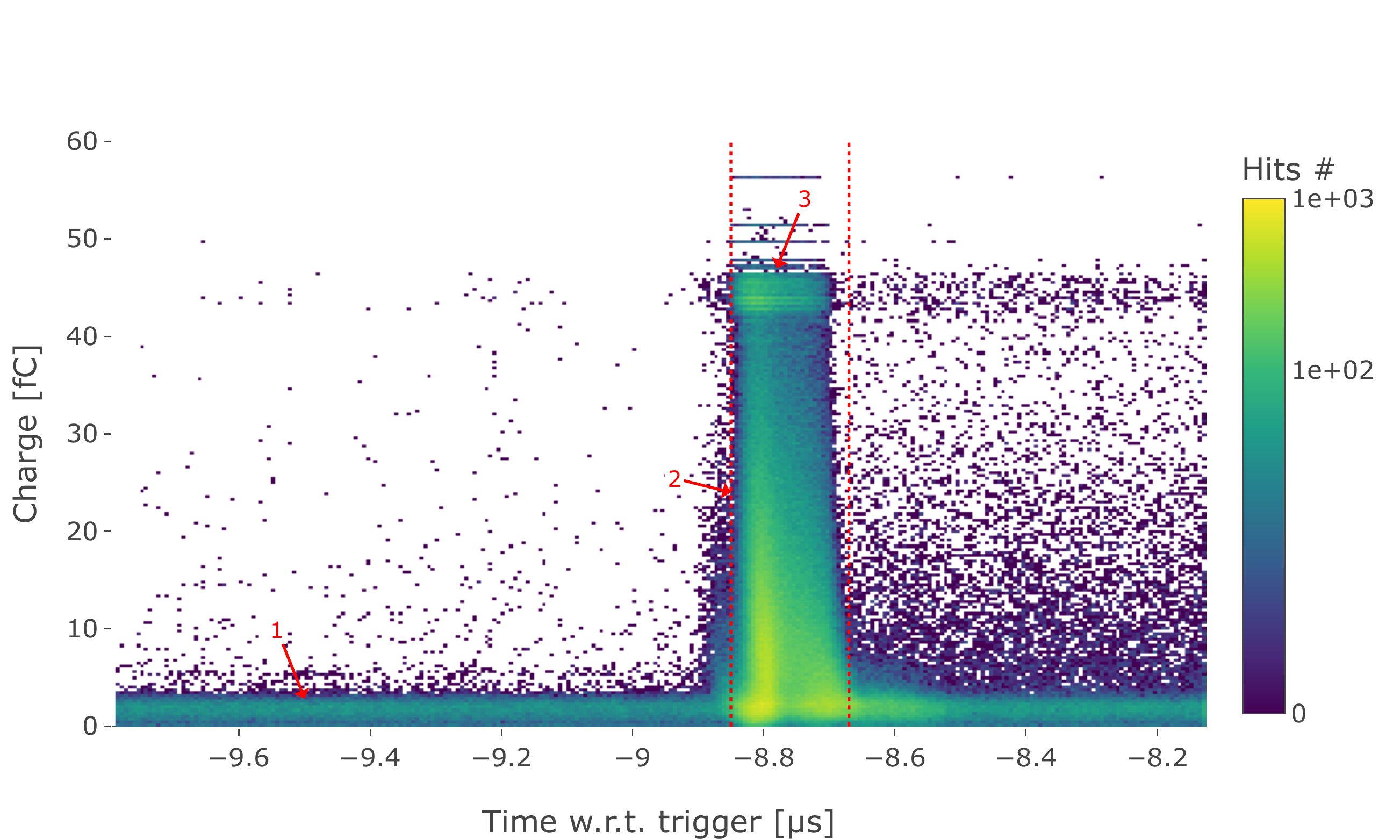}
\caption{Example plot of charge versus time. Each entry in the plot is a hit. Time is expressed with respect to the arrival time of the trigger. The Z axis (color) is in logarithmic scale.}
\label{fig:chargetime}
\end{figure}

\begin{figure}[h!]
\centering
\includegraphics[width=0.9\textwidth]{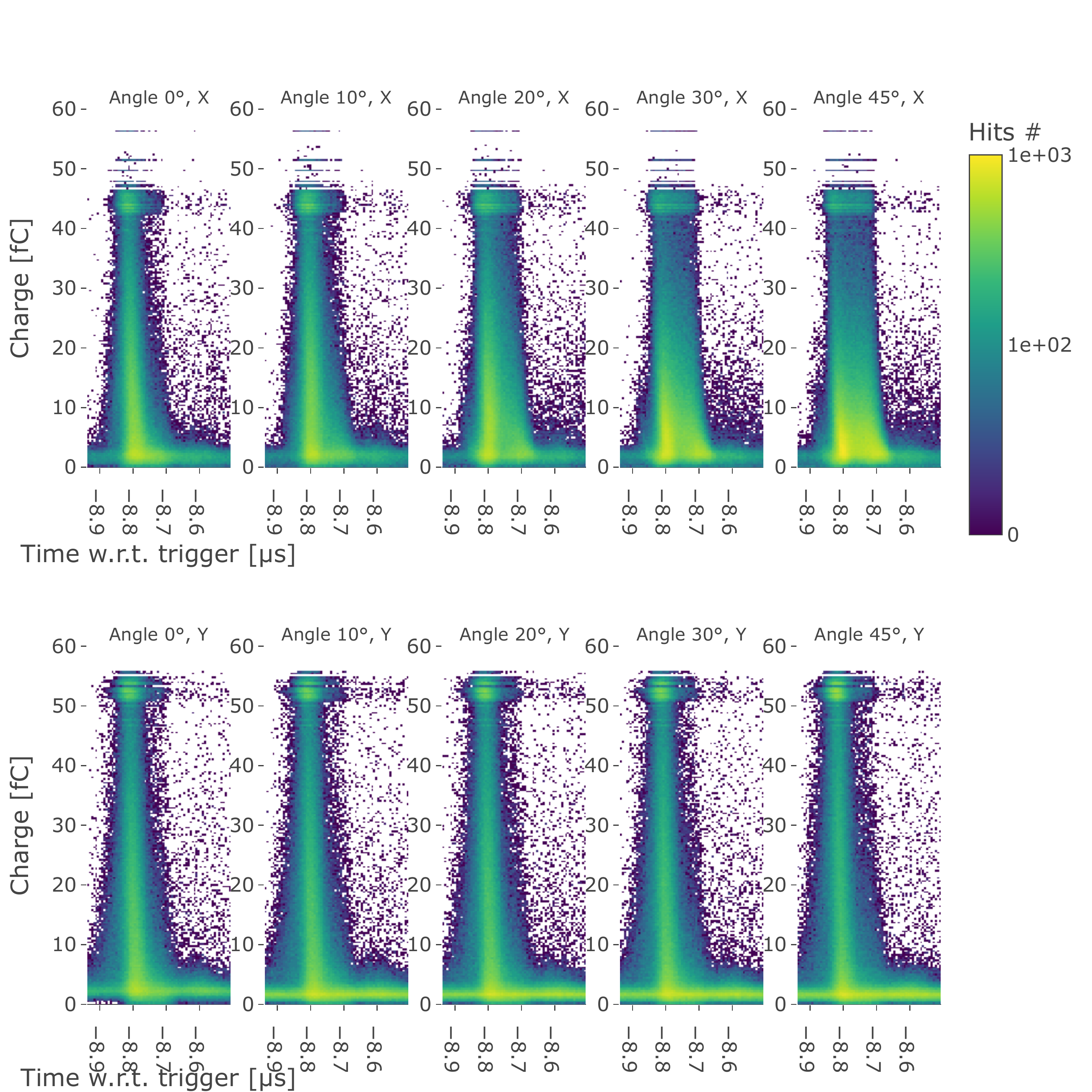}
\caption{Charge versus time, expressed in terms of arrival time of the trigger, as a function of beam angle. Angle and strip view are given at the top of each graph. }
\label{fig:angle_vs_chargetime}
\end{figure}

\begin{figure}[h!]
\centering
\includegraphics[width=0.9\textwidth]{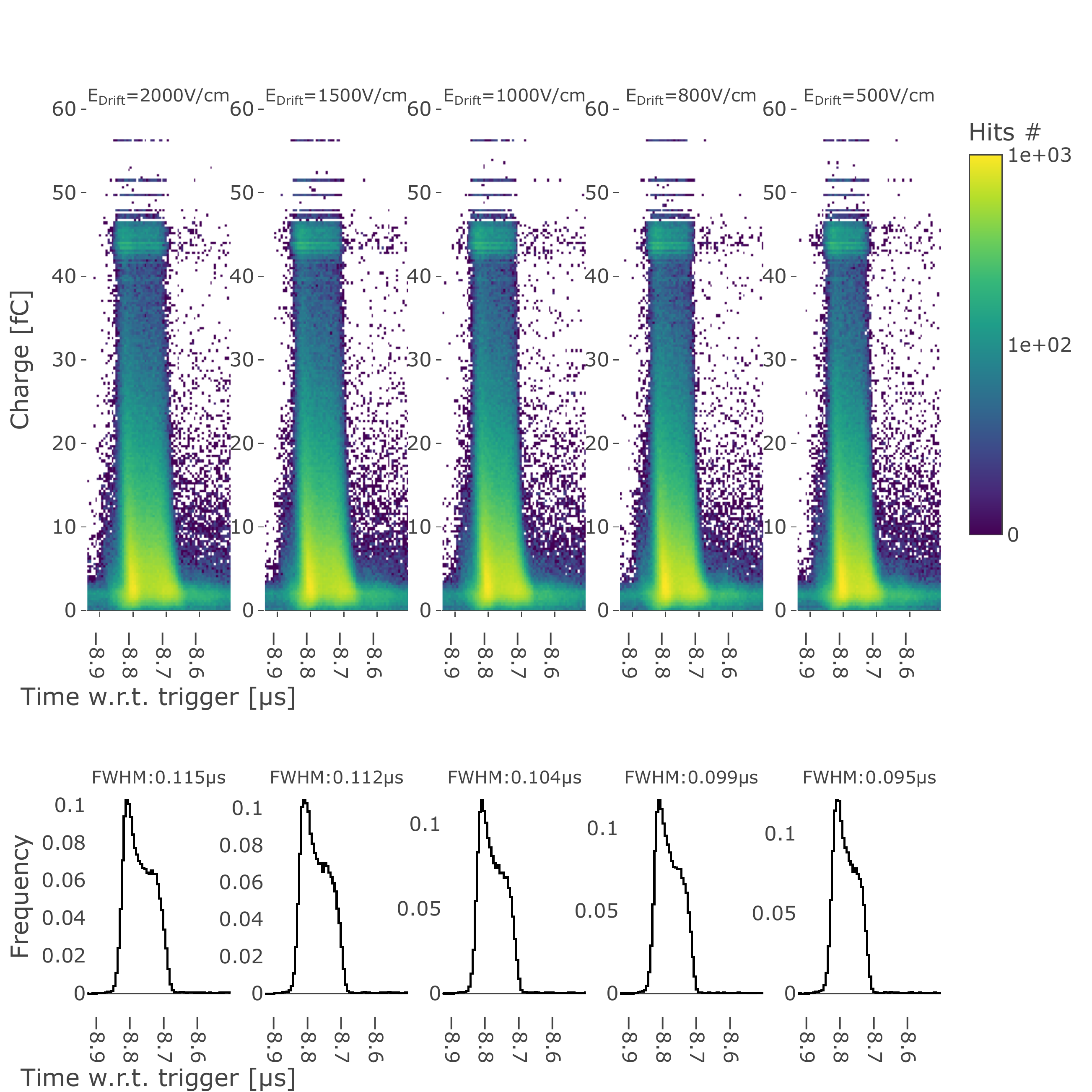}
\caption{Top: charge versus time, in terms of arrival time of the trigger, for the X view. Top of each graph shows the drift field settings. Bottom: time distributions applying a charge cut (>\SI{10}{\femto \coulomb}). The FWHM gives an indication of the varying drift velocity. }
\label{fig:drift_vs_chargetime}
\end{figure}

\begin{figure}[h!]
\centering
\includegraphics[width=0.8\textwidth]{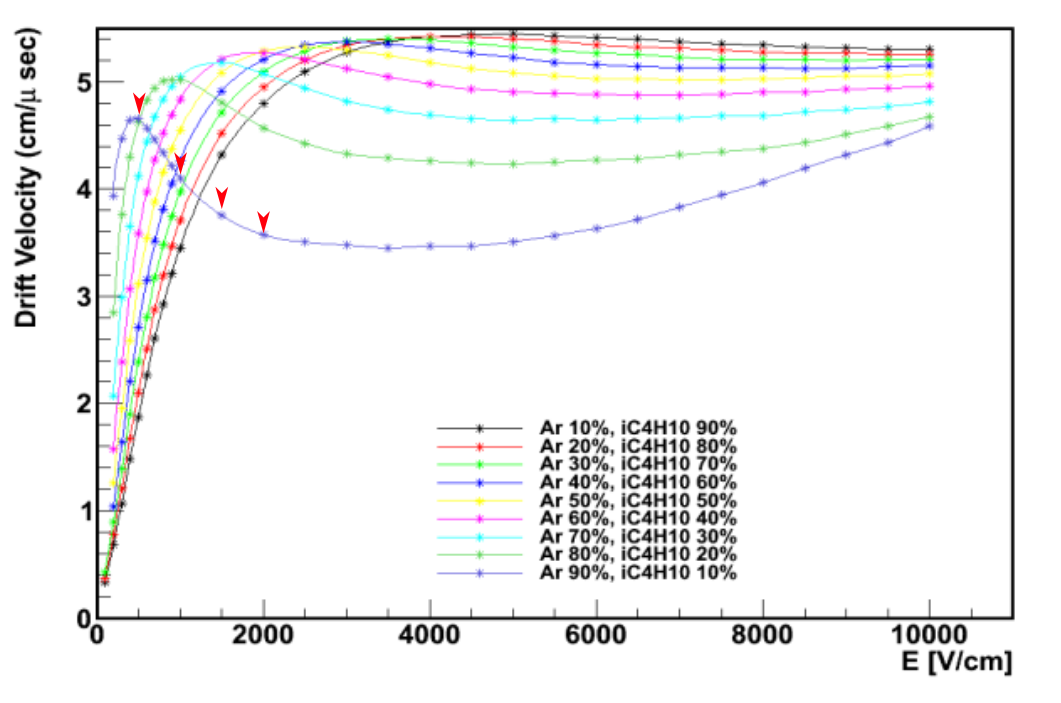}
\caption{Drift velocity in an Ar:i\ch{C_4H_10} gas mixture calculated from simulations \cite{drift_speed}. The red arrows indicate the values scanned during the test beam.}
\label{fig:drift_speed_ref}
\end{figure}

\begin{figure}[h!]
\centering
\includegraphics[width=1\textwidth]{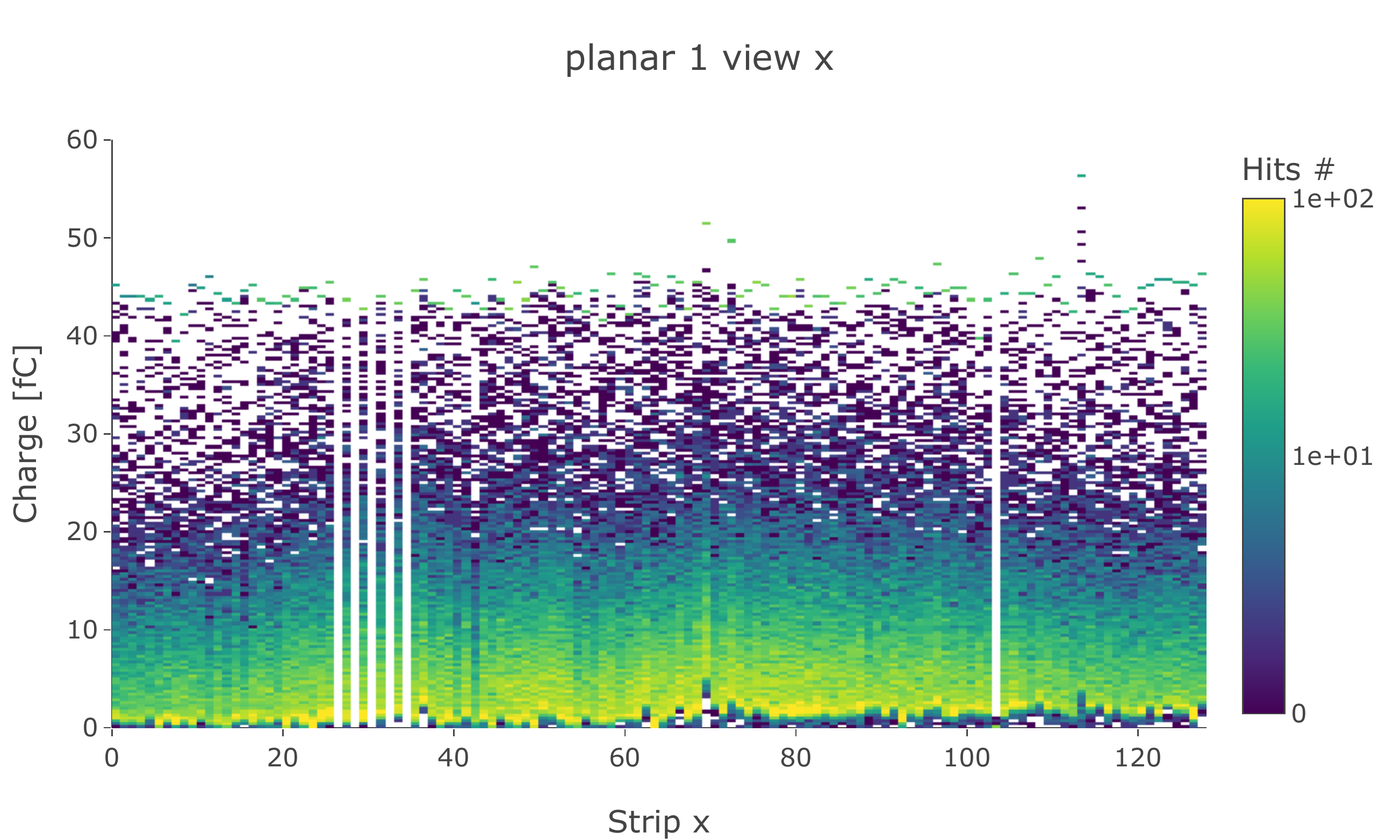}
\caption{ Histogram of charge versus strip. The data are selected in the signal region.}
\label{fig:strip_charge}
\end{figure}

\begin{figure}[h!]
\centering
\begin{subfigure}{.5\textwidth}
  \centering
  \includegraphics[width=1\linewidth]{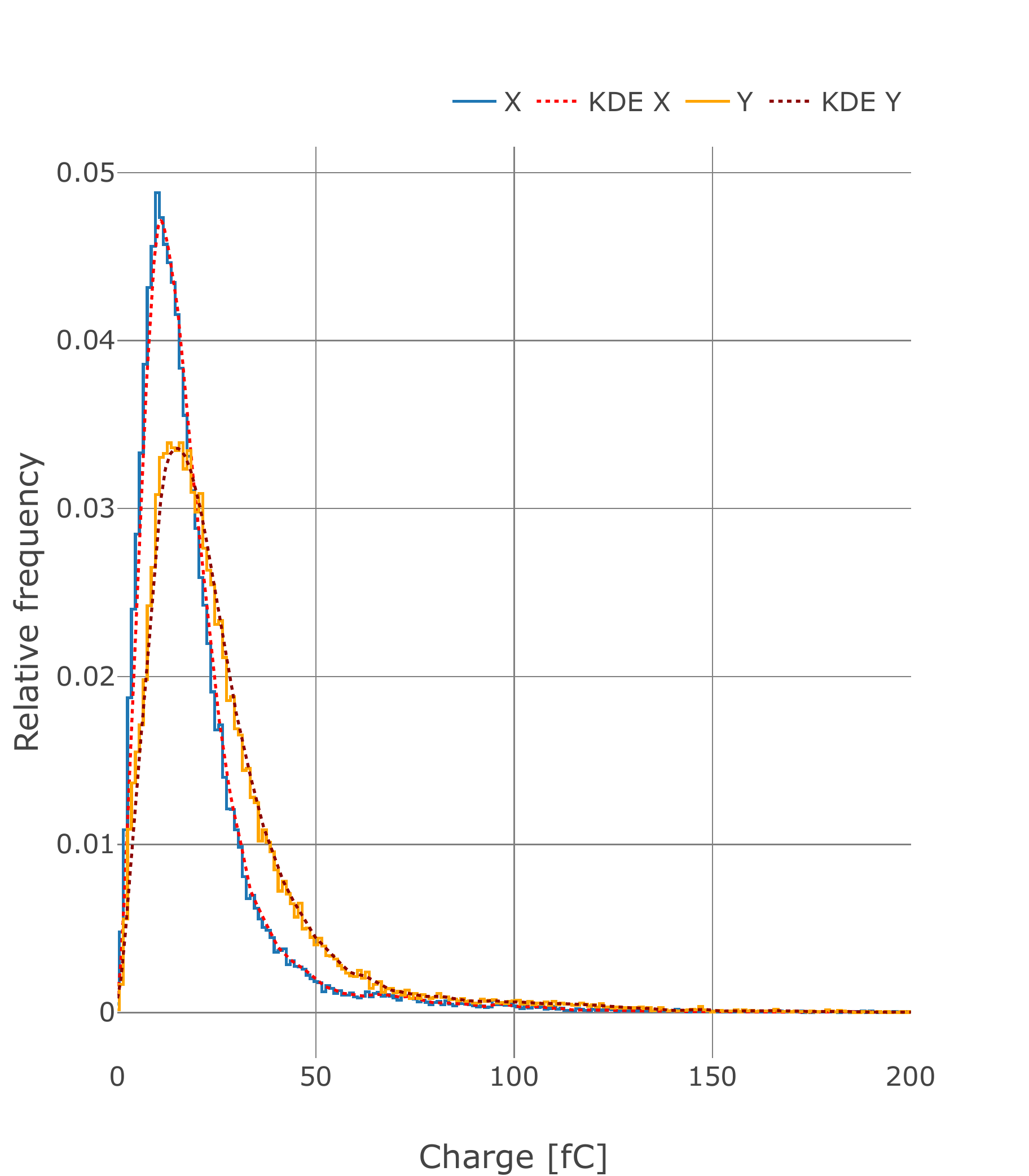}
\end{subfigure}%
\begin{subfigure}{.5\textwidth}
  \centering
  \includegraphics[width=1\linewidth]{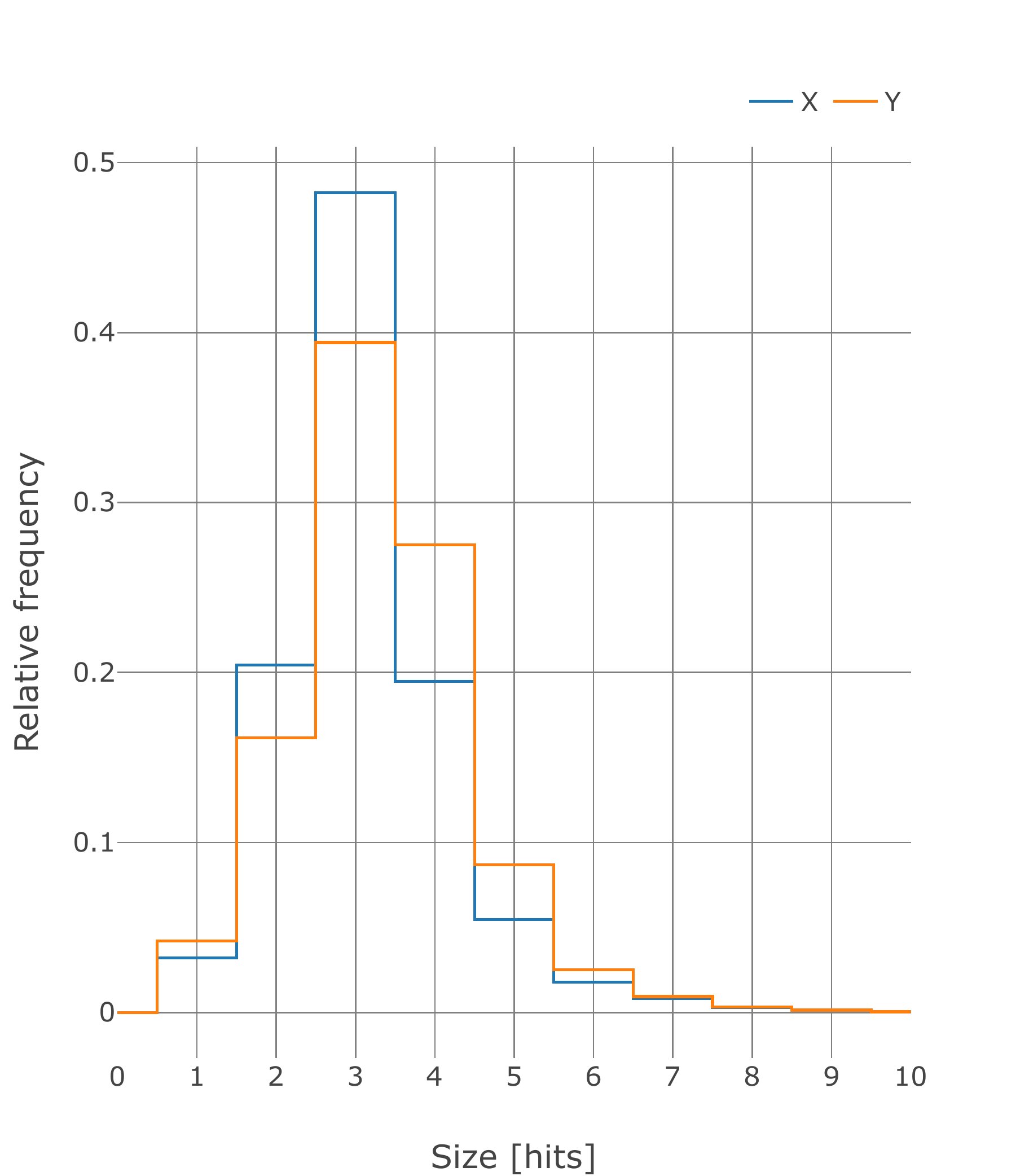}
\end{subfigure}%
 \caption{Cluster charge and cluster size distribution for a standard run. Histogram integral is normalized to one.}
\label{fig:ex_plot}
\end{figure}

\begin{figure}[h!]
\centering
\begin{subfigure}{.5\textwidth}
  \centering
  \includegraphics[width=1\linewidth]{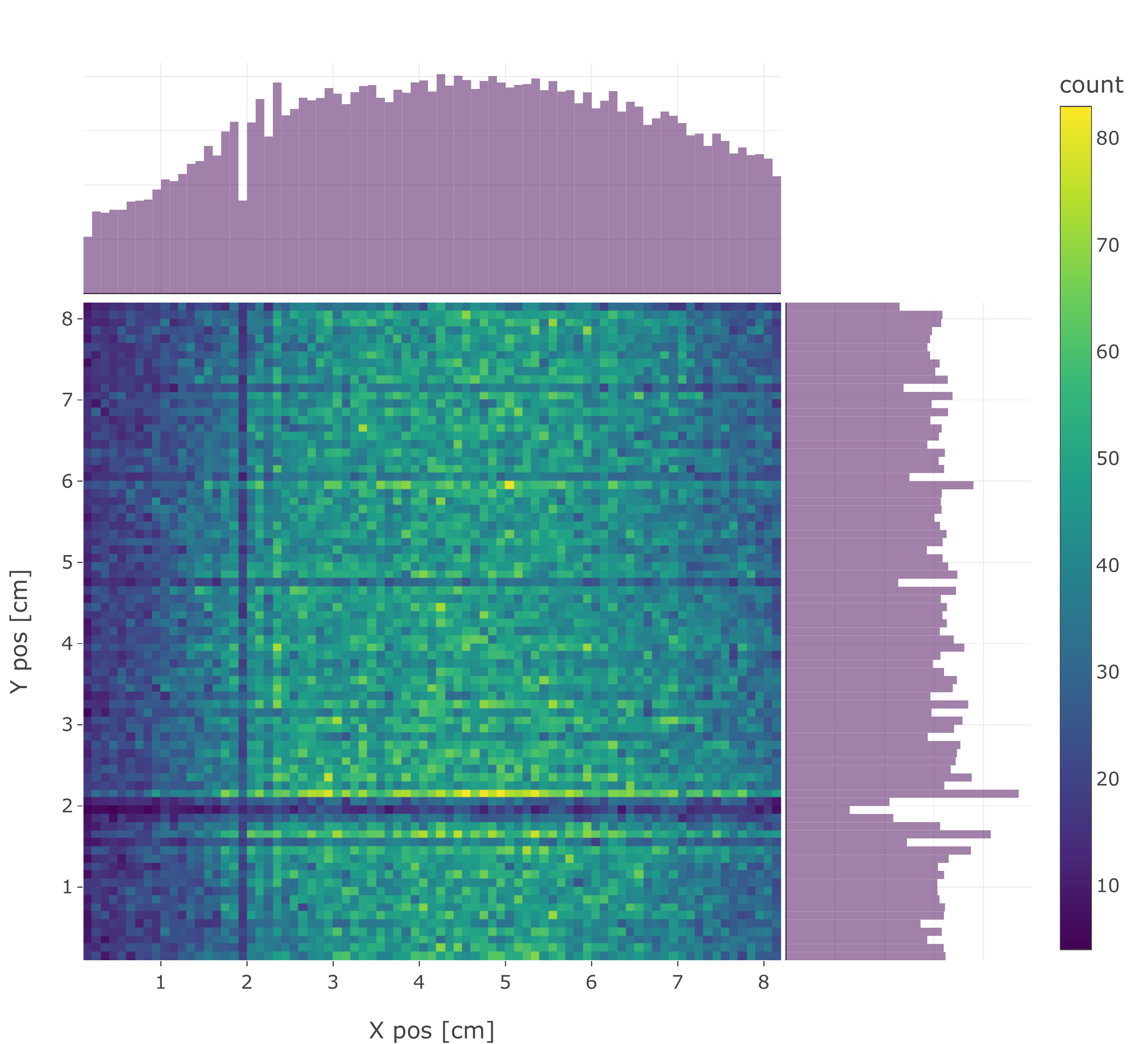}
\end{subfigure}%
\begin{subfigure}{.5\textwidth}
  \centering
  \includegraphics[width=1\linewidth]{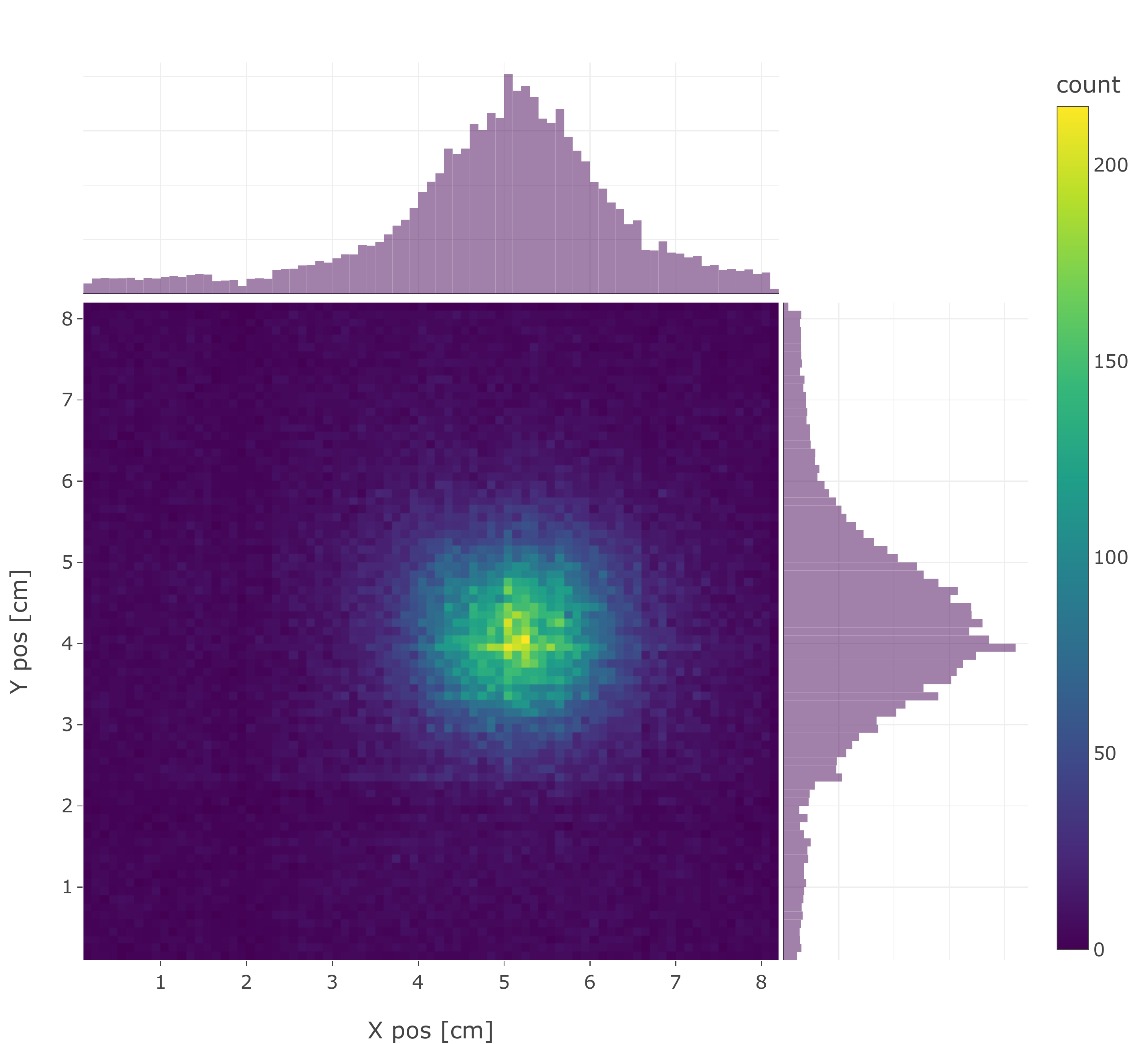}
\end{subfigure}%
 \caption{Beam profiles of muons (left) and pions (right). Projections of the profile are shown on the axes.}
\label{fig:profiles}
\end{figure}

\begin{figure}[h!]
  \centering
  \includegraphics[width=1\linewidth]{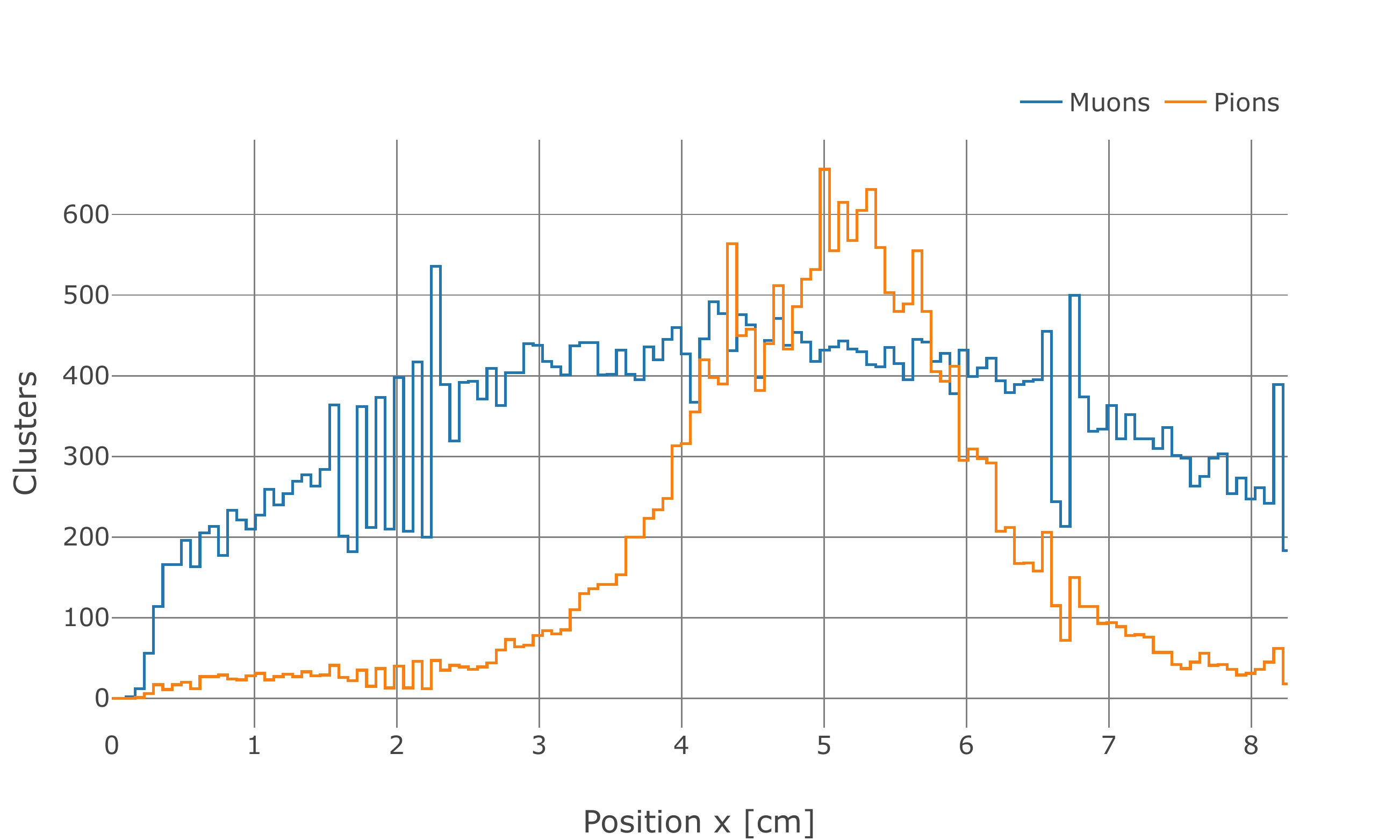}
 \caption{Cluster position in the X view for both beams in standard conditions. The effect of the missing strips (figure \ref{fig:strip_charge}) can be seen.}
\label{fig:profile_x_track}
\end{figure}

\begin{figure}[h!]
  \centering
  \includegraphics[width=1\linewidth]{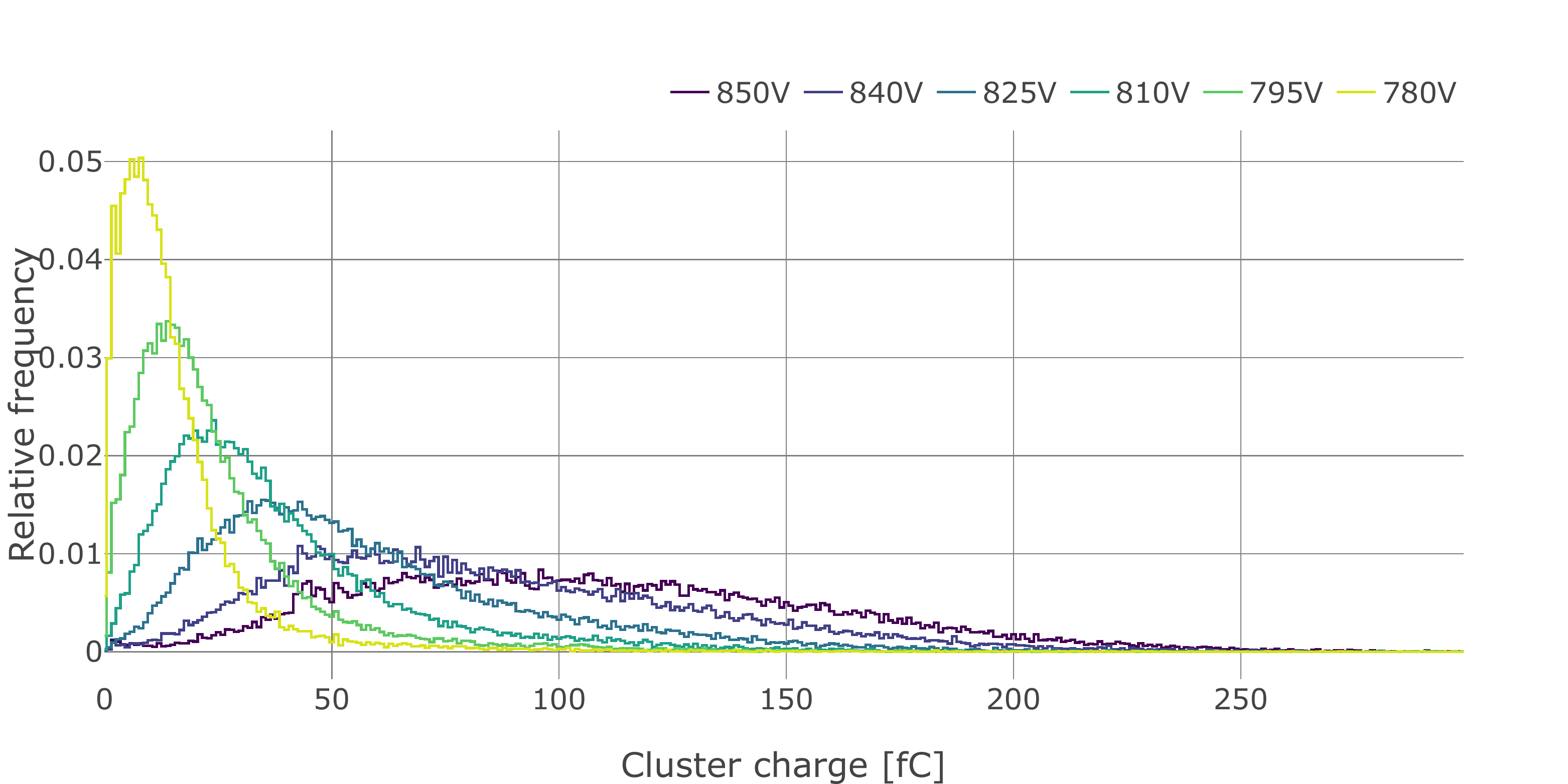}
 \caption{Cluster charge distribution at different gain values. Scan performed at \SI{30}{\degree}. Histogram integrals are normalized to one.}
\label{fig:cl_charge_gain}
\end{figure}

\begin{figure}[h!]
  \centering
  \includegraphics[width=1\linewidth]{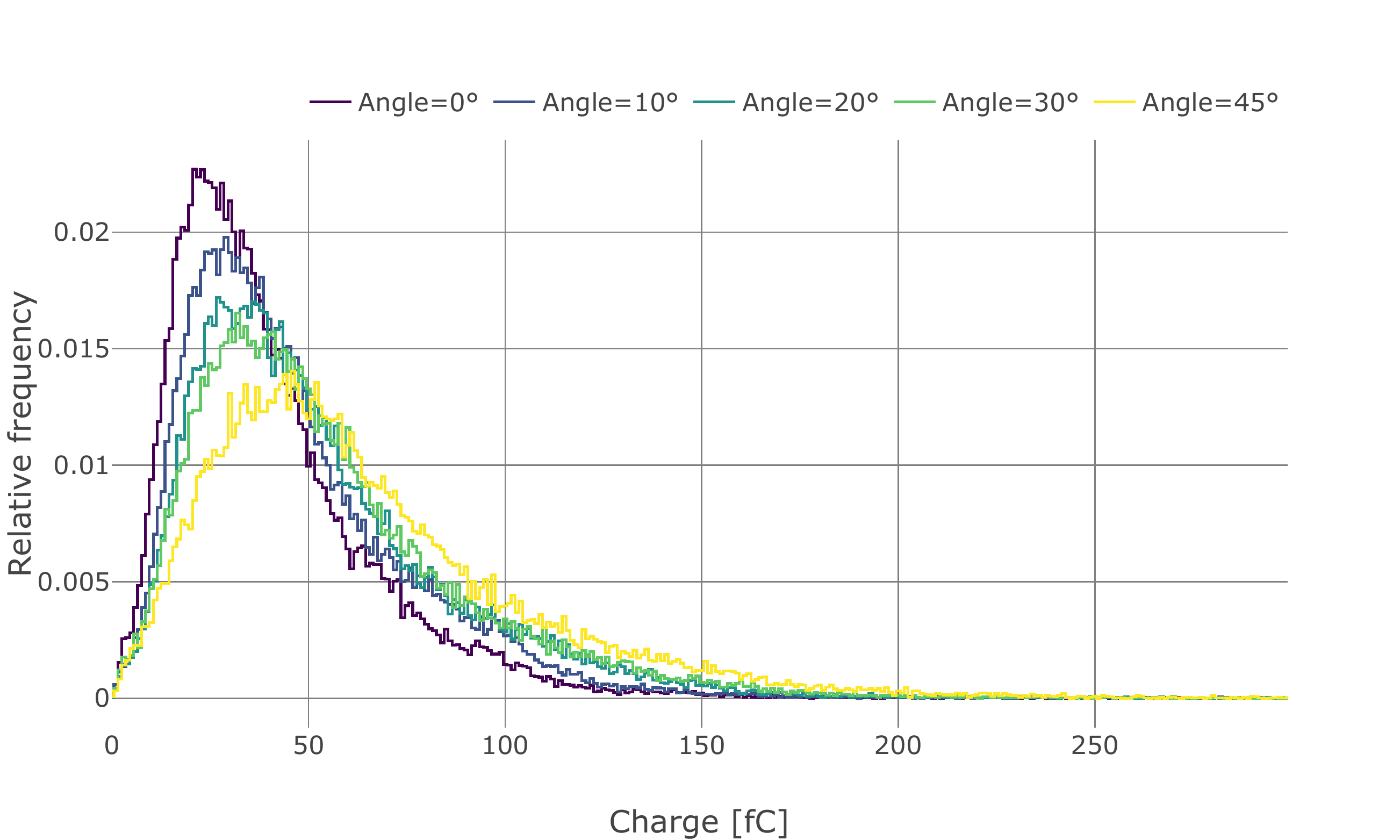}
  \includegraphics[width=1\linewidth]{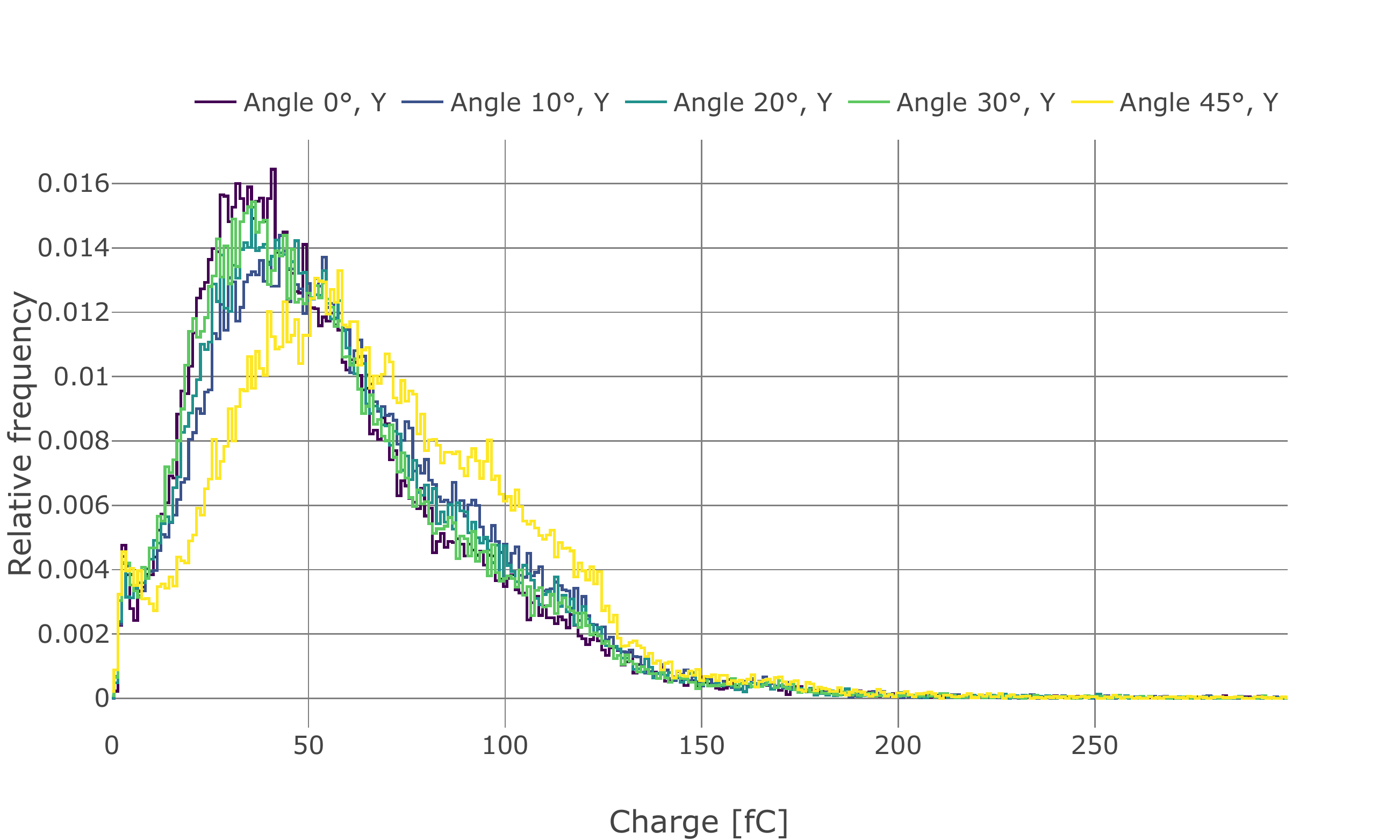}
 \caption{ Distribution of cluster charge at different beam angles in the X view (top) and Y view (bottom). Histogram integrals are normalized to one.}
\label{fig:cl_charge_angle}
\end{figure}

\begin{figure}[h!]
  \centering
  \includegraphics[width=1\linewidth]{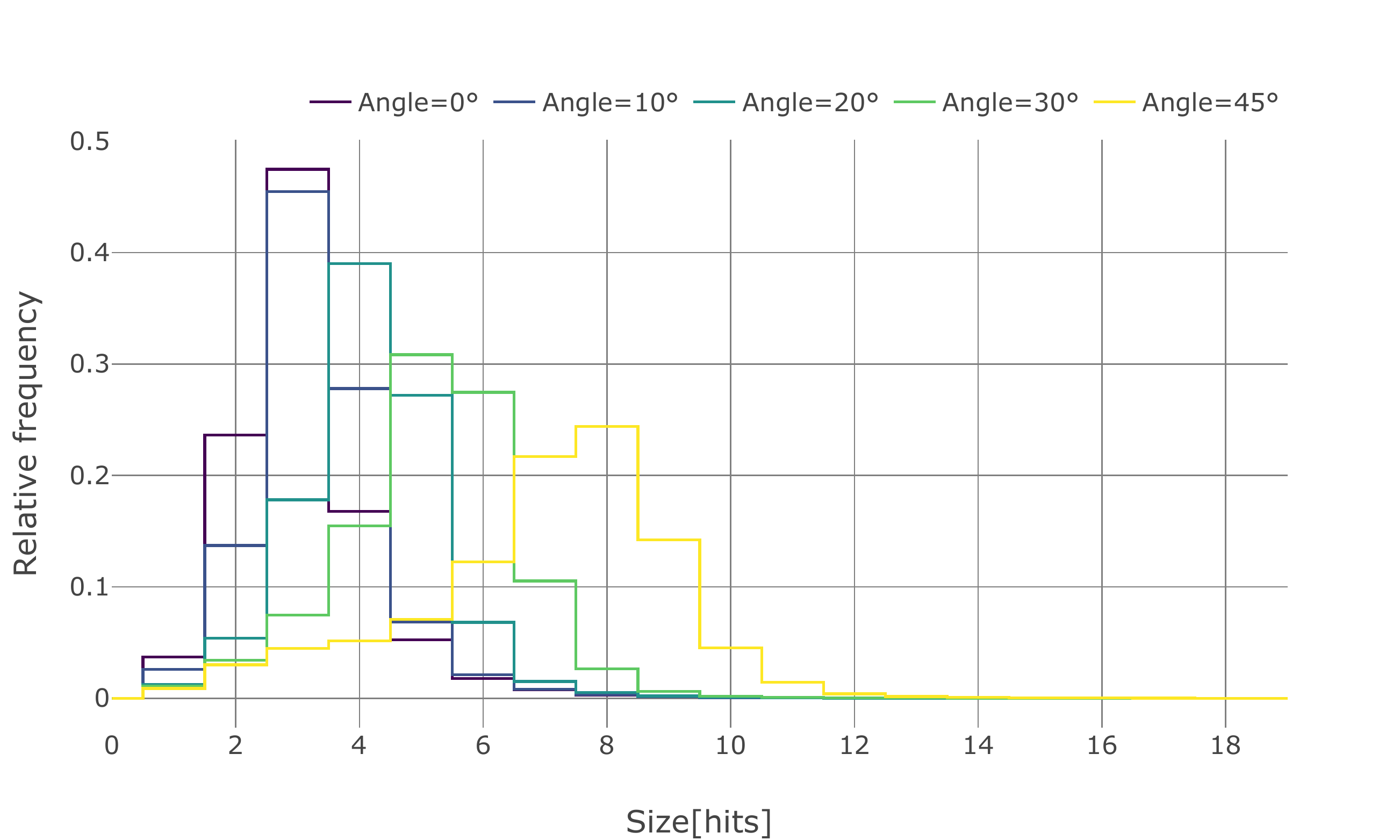}
  \includegraphics[width=1\linewidth]{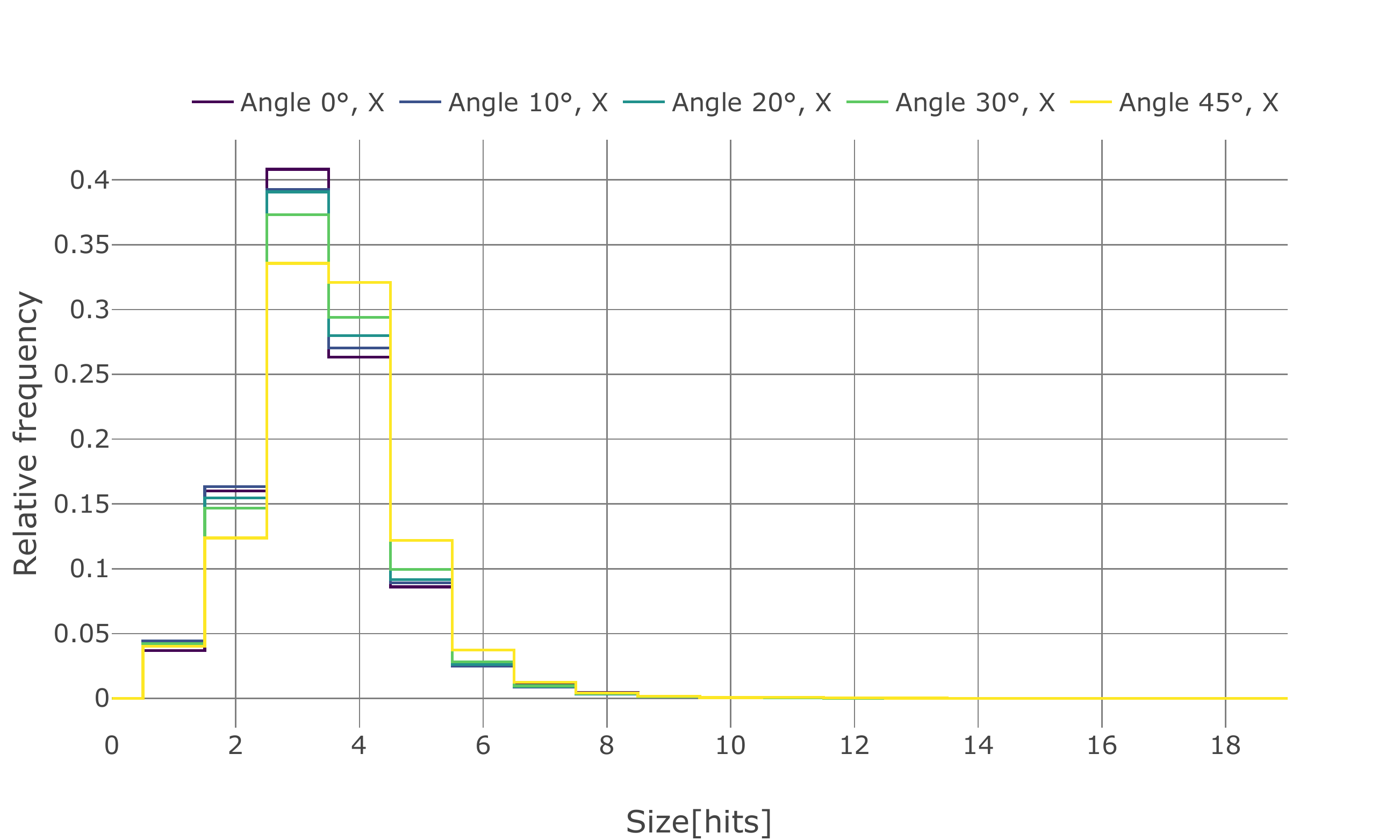}
 \caption{Distribution of cluster size at different beam angles in the X view (top) and Y view (bottom). Histogram integrals are normalized to one.}
\label{fig:cl_size_angle}
\end{figure}

In this chapter, we present some results from the CERN test beam with planar GEM detectors, carried out at the end of July 2021. The results are based on a sample of \SI{10}{\percent} of the total statistics, no alignment or other geometric correction was applied, so they should  be considered very preliminary.\\
The detectors were interfaced to the TIGER+GEMROC readout chain, managed and acquired with the GUFI software, and the data were analyzed online and offline with the CIVETTA software.\\
During the test beam, the following scans were mainly acquired:
\begin{itemize}
    \item Angle scan: when the angle between the beam and the planar detector was changed, see section \ref{setup_fe};
    \item Gain scan: when the gain of the detector was changed by changing the voltage on the GEM electrodes. The voltage at each GEM is displayed as G3/G2/G1;
    \item Drift scan: when the drift field was changed;
\end{itemize}
along with other scans described in sections \ref{sec:tb_thr_setting} and \ref{sec:integrationtime}. The results of the scans are presented qualitatively, from the hit level to more complicated evaluations. Unless  otherwise stated, the data from planar detector 1 are shown, counting from the first to interact with the beam (0) to the last (3). \\
The procedure for analyzing data at hit level is described in section \ref{civetta}.
The plot in figure \ref{fig:chargetime} is very useful for understanding the state of the detector. It shows the charge of each hit as a function of time with respect to the arrival time of the trigger. The numbers indicate the main features that can be derived from such plots:\\
\begin{figure}[h!]
\centering
\includegraphics[width=0.9\textwidth]{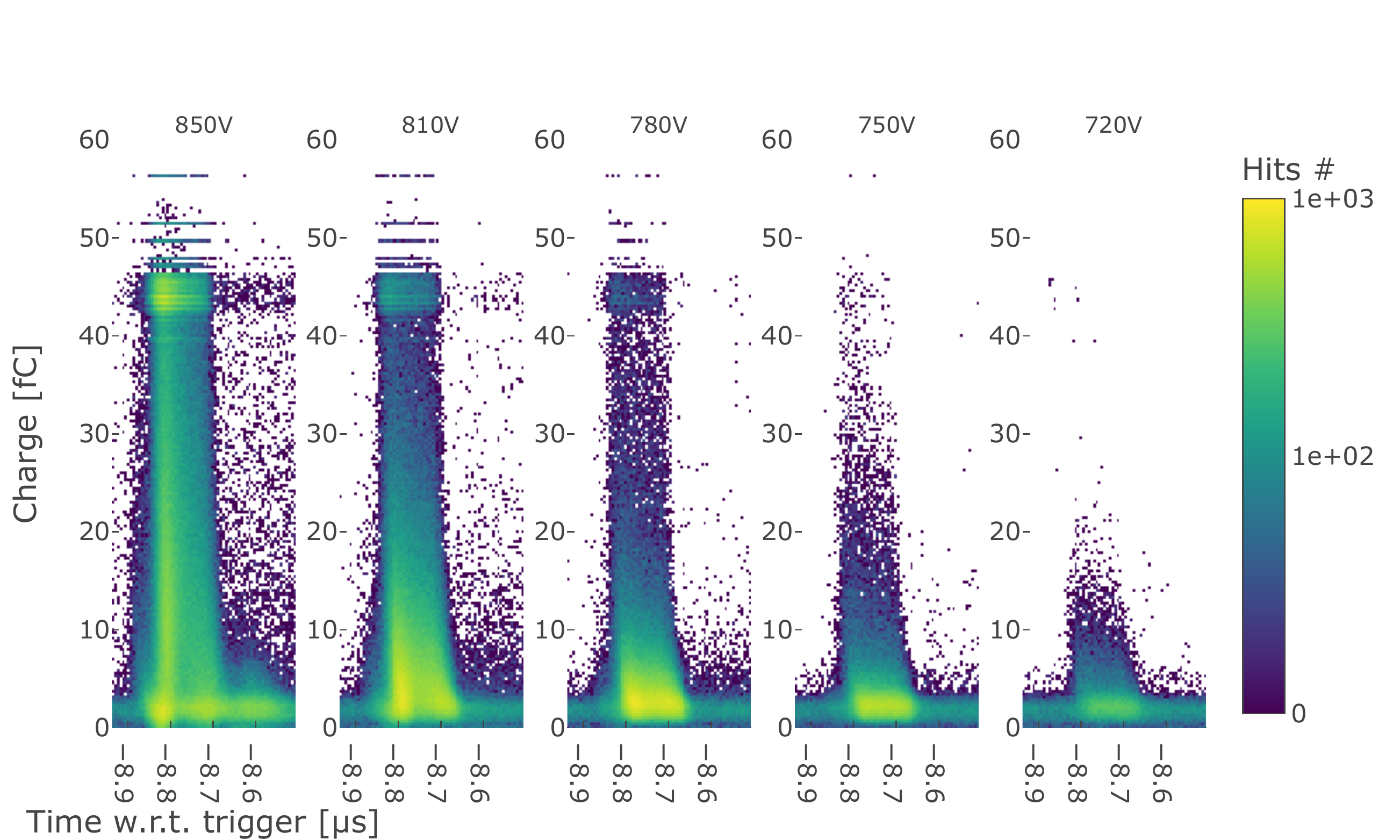}
\caption{Charge versus time, expressed in terms of arrival time of the trigger. At the top of each plot is the total sum of GEM voltages.}
\label{fig:gain_vs_chargetime}
\end{figure}
\begin{enumerate}
    \item Noise band: the area of low charge, without temporal correlation, gives a hint of the noise in the system. The color ratio to the signal range is an indication of the noise/signal ratio, while the height of the band is a measure of the noise amplitude.
    \item Signal area: the area between the two dotted lines is the area where the detector signal is located. The width of this zone depends on the volume traversed by the particle and the drift velocity. 
    \item Saturation: the different  saturation lines are due to the different saturation charge of each channel.
\end{enumerate}
Figure \ref{fig:gain_vs_chargetime} shows the charge versus time with different gain settings. The data were taken at \SI{30}{\degree} with a standard drift field (\SI{1500}{\volt \per \centi \meter}). The effects can be seen on both the saturation peak and the overall signal.\\
\newpage
\FloatBarrier
Figure \ref{fig:angle_vs_chargetime} shows charge versus time at different angles. The data were taken with a total voltage across the GEMs of \SI{825}{\volt}  (hereafter defined as standard HV settings) with a standard drift field \SI{1500}{\volt \per \centi \meter}. The plot shows both the X (top) and Y (bottom) views. Since the rotation is on the Y axis, the projection of the track becomes longer in X, while it remains the same in Y. For this reason, the signal arrival time in the X direction becomes more uniform at larger angles, while it remains peaked in the Y view . \\
Figure \ref{fig:drift_vs_chargetime} shows charge versus time at different drift fields. The data were taken with a total voltage across the GEMs of \SI{825}{\volt} and an angle of \SI{45}{\deg}. The drift velocity follows the expected trend for Ar:i\ch{C_4H_10} 90:10 (figure \ref {fig:drift_speed_ref}): the maximum velocity is reached at a field around \SI{500}{\volt \per \centi \meter}.\\
Another useful plot for checking the system status is the charge versus strip plot (figure \ref {fig:strip_charge}), where only the clusters in the time signal region have been selected. The example plot (figure \ref{fig:strip_charge}) shows a disconnected channel on the bond plane of the ASIC (strip number 103), while strips 26,28,30,32 and 34 are disconnected due to mapping incompatibility (see \ref{setup_fe}). The impact of the disconnected channels is mitigated thanks to cluster hit multiplicity, as can be seen in the cluster analysis described below.\\
%
%
\FloatBarrier
Clusters are built as described in section \ref{civetta}. A cluster can then be selected by pairing the cluster with  the highest charge in a view with the cluster with the highest charge on the other, or by fitting the track and using the residuals.\\
\begin{figure}[h!]
\centering
\includegraphics[width=1\textwidth]{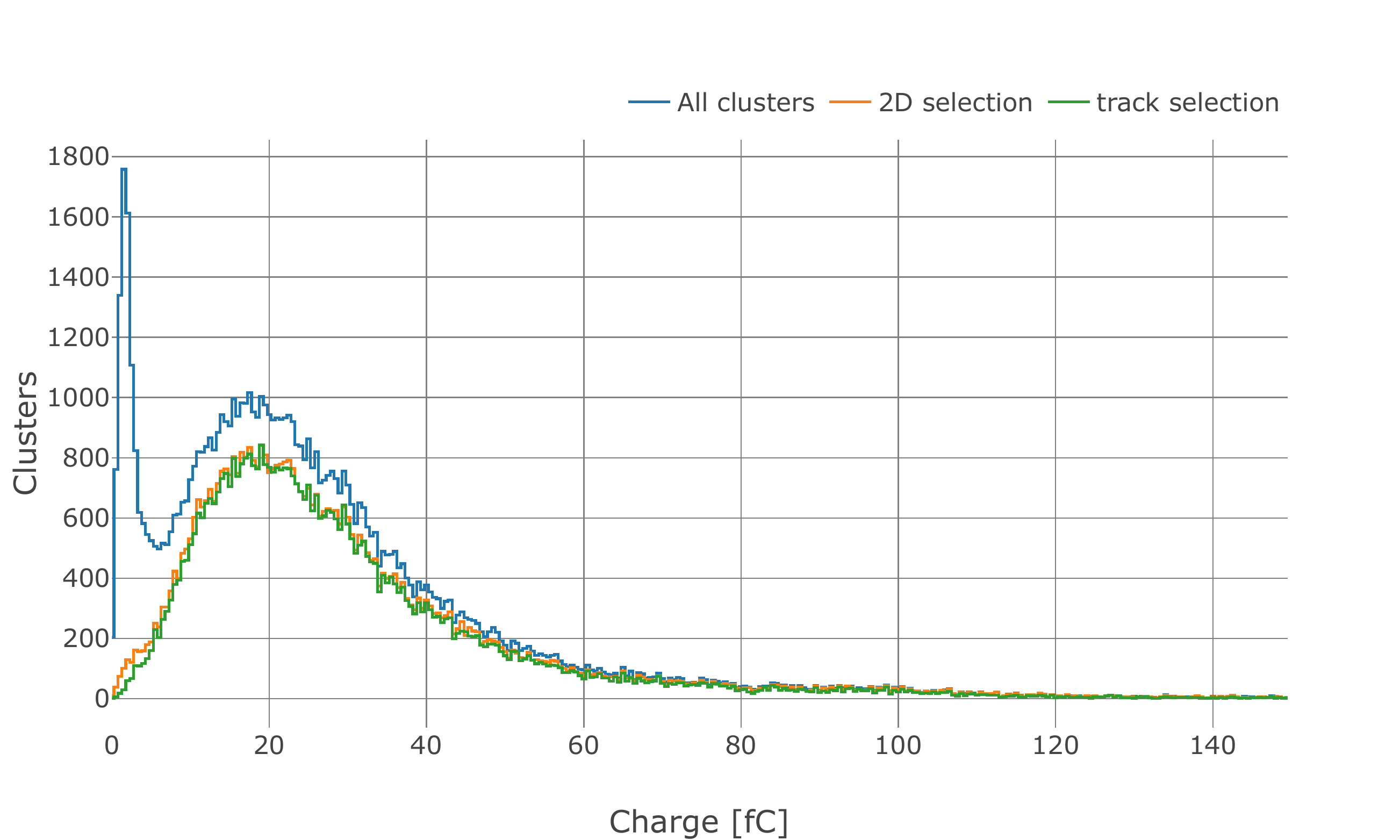}
\caption{Examples of cluster selection in the cluster charge distribution (planar detector 1, X view). Data were taken  with the default HV settings and a beam angle of \SI{0}{\degree}. }
\label{fig:drift_speed}
\end{figure}
The distribution of cluster charge for different selections is shown in figure \ref{fig:drift_speed}. As can be seen, the prominent noise peak at low charge is greatly reduced by the selection, while the statistics of the signal region is not affected too much. The 2D selection is only used to generate the beam profile plots, while the  other features are evaluated using the track selection.\\ Figure \ref{fig:ex_plot} shows the distributions of cluster charge and size in standard HV settings and at a beam angle of \SI{0}{\degree}. The kernel density estimators used to extract the most probable value (MPV) are also shown. The plot also shows the difference between the X and Y views. In this example, the MPV value for the cluster charge value in the X view is \SI{28}{\femto \coulomb} (\SI{57.4(2)}{\femto \coulomb} average), while for the Y view is \SI{36}{\femto \coulomb} (\SI{47.1(2)}{\femto \coulomb} average). The average cluster size is also larger for the Y view: 3.348$\pm$0.006 strips versus 3.156$\pm$0.005 for the X view.\\
As for the spatial uniformity of the detector, the beam profiles for muon and pion beams are shown in figure \ref{fig:profiles}. The muon beam has an oval shape and covers  the entire detector in the Y direction and almost the entire detector in the X direction. The pion beam is more collimated and circular. Figure \ref{fig:profile_x_track} shows the position of the clusters selected by tracks in the X view. The clusters on the missing strips are still reconstructed if their multiplicity is greater than one, which partially compensates for the missing strips, at least on the efficiency side.\\
\FloatBarrier
The cluster distributions are affected by many scanned parameters. Figure \ref{fig:cl_charge_gain} shows the  variation of cluster distribution with gain, while figures \ref{fig:cl_charge_angle} and \ref{fig:cl_size_angle} show the dependencies of cluster charge and multiplicity on the beam angle.\\

\begin{figure}[h!]
  \centering
  \includegraphics[width=1\linewidth]{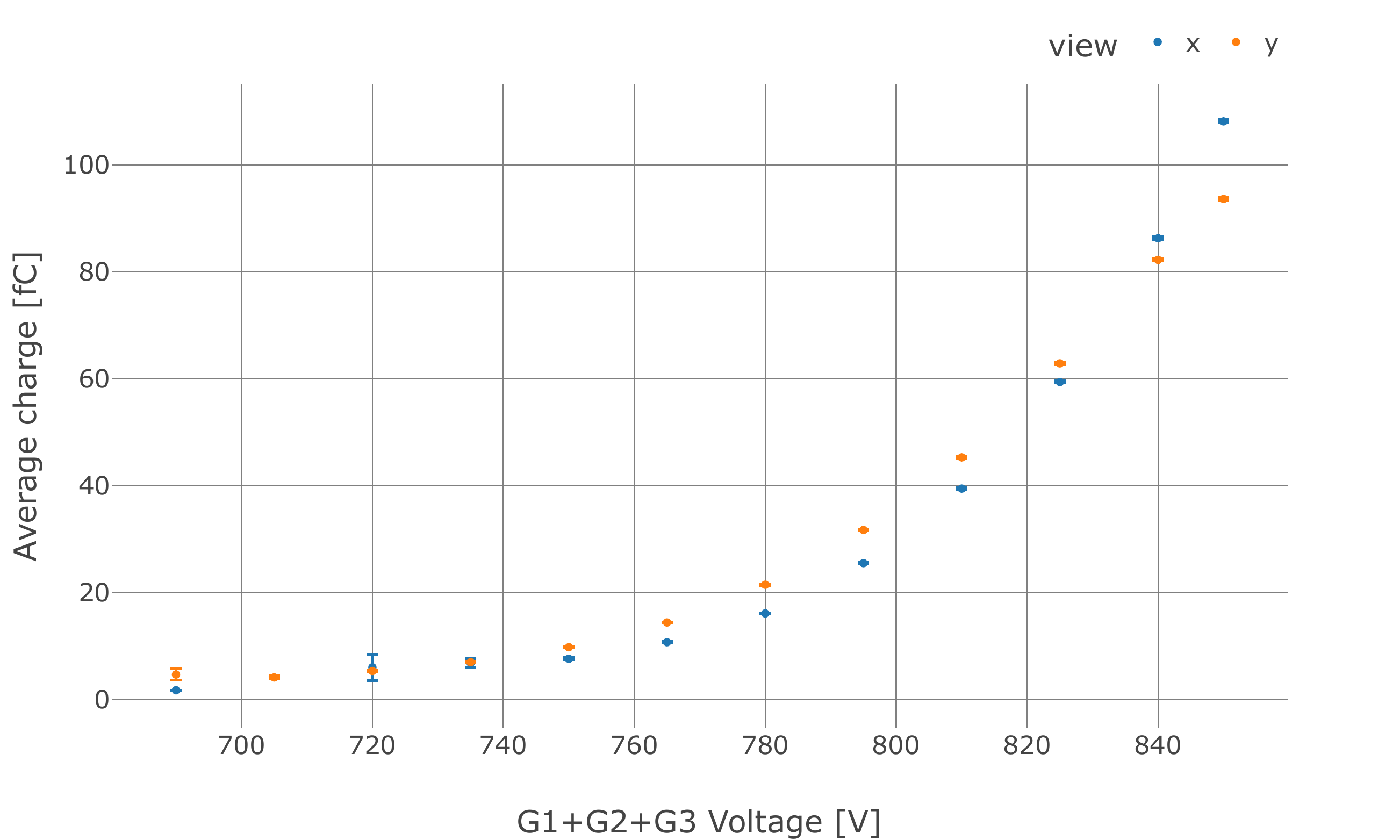}
  \includegraphics[width=1\linewidth]{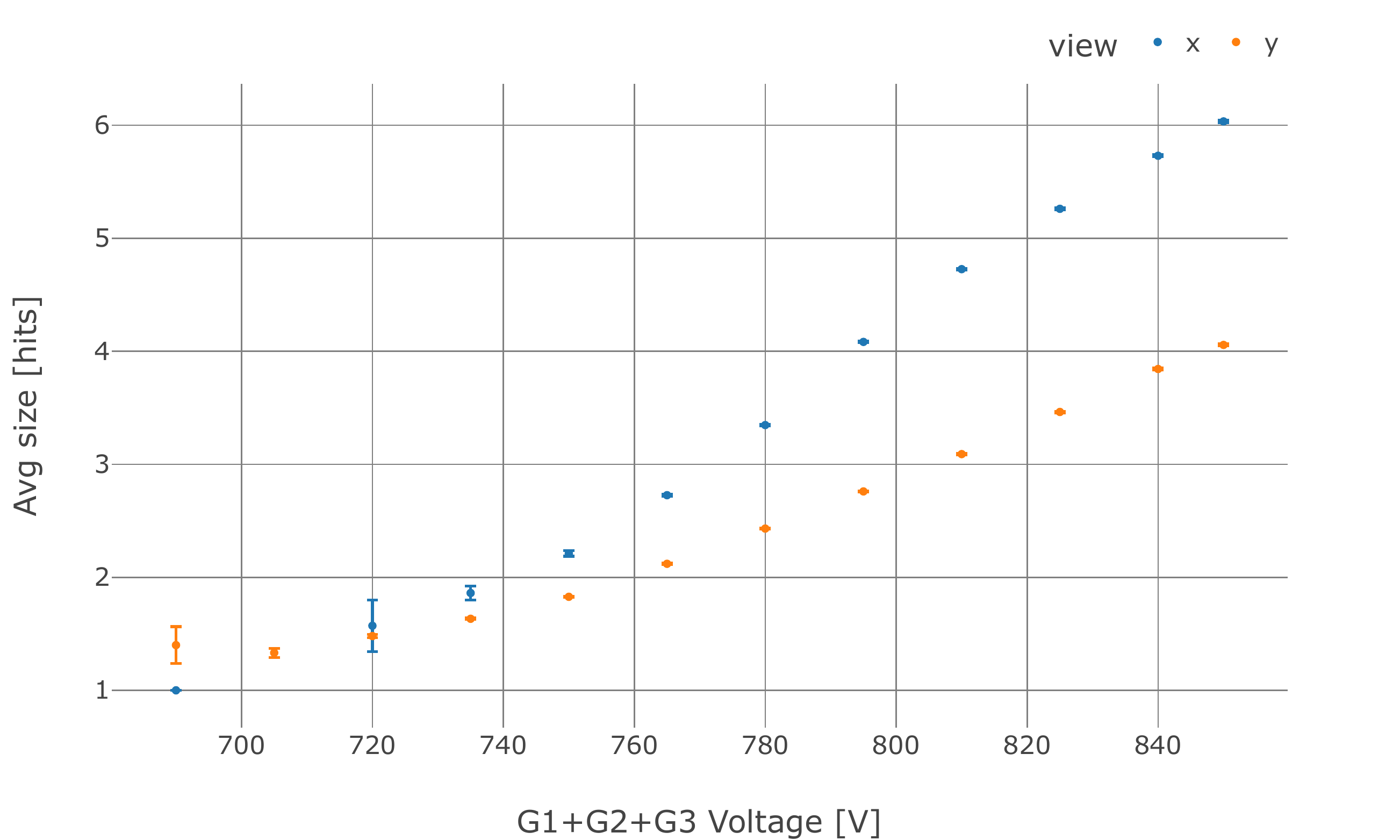}
 \caption{Average values of cluster charge (top) and size (bottom) at different gains. The uncertainties are not shown since they would be barely visible.}
 \label{fig:size_charge_vs_gain}
\end{figure}

\begin{figure}[h!]
  \centering
  \includegraphics[width=1\linewidth]{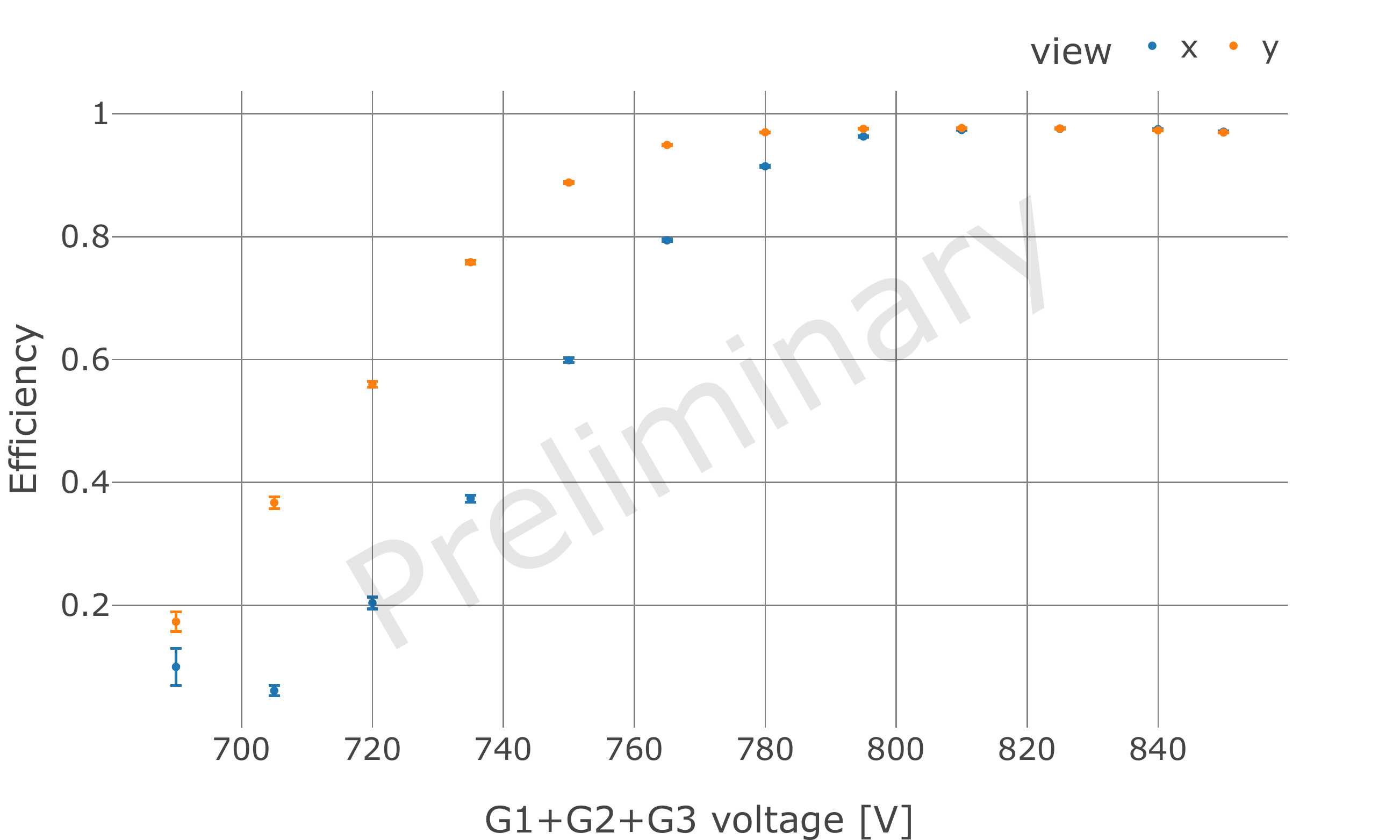}
 \caption{Efficiency with respect to total voltage on triple-GEM.}
 \label{fig:eff_gain}
\end{figure}

The average and most probable value can be extracted from the distributions to evaluate the trend of the cluster distribution with respect to different parameters. Figure \ref{fig:size_charge_vs_gain} shows the trend of average charge and size with respect to the voltages on the GEM foils.
The error of the average is calculated as $\sigma_{avg}=\frac{std}{\sqrt{N}}$ where $std$ is the standard deviation of the sample distribution and $N$ is the total number of samples.\\
One of the most important performance parameters is the efficiency of the detector. At this stage of the analysis, only a fast estimate of the efficiency is calculated after fitting the tracks for each view and plane as follows:
$$
\eta=\frac{N_d}{N_t}
$$
where $N_t$ is the number of events where the residual on the three tracking detectors\footnote{Here with "tracking detectors" we intend the detectors which are not under test} is within \SI{2}{\milli \meter} of the residual mean, while $N_t$ is the number of events where all residuals on the four detectors are below \SI{2}{\milli \meter} of the residual mean. This method allows for quick calculation of the detector efficiency without the need for alignment and with low computation power, so that these data are available even during acquisition.\\
The error in efficency is then calculated as follows:
$$
\sigma_\eta=\sqrt{\frac{\eta\left( 1-\eta \right)}{N_t}}
$$
A preliminary estimate of efficiency is plotted with respect to total voltage on triple-GEM (i.e., the detector gain) in figure \ref{fig:eff_gain}. Note that the efficiency curve shows a plateau at about \SI{800}{\volt}.
\FloatBarrier

\section{Studies on threshold settings}
\label{sec:tb_thr_setting}
\begin{figure}[h!]
  \centering
  \includegraphics[width=0.9\linewidth]{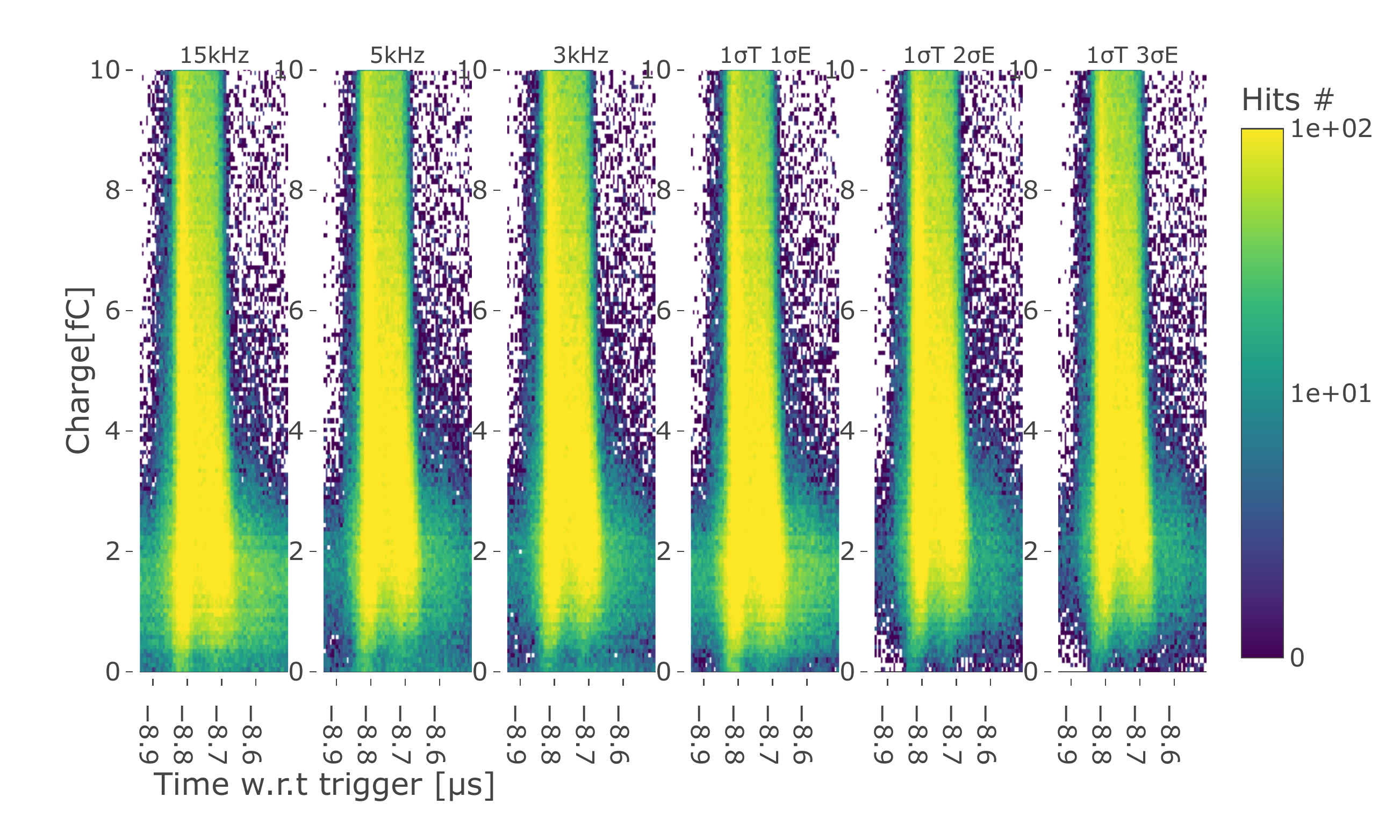}
  \includegraphics[width=0.9\linewidth]{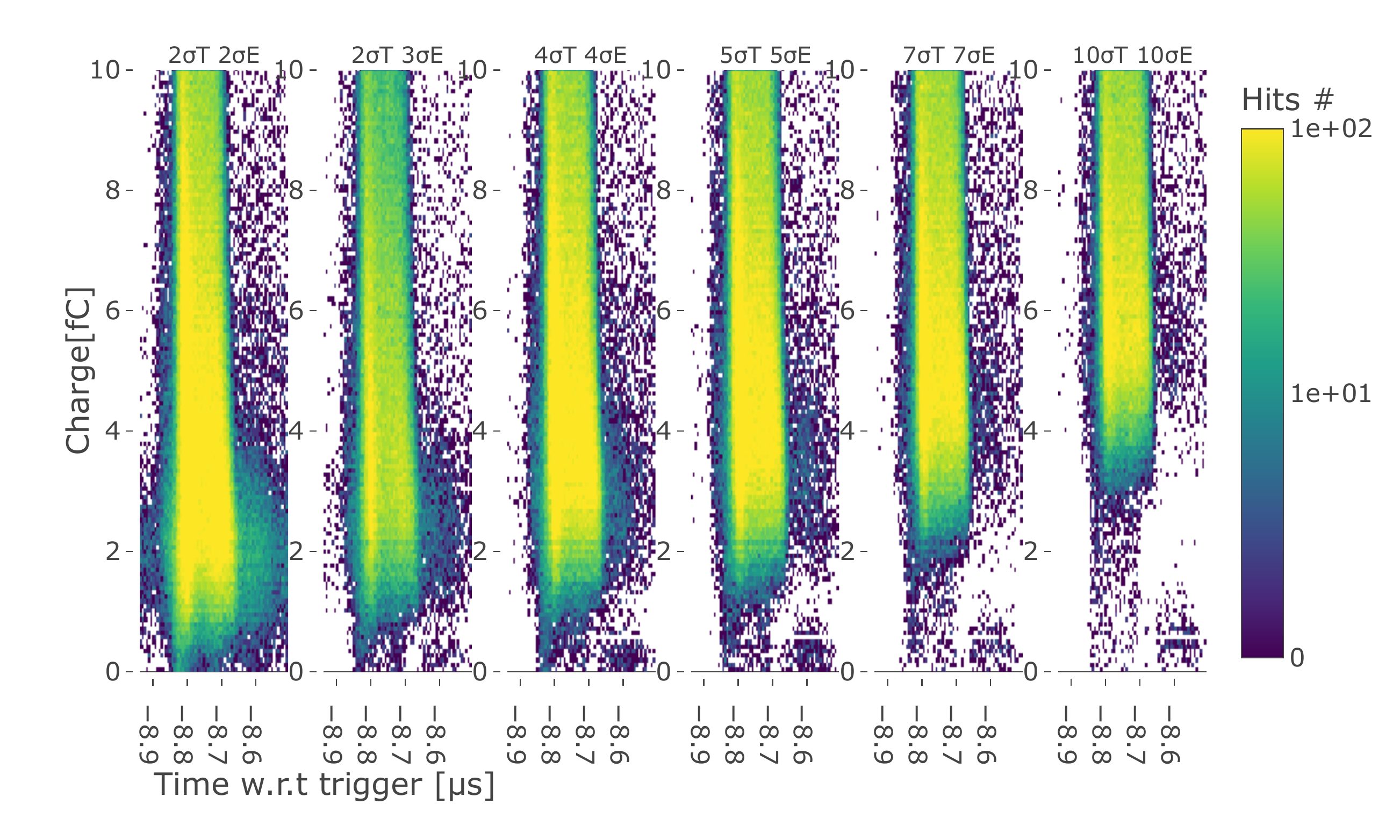}
 \caption{Charge versus time in terms of arrival time of the trigger for different threshold settings. The plot shows the hits below \SI{10}{\femto \coulomb} to underline the effect. }
\label{fig:thr_cut_charge_time}
\end{figure}

During the test beam, it was possible to test different threshold settings and their effects on performance. The different threshold settings were described in section \ref{thr_setting}. Both setting modes were tested, with three values of rate tuning and nine different sets of thresholds determined by the fit method.\\
\begin{figure}[h!]
  \centering
  \includegraphics[width=0.9\linewidth]{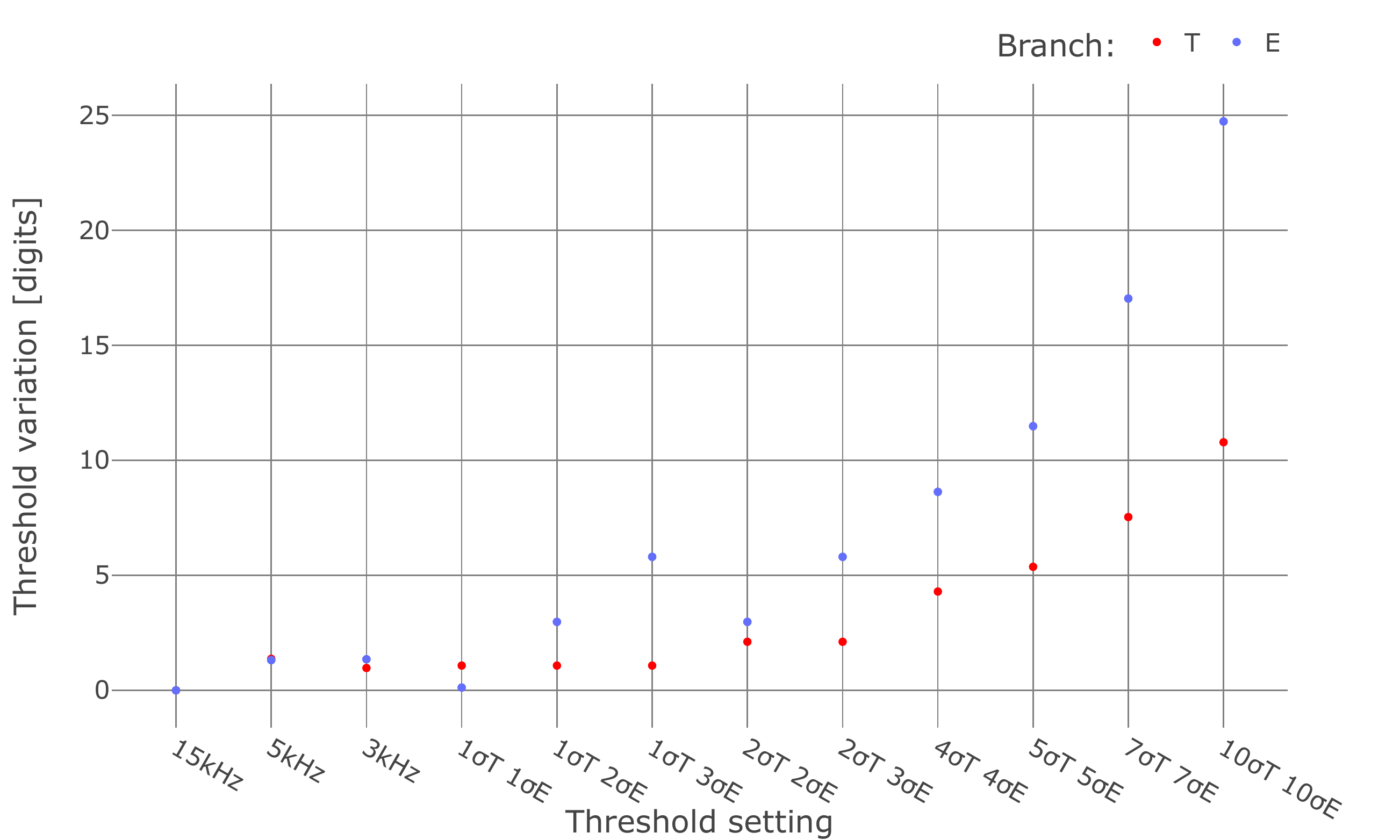}
 \caption{Threshold difference with respect to the lowest threshold set (\SI{15}{\kilo \hertz})}
 \label{fig:thr_values}
\end{figure}
The difference in digits between the lowest threshold (\SI{15}{\kilo \hertz}) and the other settings is shown in figure \ref{fig:thr_values}.\\
The different threshold settings affect the hit charge distribution, as can be seen in figure \ref{fig:thr_cut_charge_time}, by cutting the hits with lower charge values. Note that the measured charge and the threshold are not fully correlated. For a given signal shape, the charge sampled by Sample-and-Hold circuit may be less than the effective charge threshold of the discriminators. This rare effect results in few points of measured charge below the effective charge threshold, as shown figure \ref{fig:thr_cut_charge_time} for the $10\upsigma$-$10\upsigma$ threshold.

\begin{figure}[h!]
  \centering
 \includegraphics[width=0.9\linewidth]{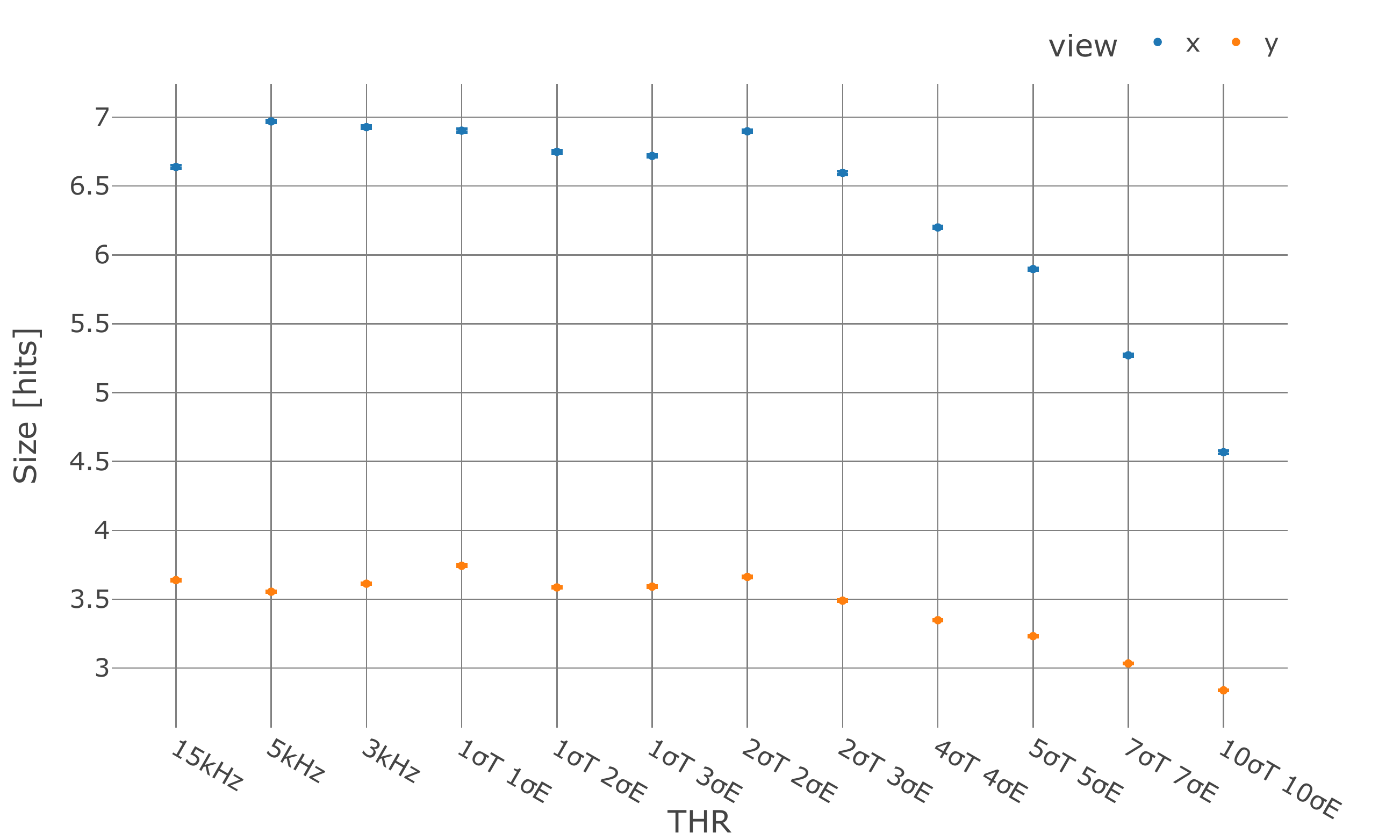}
 \caption{Average cluster size at different threshold settings. The statistical error is barely visible. }
\label{fig:thr_stats_clusters}
\end{figure}
\begin{figure}[ht!]
  \centering
 \includegraphics[width=0.9\linewidth]{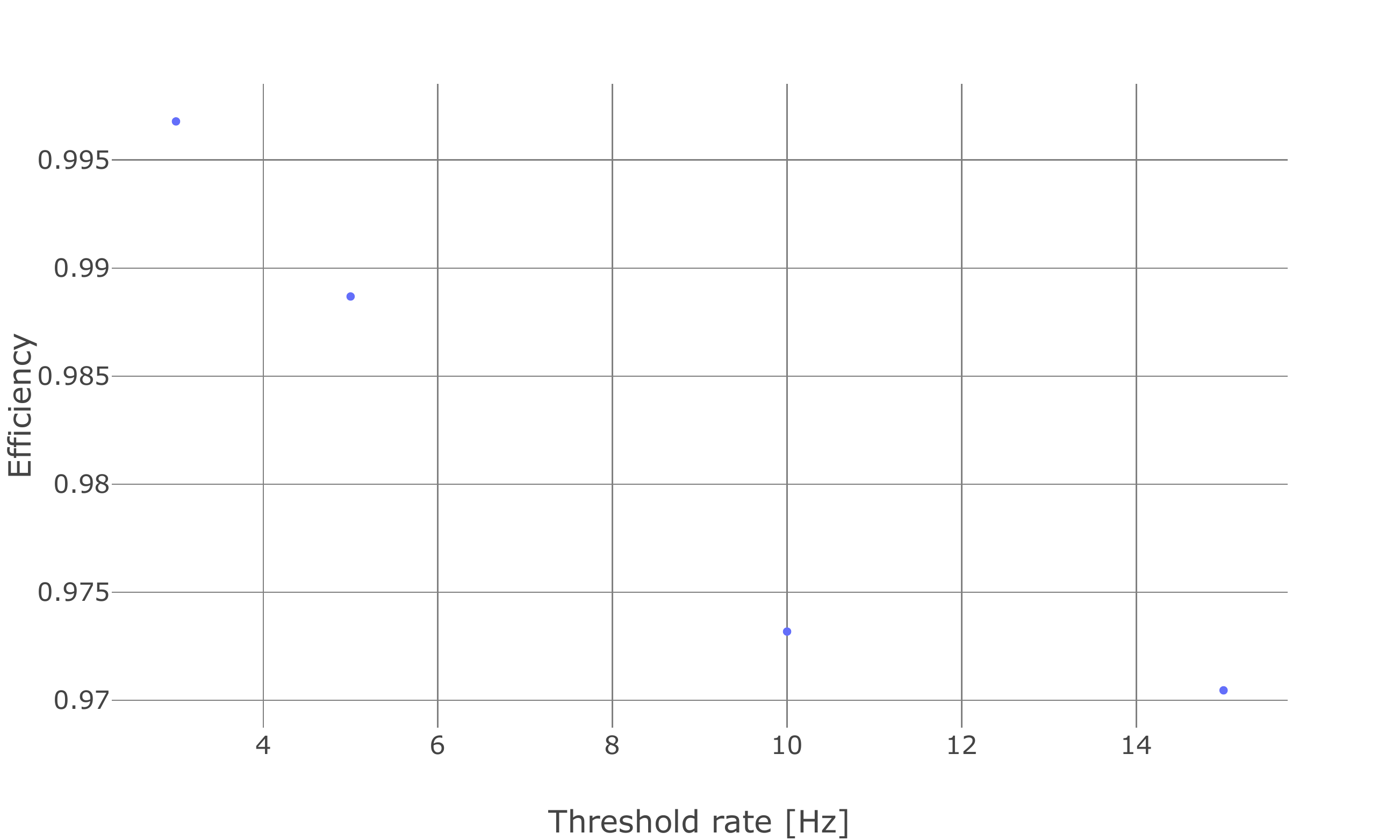}
 \caption{Efficiency of detection of test pulses compared to set threshold rate }
\label{fig:eff_tp}
\end{figure}

The scan was performed using the default setting of HV and at a beam angle of \SI{45}{\degree}, so the effects on cluster size and charge are different for the two views. Figure \ref{fig:thr_stats_clusters} shows the average cluster size at different threshold settings. Note that the cluster size is larger in the X view  than in the Y view, and both sizes decrease as the thresholds increase, because the low charge hits in the clusters get lost.\\
\begin{figure}[h!]
  \centering
 \includegraphics[width=0.9\linewidth]{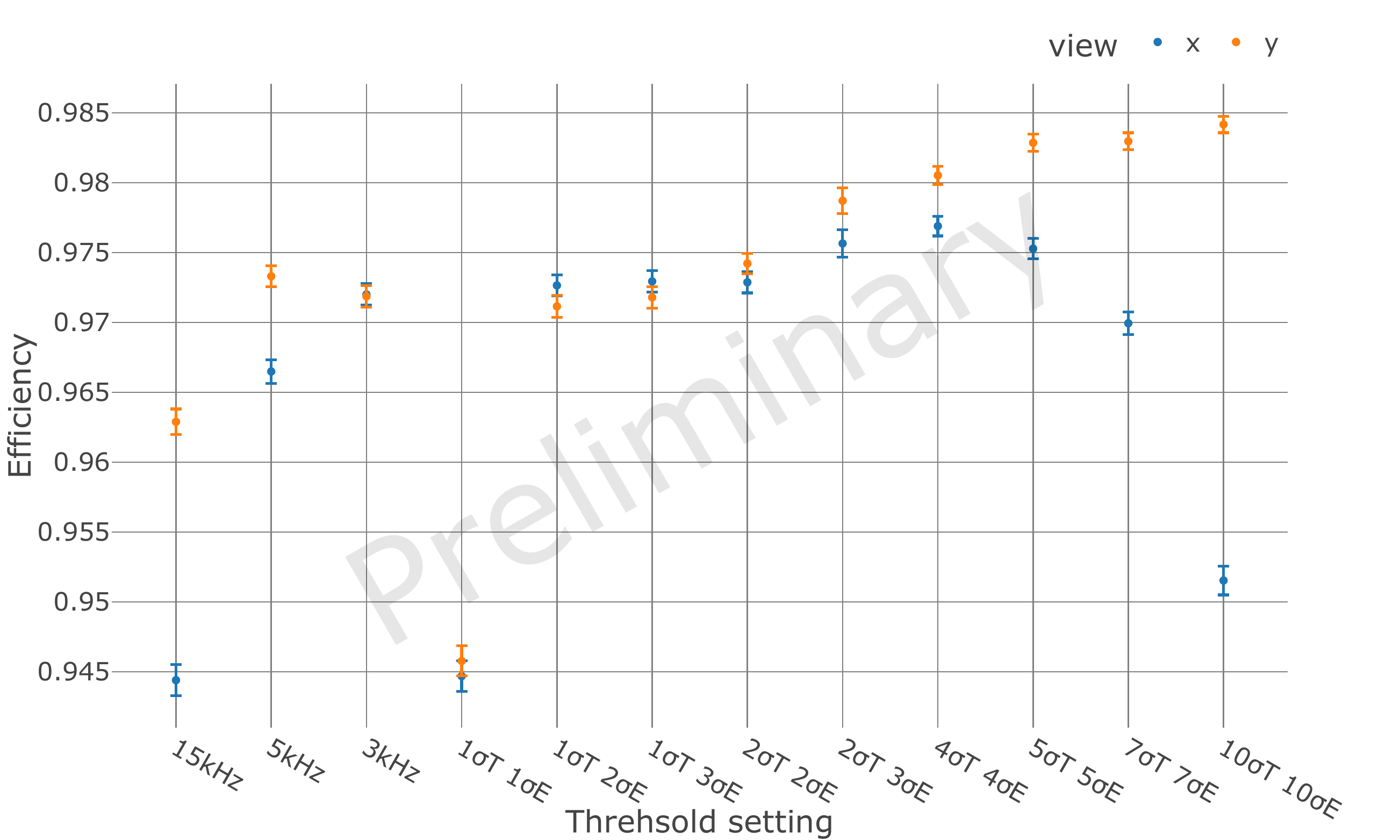}
 \caption{Detector efficiency at different threshold settings. }
\label{fig:eff_thr}
\end{figure}
When you measure efficiency as described in section \ref{civetta}, an interesting pattern emerges (figure \ref{fig:eff_thr}). Efficiency in the X view increases as the threshold increases until it decreases at thresholds higher than $4\upsigma$-$4\upsigma$ . Efficiency in the Y view, on the other hand, continues to increase with the threshold values. The increase in efficiency can be explained by studying the GEMROC trigger-matching. An efficiency test can be performed by generating  a test pulse in one channel with different noise levels on the other channels and measuring the number of such test pulses within the trigger-matched packets. The results of this test are shown in figure (\ref{fig:eff_tp}). The reduction of efficiency is not yet fully understood and may be due to inefficiency in the trigger-matching processing.\\
The loss of efficiency in the X view at higher thresholds can be explained by considering the angle of incidence of the beam. Due to the angle of incidence, the X clusters are indeed wider and with more distributed charge, so whole clusters may be lost at higher thresholds.\\
Even though the thresholds in the $1\sigma-1\sigma$ configuration are similar to the thresholds at \SI{15}{\kilo \hertz}, the efficiency is much lower in the Y view. This can be due to the poor rate optimization of this threshold setting, which may cause some channels to have a very high rate, degrading trigger-matching performance.\\
Further studies will clarify this point,  to spot the flaws in GEMROC firmware or to find the optimal threshold configuration for CGEM-IT.\\
\FloatBarrier
\section{Studies on integration time}
\label{sec:integrationtime}
\begin{figure}[h!]
  \centering
 \includegraphics[width=0.9\linewidth]{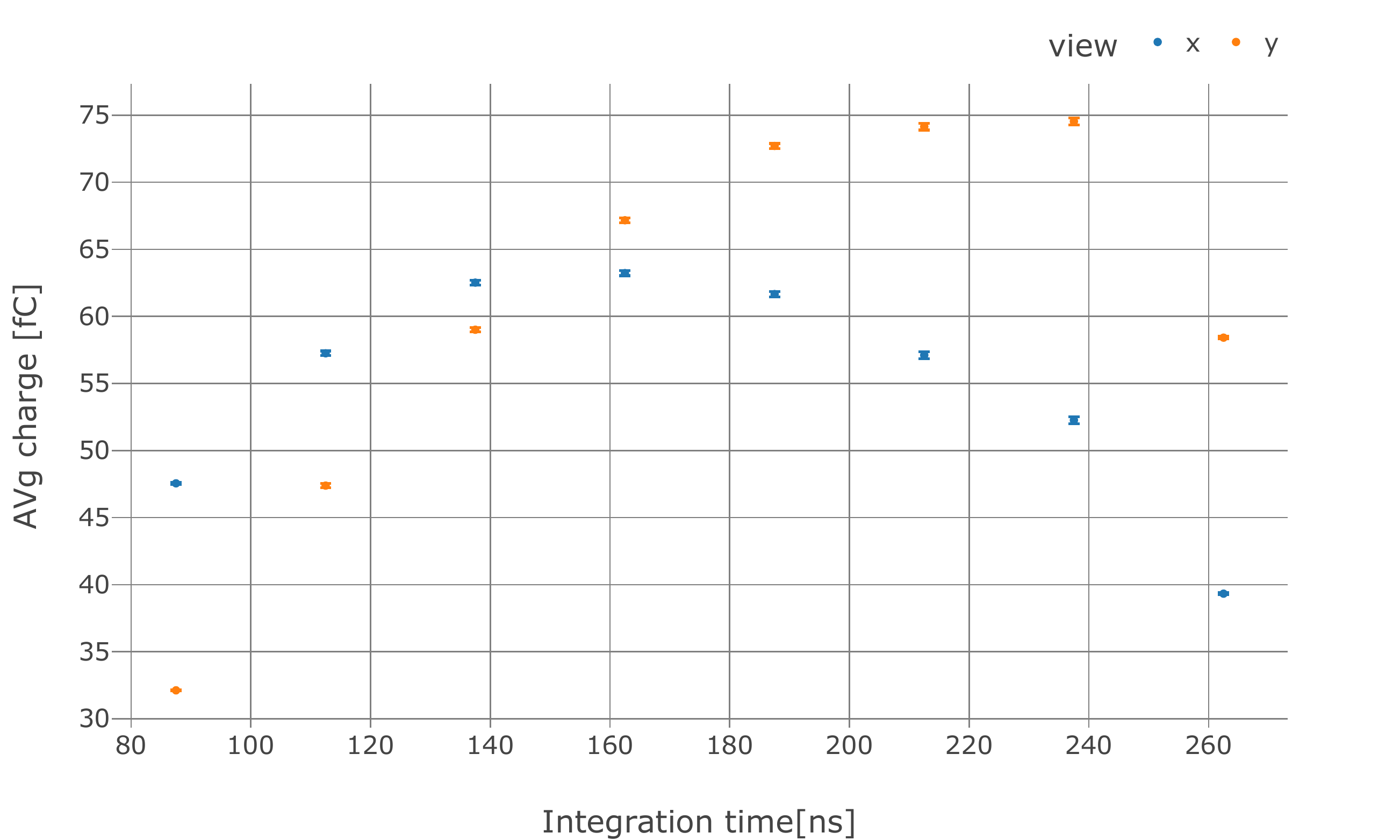}
 \caption{Average cluster charge versus integration time.}
\label{fig:avg_it}
\end{figure}

As mentioned in section \ref{asic_arch}, in the configuration chosen for the CGEM-IT readout, the TIGER ASIC extracts the hit charge information using a Sample-and-Hold circuit. This circuit samples the peak, storing after a programmable time its value in a capacitor, and then converts this value. When we change the programmable value, we can monitor the measured charge and then reconstruct important information about the signal duration.\\
The signal is sampled after 2+4*\textit{integ\_time} clock cycles, where \textit{integ\_time} is a programmable field in the ASIC registers.\\
Figure \ref{fig:avg_it} shows a scan of the available \textit{integ\_time} between 3 and 10, performed with the default settings of HV and at a beam angle of \SI{45}{\degree}.
The plot shows that in the X view  charge peaks were measured with an integration time of about \SI{160}{\nano \second}, while the Y view charge peaks were measured with a very long integration time of about \SI{230}{\nano \second}. Given the rising time of the E branch of \SI{170}{\nano \second}, we can deduce that the signal in the X view has an average development time around the design reference (\SI{50}{\nano \second}), while the signal induced in the Y view has a longer development. This can be explained by considering the beam angle. Since the geometric projection of the particle path in the drift region covers fewer Y strips, it is more likely that  multiple ionization generate signals on the same strip, and therefore the signals are longer.\\

\chapter{Conclusions}
The CGEM-IT is a novel, lightweight detector designed to upgrade the current inner tracker of the BESIII spectrometer. This detector is based on the GEM technology and represents an improvement over the current inner track\-er, especially in terms of resolution along the z-axis and reconstruction of secondary vertexes.\\
The working principle of this detector was presented together with its design features and expected performance.\\
A complete readout chain has been successfully developed to meet the CGEM-IT requirements. Dedicated on-detector and off-detector electronics have been designed according to the required performance and are described in this thesis.  \\
The on-detector electronics consists of a mixed-signal ASIC called TIGER, which directly reads the detector strips and provides the full digitized charge and time information. Its specifications are tailored to modern micropattern gas detectors. The ASICs were produced, characterized, and calibrated, and have been already installed on two layers of the CGEM-IT. The third layer is under construction.\\
The GEMROCs are FPGA-based modules designed to provide power, timing, and control signals to the TIGERs and receive and process the output data from TIGER ASICs. Since TIGER sends a continuous data stream, the GEMROC firmware is responsible for matching the information on each hit with the arrival time of the trigger and creating the event data packets.\\
The key features of the GEMROC firmware were explained, as well as the patches and updates that have been implemented to ensure proper operation of the devices.\\
The VME-based GEM Data Concentrators downstream the GEMROC modules receive the trigger-matched data packets via optical links and assemble them into full events which are then provided to the VME-based BESIII DAQ system.\\
The entire electronics chain has been studied in terms of its characteristics and performance in relation to the detector and the environmental conditions.\\
All components have already been tested and proven to perform as required. The detector electronics is now in the final stages of commissioning and optimization.\\
During my PhD, a complete software suit was developed specifically for the needs of the CGEM project. The GUFI software is a complete interface to both the on-detector and off-detector electronics. It is able to configure the devices, measure their status and performance and control the data acquisition. CIVETTA is an online and offline analysis tool that allows to measure the status and performance of the detector and electronics. Both software programs are mature and tested and can be easily adapted to different setups and scenarios. \\
The entire chain, from detector to software, has been tested in two different setups, one with two over three layers of CGEM-IT and one with four planar GEM detectors. The tests were performed with cosmic rays and during a test beam. Some preliminary results of the tests have been presented in this thesis.\\
The whole readout system has been developed with a view to great modularity and could be used for other innovative detectors.
\subsection*{Acknowledgment}
We acknowledge the support of the European Commission in the RISE Project 645664-BESIIICGEM, RISE-MSCA-H2020-2014 and in the RISE Project  872901-FEST, H2020-MSCA-RISE-2019.\\


\appendix 

\chapter{Gaseous radiation detectors working principles}
\label{working_principles}
Gaseous radiation detectors are extensively used in high-energy and par\-ti\-cle physics. The development of detectors with the underlying physical process of the ionization of a gas-filled region began in the early twentieth century and continues today. Although the technology has evolved in many forms over the years, the basic physical principles used by these detectors remain the same. This appendix provides a brief introduction to these principles, specifically how GEM detectors work.  For more information see \cite{Knoll} and \cite{sauli_2014}.
\subsubsection{Ionization}
\begin{figure}
    \centering
    \includegraphics[width=1\textwidth]{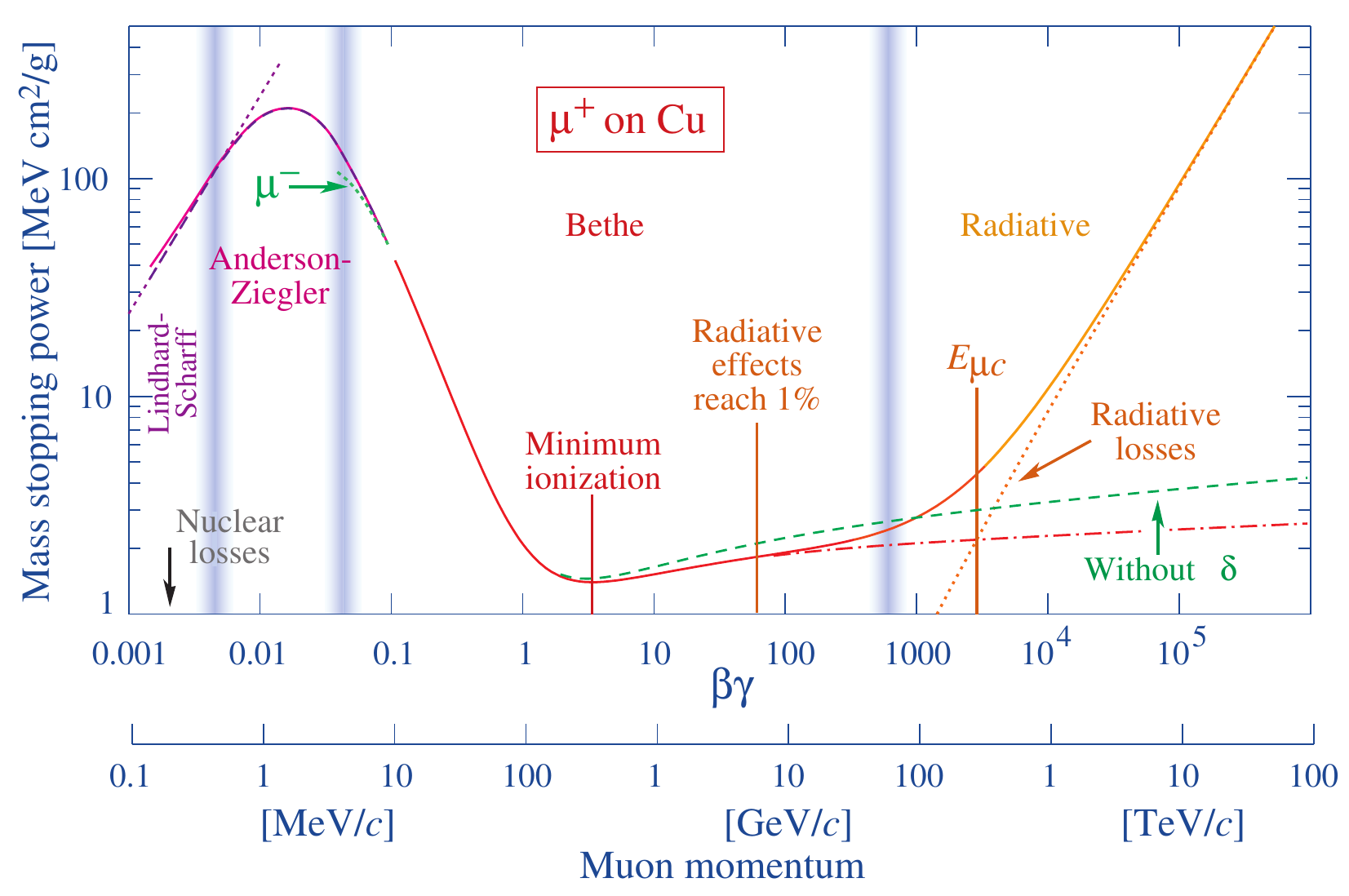}
    \caption{Mass stopping power (= $\langle -$dE/dx$\rangle$) for positive muons in copper as a function of $\upbeta\upgamma$ = p/Mc over nine orders of magnitude of momentum (twelve orders of magnitude of kinetic energy). The solid curves show the total stopping power\cite{PDG}.}
    \label{fig:stopping_power}
    \end{figure}

In all detector architectures, detection of nuclear radiation is always by transferring some  or all the energy to a suitable medium in which direct or indirect effects of the interaction can be detected.\\
For charged particles, most of the energy released in matter is due to electromagnetic interactions. Except for particles approaching the end of their range, where mechanical elastic collisions become relevant, the energy loss in matter is mainly due to inelastic processes of ionization and excitation, whose probability is a function of the energy transfer involved.\\
The energy lost by the particle in the material and the process that dominates the loss depend on the type of radiation, the velocity of the particle and the material. Figure \ref{fig:stopping_power} shows the mass stopping power\footnote{The stopping power is the retarding force acting on charged particles due to their interaction with matter, resulting in a loss of particle energy. The mass stopping power is the stopping power divided by the density of the material and therefore depends very little on the density of the material.} for positive muons in copper as a function of $\upbeta\upgamma$ = p/Mc over nine orders of magnitude of momentum (twelve orders of magnitude of kinetic energy). Among the wide variety of processes at high or low energy, for this type of application and the purpose of this appendix, we will consider the processes that dominate the "Bethe" region of the graph.\\
In this region, in the range between 0.1 $\lesssim \upbeta\upgamma \lesssim $ 1000, the average energy loss due to multiple Coulomb interactions is described by the Bethe-Bloch expression:
$$
\frac{\upDelta E}{\upDelta x } = -\uprho \frac{2KZ}{A \beta^2} \left[ln \frac{2mc^2\beta^2}{I(1-\beta^2)}- \beta^2 -\frac{C}{Z} - \frac{\delta}{2} \right] \qquad K = \frac{4\pi N e^2}{mc^2}
$$
where $\beta$ is the particle velocity, $e$ and $m$ are the charge and mass of the electron, $Z$,$A$,$\uprho$ are the atomic number, mass number and density of the medium, $N$ is the Avogadro's number. The term $\frac{C}{Z}$ represents the so-called inner shell corrections that account for a reduced ionization efficiency in the deepest electronic levels, and $\frac{\delta}{2}$ is a density effect correction that results from a collective interaction between the medium and the Coulomb field of the particle. There is no simple analytical expression for these factors. They are usually empirically derived and tabulated.\\
In practice, most relativistic particles, such as cosmic-ray muons, have an average energy loss rate that is close to the minimum, and they can be considered "minimum-ionizing particles" or MIPs.\\
The result of Coulomb collision in this momentum range is usually the production of excited species or the production of positive ions with the ejection of free electrons into the medium (primary ionization). The photon produced by the de-excitation of the excited species and the electrons with a high energy transfer yield (often called delta electrons) can further interact with the medium and produce more ionizations (secondary ionization). \\
The number of primary ionizations, which are the result of independent events, follow Poisson-like statistics.
$$
P^{N_p}_k = \frac{N_p^k}{k!}e^{-N_p}
$$
$P^{N_p}_k$ is the probability that $k$ primary ionizations occur in a region with an average of ${N_p}$ primary ionizations. The numerical value of $N_p$ depends directly on the primary ionization cross section $\sigma_p$ and the density of interaction centers. Since $\sigma_p$ is the sum of the cross sections of many processes, it is quite difficult to determine it from theoretical calculations.\\
The theoretical detector efficiency, defined as the probability that there is at least one interaction, is then:
$$
\epsilon = 1 - P^{N_p}_0= 1- e^{N_p}
$$
This efficiency represents an upper limit to the detector performance. Many processes of capture and recombination lower this efficiency, in addition to the inefficiency due to the next steps of detection.\\ 
The minimum energy required for ionization, called the First Ionization Potential, is not a value of great interest in the development of detectors since the average energy per ion/e$^-$ couple (Effective Ionization Energy  $W_i$) is higher due to energy losses during excitation phenomena.\\
Knowing both $\upDelta E$ and $W_i$, the total number of ion pairs per unit length can be calculated as follows:
$$
N_T=\frac{\upDelta E}{W_i} 
$$
The number of these primary and secondary ionizations depends on the medium traversed,  and the values for MIP can be found in the literature. For composites and for gas mixtures, a composition law based on relative concentrations can be used with good approximation.\\
\subsubsection{Charge transport}
\begin{figure}
    \centering
    \includegraphics[width=1\textwidth]{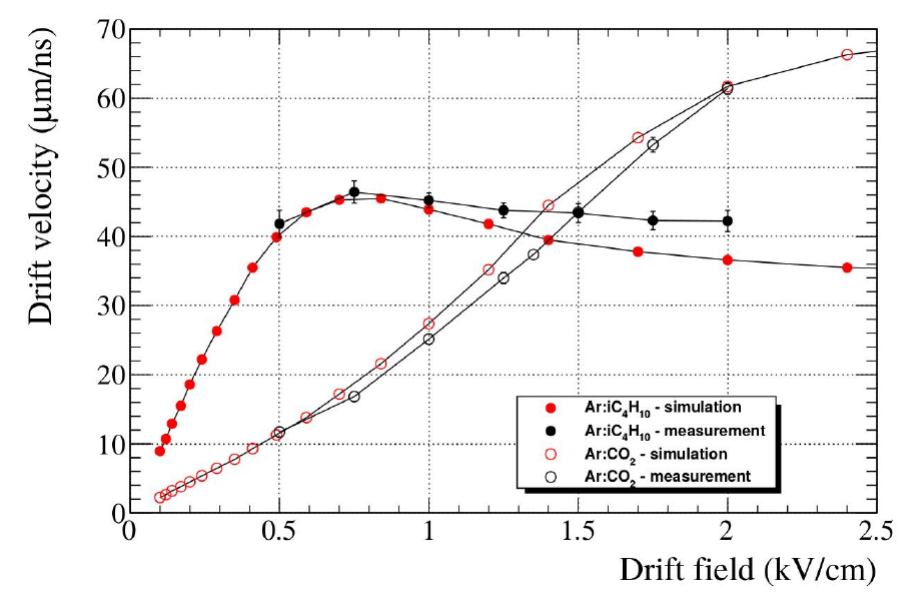}
    \caption{Simulated drift velocity compared to the measured drift velocity using a triple-GEM with \SI{5}{\milli \meter} drift gap and an incidence angle of \SI{30}{\degree}. Gas mixtures are Ar:i\ch{C_4H_10} (90:10) and Ar:\ch{CO2} (90:10) \cite{riccardo_tesi}.
}
    \label{fig:drift_vel}
    \end{figure}

Ions and electrons released in a gas by ionizing encounters rapidly lose their energy through multiple collisions with the surrounding molecules and acquire the thermal energy distribution of the medium. Under the action of moderate external electric fields, the charges move through the medium and diffuse, until they are neutralized either by recombination in the gas or on the walls. \\
In the absence of external fields and inelastic collision processes, the ions and electrons released in the gas behave like neutral molecules, whose properties are described by classic kinetic theory of gasses.  Maxwell-Boltzmann law gives the probability that an atom or molecule has an energy $\upepsilon$:
$$
F(\upepsilon) = 2 \sqrt{\frac{\upepsilon}{\pi(kT)^3}}e^{\frac{\upepsilon}{kT}}
$$
Where $k$ is the Boltzmann constant. The average thermal density is obtained by integrating the distribution:
$$
\Bar{\upepsilon}_T=kT
$$
The corresponding distribution of velocity $f \left(v\right)$ for a particle of mass $m$ is:
$$
f(v) = 4\pi \left( \frac{m}{2\pi kT} \right)^{\frac{3}{2}}v^2e^{-\frac{mv^2}{2kT}}
.$$
This yields an average velocity by integration:
$$
\Bar{v}= \sqrt{\frac{8kT}{\pi m}}
$$
A localized distribution of particles diffuses symmetrically through multiple collisions, and follows a Gaussian law:
$$
\frac{\mathrm{d} N}{N}= \frac{1}{\sqrt{4\pi Dt}}e^{-\frac{x^2}{4Dt}}dx
$$
where $\frac{\mathrm{d} N}{N}$ is the fraction of particles located in the element $dx$ at a distance $x$ from the origin after a time $t$, while $D$ is the diffusion coefficient.\\
When an electric field is applied to the gas volume, a net movement of ions and electrons along the field direction is observed. The average velocity is called drift velocity $w^+$.
This velocity is linearly proportional to the electric field up to very high values of the electric field $E$. We can define a quantity $\mu$, the ion mobility, as:
$$
\mu=\frac{w^+}{E}
$$
The value of mobility is specific to each ion moving in a given gas and depends on pressure and temperature by the expression:
$$
\mu(P,T)=\frac{T}{T_0}\frac{P_0}{P}\mu(P_0, T_0)
$$
As for the electrons, due to their small mass, they can considerably increase their energy between collisions with the gas molecules, and therefore their mobility with respect to the electric field is not constant .\\
A simple formulation of Townsend allows to write the drift velocity of electrons as follows:
$$ 
w^- = k \frac{eE}{m}\tau
$$
Where $\tau$ is the mean time between collisions and $e$ and $m$ are the charge and mass of the electrons. The value of the constant $k$ depends on the assumptions about the energy distribution of the electrons. Since the value of $\tau$ depends on the gas and the field, it is not easy to calculate the $w^-$ from this formula. Figure \ref{fig:drift_vel} shows the drift velocity of electrons as a function of electric field for two different gas mixtures.\\
\subsubsection{Multiplication}
The processes involved in the multiplication can be divided into three categories:\\
\textbf{–Inelastic scattering}\\
When a strong electric field is generated in region of the detector, the kinetic energy of electrons may increase above the average ionization energy $W_i$, allowing the ionization electrons to ionize again. The first Townsend coefficient $\alpha$ is the number of electron-ion pairs produced by one electron per path length:
$$
\alpha = \sigma_i N_m
$$
where $\sigma_i$ is the cross section of the ionization events and $N_m$ is the number of molecules per unit volume. $\sigma_i$   actually depends on the electric field, which is strongly non-uniform in the  GEM holes. Therefore, in order to calculate the gain accurately, it is necessary to perform a complex reconstruction of the electric field and then simulations of the process.\\
The total number of electrons produced by this process is:
$$
N = N_0  e^{\int \alpha(x)d(x)} = N_0  G_f
$$
where $G_f$ is the gain due to the first Townsend coefficient and $N_0$ is the number of
electrons entering the multiplication process.\\
\textbf{–Photoelectric processes}\\
After ionization, atoms may be in an excited state and subsequently emit photons in the UV-visible range, producing additional electrons through the photoelectric effect.\\
The probability with which an electron produces a photoelectron is called the second Townsend coefficient $\gamma$. Taking into account the generation of $N =N_0 G_F$ electrons, they will produce $γN_0 G_F^2$ photoelectrons and so on:
$$
N (x) = N_0 G_F + \gamma N_0 G_F^2 + \gamma^2 N_0 G_F^3 + ... = N_0 G^f \cdot \sum^\infty_{k=0} (\gamma G_F)^\gamma = \frac{N_0 G_F}{1-\gamma G_f} 
$$
If $\gamma G_F \xrightarrow{}1$, the signal becomes independent of primary ionization and enters the Geiger-Muller operation regime. This photoelectric process must also be limited because photons can cause ionization outside the multiplication zone, resulting in spurious signals and degradation of spatial resolution.\\
\textbf{–Penning effect}\\
In a gas mixture, when the metastable excited state of one gas component is energetically higher than the ionization energy of the second gas, the excited gas atoms/molecules ionize the second gas by collisions, effectively increasing the number of electron-ion pairs.
$$
A^* + B \xrightarrow{} A + B^+ + e^-
$$
\subsubsection{Signal induction}
Electrons and ions released in a gaseous counter by ionizing events drift to the anode and cathode, respectively, under the action of the applied electric field. \\
The instantaneous current $i$ generated at an electrode by the charge $q$ moving with velocity $v$ can be calculated using the Shockley-Ramo theorem:
$$
i=q \Bar{v} \cdot \Bar{E}_0
$$
where $\Bar{E}_0$ is the so-called weighting field. An equivalent formulation of the same principle is that the induced charge on the electrode ($Q$) is given by the product of the charge on the carrier ($q$) multiplied by the difference in the weighting potential $\phi_0$ from the beginning to the end of the carrier path:
$$
Q = q \Delta \phi_0
$$
To calculate the weighting potential $\phi_0$,  Laplace's equation for the detector must be solved with some additional boundary conditions:
\begin{itemize}
    \item  The voltage at the electrode for which the induced charge is to be calculated is set equal to one.
    \item The voltage at the other electrodes is set to zero.
    \item The charge in the detector volume is ignored in the calculation.
\end{itemize}
The weighting potential gradient gives the weighting field. The carrier path and velocity must be determined from the actual electric field lines. 

 \chapter{Test setups}
\label{setups}
The software, electronics upgrades, and methods described in this thesis were largely developed and tested using two test setups.
\subsection{Cosmic ray test stand in Beijing}
\label{setup_beijing}
In order to characterize and commission the CGEM-IT prior to its installation, a laboratory was set up in the Chinese Institute for High Energy Physics buildings in Beijing.\\
The setup consists of CGEM-IT layer 1 and layer 2 assembled together and fully instrumented. The system was built in late 2019, a few months before the COVID-19 outbreak.\\
This setup was intended to replicate the conditions of the final installation, so the HV and LV power supplies and the electronics chain were installed as described in chapter \ref{cap:readout_chain}. The cooling water is provided by an SMC-HRS0 chiller\cite{Chiller}.\\
In order to measure the performance of the detector with physical signals, a cosmic ray trigger setup was built. The setup originally consisted of three plastic scintillators and photomultipliers (now reduced to two due to the failure of one photomultiplier). Cosmic rays provide a constant rate of particles, mainly muons with an average energy around \SI{4}{\giga \electronvolt}, essentially MIPs\footnote{A Minimum Ionizing Particle (MIP) is a particle whose average rate of energy loss through matter is close to the minimum. Because of the broad minimum peak of the Bethe-Bloch formula, the energy loss can be assumed to be constant.}, at a rate of about \SI{0.4}{\hertz}. \\
\begin{figure}
    \centering
    \includegraphics[width=1\linewidth]{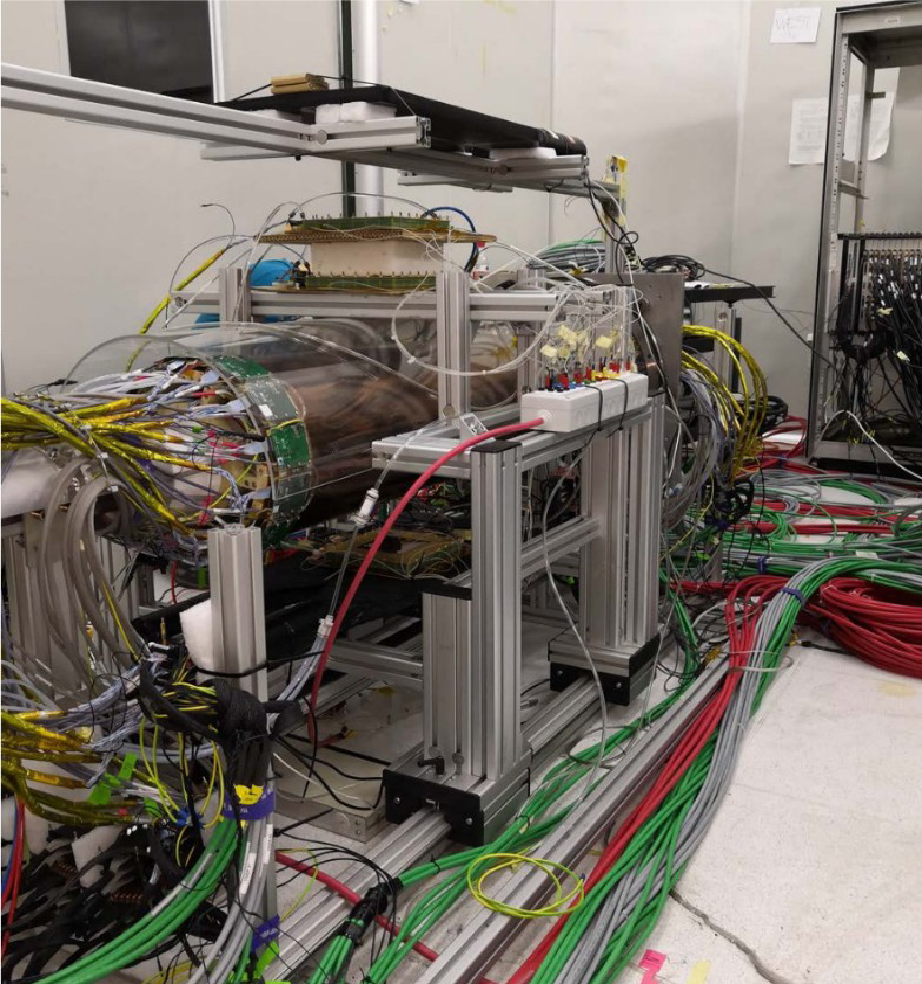}
    \caption{Setup for cosmic rays acquisition with the CGEM-IT at the IHEP laboratory in Beijing. The planar detectors in the picture were sent back to Italy to be used in the setup described in \ref{setup_fe}. }
    \label{fig:setup_photo}
\end{figure}
A picture of the setup can be seen in figure \ref{fig:setup_photo}. Unfortunately, the structure that hosts the layers has a metal pole in the middle. Indeed, the structure was designed to facilitate the insertion of layer 1, not for cosmic rays detection. The effects of multiple scattering on the performance are not negligible and a new structure will be mounted as soon as possible.\\
Due to the outbreak of the pandemic, this setup cannot be controlled onsite. Only absolutely necessary operations are carried out by our Chinese colleagues. The tools described in chapter \ref{cap:software} allow the collaboration to continue to operate the detector and collect data remotely.
\subsection{Test stand with planar GEM detectors}
\label{setup_fe}
Since it was not possible to physically operate on the facility in Beijing, a small test setup with planar detectors was assembled in the laboratory of INFN Ferrara. The setup consists of four 10$\times$10 \si{\centi \meter \squared} triple GEM planar detectors with an active area of 8.32$\times$8.32 \si{\centi \meter \squared} each. These detector have the same spacing between the electrodes of the CGEM-IT, but are read from two adjacent sides with an x,y \SI{90}{\degree} angled layout with 128 strips per view.\\
\begin{figure}
    \centering
    \includegraphics[width=1\linewidth]{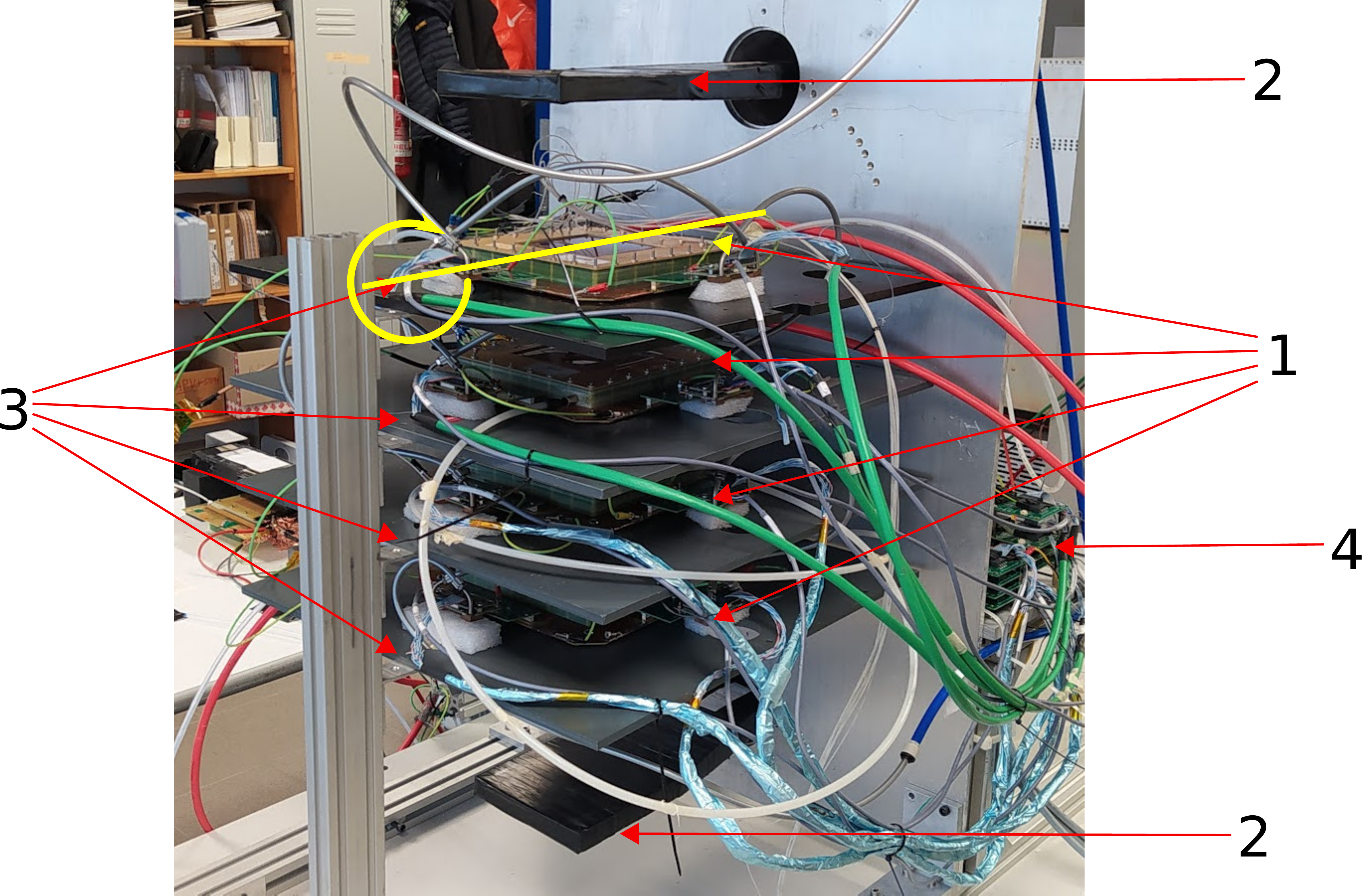}
    \caption{Setup for the cosmic ray acquisition in the laboratory of INFN Ferrara. }
    \label{fig:setup_photo_fe}
\end{figure}
The setup is shown in figure \ref{fig:setup_photo_fe}:
\begin{enumerate}
    \item Planar GEM detector;
    \item Plastic scintillators read by photomultipliers to provide the trigger;
    \item Plastic supports mounted on the aluminium structure. The supports and detector can rotate together around the axes (yellow line) to change the incidence angle of the particles;
    \item Data Low Voltage Patch Cards;
\end{enumerate}
For low voltage, the same devices and structure are used as described in chapter \ref{cap:readout_chain}, while for the high voltage is used a CAEN A1515 board \cite{A1515}, powering two detectors together through the same power supply channels.\\
\begin{figure}
    \centering
    \includegraphics[width=1\linewidth]{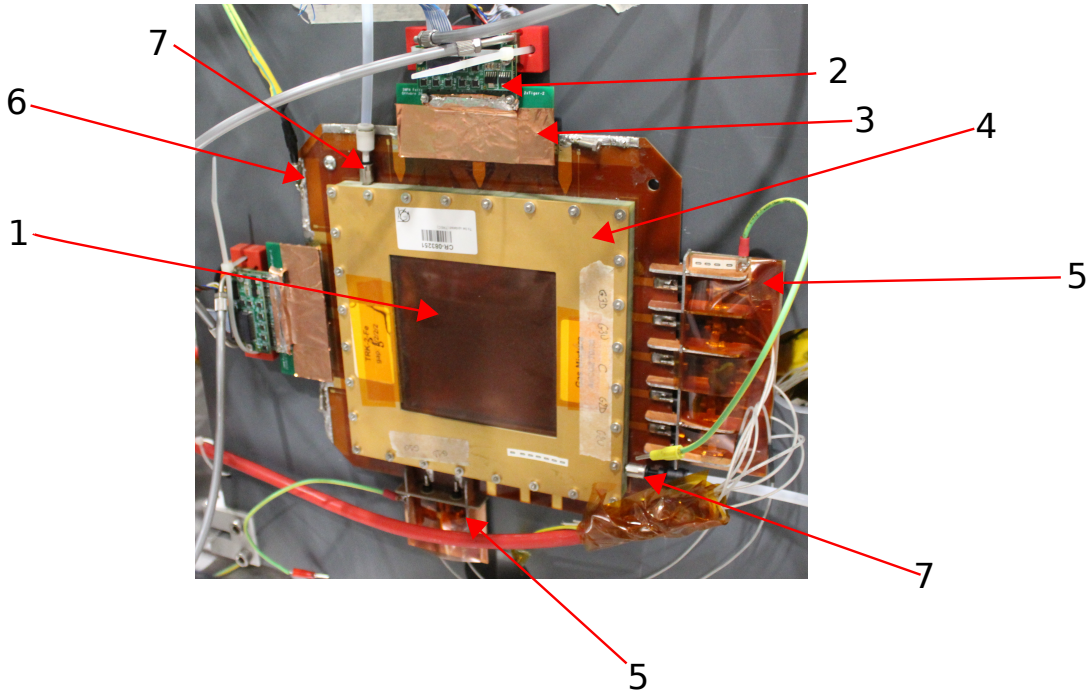}
    \caption{Planar GEM detectors equipped with readout electronics and ready to acquire.}
    \label{fig:planar_photo}
\end{figure}
A picture of a GEM detector equipped with the electronics is shown in figure \ref{fig:planar_photo}:
\begin{enumerate}
    \item Active area of the planar GEM detector; 
    \item Front End Board, mounting two TIGERs for the readout of a detector view; layer 3 FEBs were used because they were not used in the Beijing setup. 
    \item Transition boards to match the FEB connector with the planar connector. The transition board was originally designed to read layer 1 FEBs, but due to their availability, layer 3 FEBs were used. Because of the mapping incompatibility between layer 3 FEBs and transition boards, strips 26,28,30,32 and 34 are not connected. Grounding between the FEB and the detector is critical to reduce the noise levels. The copper foil shown in the picture is soldered to the FEB ground pad and glued on the transition board;
    \item Detector frame;
    \item LEGO HV connection (see below);
    \item Ground reference;
    \item Gas flow connections;
\end{enumerate}
The system is mounted on a mechanical structure provided by the INFN Ferrara mechanical workshop. The structure allows to rotate the detectors and to test the setup at different incidence angles of the particles.\\
The structure was designed to be used during a test beam at CERN in July 2021. The test took place in the H4 experimental area, using a beam of muons or pions. The beams consists in spills of \SI{4.5}{\second}, containing $\sim$ 180k muons or $\sim$ 4M pions reaching the detectors, with a momentum of $\sim$\SI{80}{\giga \electronvolt \per c} for the muons and $\sim$\SI{150}{\giga \electronvolt \per c} for the pions.\\
\begin{figure}
\centering
\begin{subfigure}{.48\textwidth}
  \centering
  \includegraphics[width=0.98\linewidth]{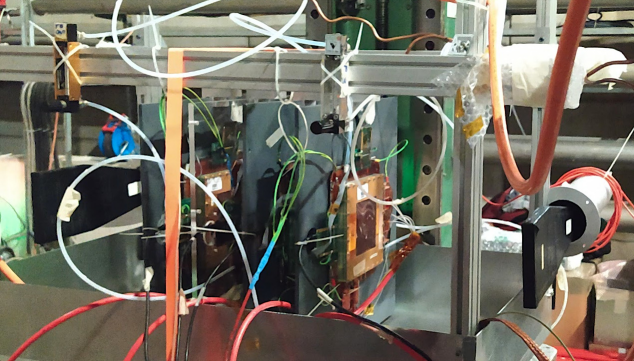}
\end{subfigure}%
\begin{subfigure}{.48\textwidth}
  \centering
  \includegraphics[width=0.98\linewidth]{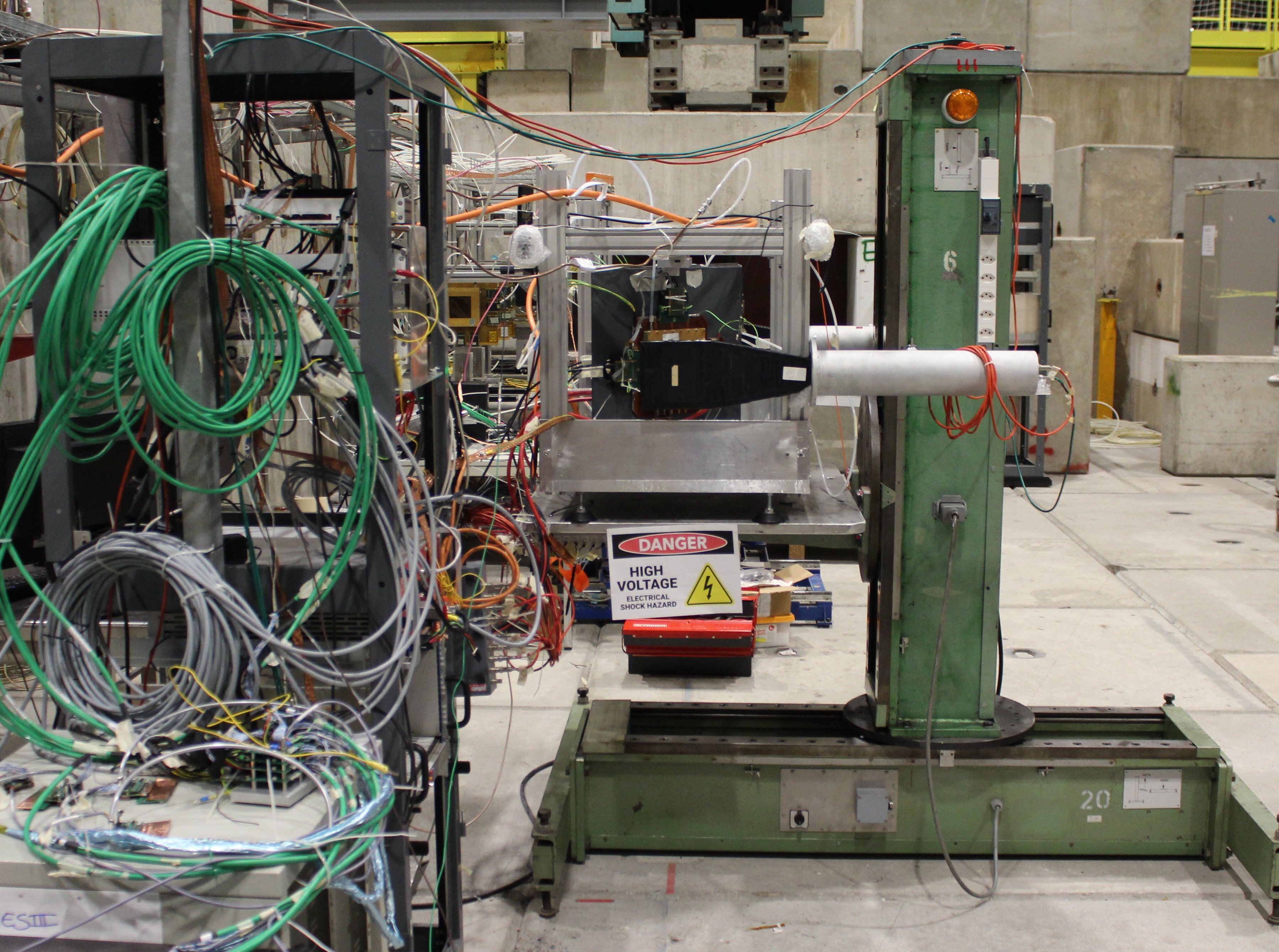}
\end{subfigure}%
  \caption{Pictures of the setup with planar detectors at the CERN test beam.}
\label{fig:TB_pic}
\end{figure}
Pictures of the setup during the test beam are shown in figure \ref{fig:TB_pic} and preliminary results can be found in chapter \ref{cap:res_TB}.\\
Before the test beam, the system was tested with cosmic rays to further advance  the system validation. \\
The test of the noise level (chapter \ref{cap:noise}) suggested some improvements in the HV distribution chain. In particular, the cable shielding was connected to ground and a RC filter system was fitted to all detector electrodes. To mount the filters, provide electrical insulation between the electrodes and facilitate their connection, a mechanical fixture was used, called LEGO\footnote{Designed by Roberto Malaguti (INFN-FE)}. The noise reduction obtained on the planar detectors will lead to testing different HV filter configurations on the CGEM-IT, taking into account the differences between the two detectors.\\ 
The system can be read with either the standard CGEM-IT readout chain (TIGER + GEMROC) or with the APV/SRS system. This readout system consists of the APV25 front end chips \cite{949881} and the ADS5281 ADCs \cite{ads5281}, read by a  SRS slow control system  \cite{Martoiu_2013}. The readout devices can be easily interchanged so that data can be acquired with both to validate the electronic chains.\\
This setup was also the first testbench for the improved signal distribution system (section \ref{sec:FCS}), which consists of the Fast Control system Fanout and Fast Control system Local Fanout.
\chapter{Third-party software and libraries}
\label{cap:third_party}
This appendix lists and describes the third-party software and libraries used in the development of GUFI, CIVETTA, and the other software described in chapter \ref{cap:software}. This description does not claim  to be exhaustive, but is intended to help the reader.
\section*{Python and libraries}
Python \cite{Python} is a widely used and interpreted\footnote{Interpreted =it does not need explicit compilation} high-level programming language. All software tools described in this thesis are written in Python v. 3.7. GUFI was originally written in Python v. 2.7, but it was ported to v. 3.7 when Python v. 2.7 was no longer maintained.\\
\subsubsection*{Tkinter}
Tkinter \cite{tkinter} is the standard Python interface for the Tcl/Tk GUI toolkit \cite{tcltk}. Tk is a free and open-source cross-platform widget toolkit that provides a library of basic GUI widgets elements  for creating a graphical user interface (GUI) in many programming languages. Both Tk and Tkinter are available on most Unix platforms, including macOS, as well as  Windows systems. It has been used to create all the graphical user interfaces presented in this thesis.
\subsubsection*{Pickle}
The pickle \cite{pickle} module implements binary protocols for serializing and de-serializing a Python object structure. This module is often used to store Python objects into a byte stream. A lot of information and configurations used by the presented software are stored in this way.
\subsubsection*{Pandas}
Pandas \cite{pandas} is a Python library for data manipulation and analysis. It provides data structures and tools to manipulate them. In particular, it provides 2-dimensional indexed and labeled data structures with columns of different types called "dataframes". These structures are used to store the decoded and analyzed data acquired from the detectors. Also the Pandas dataframes  are stored on disk using pickles.
\subsubsection*{Scikit-learn}
Scikit-learn \cite{scikit-learn} is a Python machine learning library. It provides various classification, regression and clustering algorithms. In particular, I have used the K-means algorithm for clusterization, and the kernel density estimation to extract the most probable value from distributions.
\subsubsection*{Plotly}
Plotly \cite{plotly} is a library for visualizing data in Python, R and Java. It can be used to create interactive HTML plots or integrated into an interactive "dash" page.
\section*{Other Software}
\subsubsection*{InfluxDB}
InfluxDB \cite{Influx} is an open-source database for time series. I chose it for this project because of its versatility and simplicity. It hosts the data from online monitoring.
\subsubsection*{Grafana}
Grafana \cite{Grafana} is an open-source web application for analytic and interactive visualization. It provides charts, graphs, and alerts when connected to supported data sources, such as the InfluxDB database.\\
\subsubsection*{Dash}
Dash \cite{dash} is a Python framework for building web analytics applications. It allows the creation of interactive pages where the user can select the parameters for the display and the software elaborates the desired plot and presents it in an interactive interface.
\subsubsection*{Gunicorn}
Gunicorn \cite{gunicorn} is a Python Web Server Gateway Interface (WSGI) HTTP server. It is used together with Dash to provide interactive data analysis and visualization through a web page. It has to manage the HTTP requests coming to the server and forward them to the Dash application.
\subsubsection*{Quartus Signal Tap}
Quartus \cite{Quartus} is a programmable logic device design software developed by Intel for its FPGA families. The Signal Tap Logic Analyzer module enables the creation of special memories and logic structures in the firmware project that allow to read the FPGA memory content on the fly after the firmware is flashed into the FPGA. This capability has proven very useful in the process of debugging the GEMROC firmware.


\listoffigures 

\listoftables 
\printbibliography[heading=bibintoc]


\end{document}